\documentclass[PhD]{iitmdiss}
\usepackage{etex}
\usepackage{times}
\usepackage{t1enc}
\usepackage{threeparttable}
\usepackage{graphicx}
\usepackage{epstopdf}
\usepackage{graphics}
\usepackage{hyperref} 
\usepackage{amssymb,amsmath} 
 \usepackage{subfig}
\usepackage{lscape}
\usepackage{epsfig}
\usepackage{indentfirst}
\usepackage{tikz}
\usepackage{multicol}
\usepackage{amsmath,amsthm}
\usepackage{color}
\usepackage{textcomp}
\usepackage{verbatim}
\usepackage[ruled,algonl,vlined]{algorithm2e}
\usepackage{setspace}
\usepackage{enumerate}
\usepackage{mathtools}  
\usepackage{float}
\usepackage{array}
\usepackage{booktabs}
\usepackage{adjustbox}
\usepackage{multirow}
\usepackage{caption}

\makeatletter
\DeclareRobustCommand*\textsubscript[1]{%
	\@textsubscript{\selectfont#1}}
\def\@textsubscript#1{%
	{\m@th\ensuremath{_{\mbox{\fontsize\sf@size\z@#1}}}}}
\makeatother

\DeclareMathOperator{\Tr}{Tr}

\newtheorem{Definition}{Definition}

\newcounter{example}

\usepackage{mathrsfs}
\newcommand\inv[1]{#1\raisebox{1.5ex}{$\scriptscriptstyle-\!1$}}

\usetikzlibrary{arrows,shapes,positioning,shadows,trees}

\tikzset{
   basic/.style  = {draw, text width=7cm, drop shadow, font=\sffamily, rectangle},
   root/.style   = {basic, rounded corners=4pt, thin, align=center,
                    fill=green!50},
   level 2/.style = {basic, rounded corners=6pt, thin,align=center, fill=green!60,
                    text width=12em},
  level 3/.style = {basic, thin, align=left, fill=pink!60, text width=10em}
 }

\usepackage{titlesec}
\newcommand*{\justifyheading}{\raggedleft}
\titleformat{\chapter}[display]
  {\normalfont\huge\bfseries\justifyheading}{\chaptertitlename\ \thechapter}
  {20pt}{\Huge}
 \titleformat{\section}
   {\normalfont\Large\bfseries}{\thesection}{1em}{}
 \titleformat{\subsection}
   {\normalfont\large\bfseries}{\thesubsection}{1em}{}

\usepackage{epigraph}
\begin{document}
  \title{\uppercase{Collaborative Filtering and Multi-Label Classification with Matrix Factorization}}

\author{Vikas Kumar}

\department{COMPUTER AND INFORMATION SCIENCES}

\nocite{*}
\maketitle

\newpage
\certificate
\begin{onehalfspacing}
This is to certify that the thesis entitled \textbf{``Collaborative Filtering and Multi-Label Classification with
Matrix Factorization''} submitted by \textbf{Vikas Kumar} bearing \textbf{Reg. No. 14MCPC07} in partial fulfillment of the requirements for the award of \textbf{Doctor of Philosophy} in \textbf{Computer Science} is a bonafide work carried out by him under our supervision and guidance.
\end{onehalfspacing}

\vspace{-0.3cm}
\begin{onehalfspacing}
This thesis is free from plagiarism and has not been submitted previously in part or in full to this or any other university or institution for the award of any degree or diploma. The student has the following publications before submission of the thesis for adjudication and has produced evidence for the same.
\end{onehalfspacing}
\vspace{-0.3cm}

\begin{enumerate}
 \item "Collaborative Filtering Using Multiple Binary Maximum Margin Matrix Factorizations." \textit{Information Sciences} 380 (2017): 1-11.
 \vspace{-0.22cm}
 \item  "Proximal Maximum Margin Matrix Factorization for Collaborative Filtering." \textit{Pattern Recognition Letters} 86 (2017): 62-67.
 \vspace{-0.22cm}
 \item  "Multi-label Classification Using Hierarchical Embedding." \textit{Expert Systems with Applications} 91 (2018): 263-269.
 \vspace{-0.22cm}
\end{enumerate}
\vspace{-0.5cm}
Further, the student has passed the following courses towards fulfillment of coursework requirement for Ph.D. 
\vspace{-0.3cm}
\begin{table}[ht!]
\centering
 \begin{tabular}{|llcc|}
 \hline
  \textbf{Course Code} & \textbf{Name}&  \textbf{Credits}& \textbf{Pass/Fail} \\ \hline
  \hline 
  CS 801 & Data Structures and Algorithms & 4 & Pass
 \\
  CS 802 & Operating Systems and Programming & 4 & Pass
 \\
  AI 810 & Metaheuristic Techniques & 4 & Pass
 \\
  AI 852 & Learning \& Reasoning & 4 & Pass \\  \hline
 \end{tabular}
\end{table}
\vspace{-0.4cm}
\begin{singlespacing}
	\noindent
	\begin{tabular}{lc}
		\begin{minipage}{0.6\textwidth}
		\vspace{0.4cm}
			\textbf{Prof. Arun K Pujari}\\
			Supervisor\\
		\end{minipage}
		&
		\begin{minipage}{0.6\textwidth}
		\vspace{0.4cm}
			\textbf{Prof. Vineet Padmanabhan}\\
			Supervisor\\
		\end{minipage}\\
		\multicolumn{2}{l}{
			\begin{minipage}{0.9\textwidth}
			\vspace{0.8cm} 
			   \centering
				\textbf{Prof. K. Narayana Murthy}\\
				Dean, School of Computer and Information Sciences
		\end{minipage}}
	\end{tabular}
\end{singlespacing}%

\declaration
I, \textbf{Vikas Kumar}, hereby declare that this thesis entitled \textbf{``Collaborative Filtering and Multi-Label Classification with Matrix Factorization''} submitted by me under the guidance and supervision of \textbf{Prof. Arun K Pujari} and \textbf{Prof. Vineet
Padmanabhan} is a bonafide research work. I also declare that it has not been submitted previously in part or in full to this university or any other university or institution for the award of any degree or diploma.

\vspace*{0.5in}
\begin{singlespacing}
	\hspace*{-0.25in}
	\parbox{5.7in}{
		\noindent {Date: }
	}
\end{singlespacing}

\vspace*{-0.35in}
\hspace*{2.6in}
Name: {\bf ~~~Vikas Kumar}\\
\vspace*{-0.35in}
\hspace*{2.8in}
Signature of the Student:

\vspace*{0.1in}
\begin{singlespacing}
\hspace*{2.6in}
\parbox{2.0in}{
Reg. No. {\bf 14MCPC07}
}
\end{singlespacing}

\newpage
\begin{center}
    \thispagestyle{empty}
    \vspace*{2.5in}
    \Large
 \emph{Dedicated to my parents and family members, for everything what I am today and above all, devoted to The Almighty God!}   
\end{center}

  \newpage
\acknowledgements
\vspace{-1cm}
The accomplishment of doctoral thesis and degree is a result of collective support and blessings of various people during my journey of Ph.D. I feel overwhelm and pleasure to acknowledge each one for the kind of contribution provided to me throughout my life. 

The foremost is my advisor, \textbf{Prof. Arun K Pujari} who has not only taught me science but the facts behind and beyond science. Sir has endowed me with a sacred vision to pursue productive full time research in computer science. The importance of bearing an optimistic attitude and crystal clear concept of the subject is a boon conferred by sir towards shaping my research questions during my years of Ph.D. I am extremely obliged to sir for making himself always available for rigorous scientific discussions in spite of his busy schedule. Those insightful inputs by sir are the substantial asset to build the carrier further. I wish to acknowledge my sir for being the guiding light in my dark days of research. It is also a privilege to express my sincere regards to \textbf{Mrs. Subha Lakhmi Pujari} for her care, encouragement and responsible assistance.

I sincerely gratify my other advisor, \textbf{Prof. Vineet Padmanabhan} for being both amicable and friendly supervisor with whom I was able to share the days of being a research scholar. The periodical scientific counselling and supervision when I was at the initial phase of my Ph.D. carrier is accountably assisted by sir. I genuinely acknowledge sir for taking out his valuable time for undertaking detailed analysis of the given research problem and giving me precise comments.  I thank \textbf{Mrs. Mrinalini Vineet} for her truthful blessings and generous wishes to accomplish success in my research.

I would like to thank my DRC members, \textbf{Prof. Arun Agarwal} and \textbf{Prof.\linebreak Chakravarthy Bhagvati}, for their valuable suggestions.  I extend admiration to my teachers of Pondicherry University, \textbf{Dr. K. S. Kuppusamy} and \textbf{Prof. G. Aghila} for their inspiration to choose computer science as a research field. I thank my primary mentor, \textbf{Dr. Ajit Kumar}, for guiding and encouraging my talent. He is the person who has introduced me to the computer science research related examination and hence to opt for research carrier. I am thankful to \textbf{Mr. Sudarshan} for teaching me the importace of education in life.

I would like to honestly praise my colleagues, \textbf{Mr. Venkat}, \textbf{Mr. Sandeep}, and \textbf{Mrs. Sowmini} for building an enthusiastic environment and engaging in critical discussions related to research in both inside and outside the lab. I thank all my Ph.D. classmates for showering the fun times in hostel days stay. I am thankful to my friends, \textbf{Mr. Ashutosh}, \textbf{Mrs. Surabhi} and \textbf{Mr. Sandeep}, for their considerate friendship. They have always been appreciating the passion of my profession and unboundedly keeping in touch during my bad times. I thank my best junior buddies, \textbf{Mr. Sudhanshu} and \textbf{Miss Prabha} for encouraging me whenever some unfavourable situation demands. 

I extend deepest salutation to my whole family members for availing me every needful facilities and emotional assistance. Their support imprints me the property to be flexible as water and rigid as rock to balance my twisted research life. I also thank \textbf{Miss Purnima} for always being a lovable and indulgent listener of my problems.  She understands me and tries to be an instant saviour in every possible way.

I wish to express that I am indebted and owe so much to my mother, \textbf{Mrs. Govinda Devi} and father, \textbf{Mr. Baban Pandey} such that my depth of expressing acknowledgement will always be small when compare to their contributions of parenting. The unconditional sacrifices and determined motivation is what I have learnt from them at every stage of my life and this has enforced me to become a responsible citizen at first place. With an immense pride this doctoral dissertation is dedicated to my parents for providing me the strong roots to standalone and sustain in such long journey of life. 

Finally, I bow to The Almighty for being a source of unknown power and strength in my life.

\vspace{1cm}
\begin{flushright}
Vikas Kumar
\end{flushright}

\newpage
\abstract
Machine learning techniques for Recommendation System and Classification has become a prime focus of research to tackle the problem of information overload. Recommender Systems are software tools that aim at making informed decisions about the services that a user may like. Recommender Systems can be broadly classified into two categories namely, \emph{content-based filtering} and \emph{collaborative filtering}. In content-based filtering, users and items are represented using a set of features (profile) and an item is recommended by finding the similarity between the user and item profile. On the other hand in collaborative filtering, the user-item association is obtained based on the preferences of the user given so far and the preference information of other users. 

Classification technique deals with the categorization of a data object into one of the several predefined classes. The majority of the methods for supervised machine learning proceeds from a formal setting in which data objects (instances) are represented in the form of feature vectors wherein each object is associated with a unique class label from a set of disjoint class labels. Depending on the total number of disjoint classes, a learning task is categorized as binary classification or multi-class classification. In the multi-label classification problem, unlike the traditional multi-class classification setting,  each instance can be simultaneously associated with a subset of labels. 

In both recommendation and classification problem, the initial assumption is that the input-data is in the form of matrices which are inherently low rank. The key technical challenge involved in designing new algorithms for recommendation and classification is dependent on how well one can handle the huge and sparse matrices which usually has thousands to millions of rows and which are usually noisy. Recent years have witnessed extensive applications of low-rank linear factor model for exploiting the complex relationships that exist in such data matrices. The goal is to learn a low-dimensional embedding where the data object can be represented with a small number of features. Matrix factorization methods have attracted significant attention for learning the low-rank latent factors. The focus of thesis is on the development of novel techniques for collaborative filtering and multi-label classification. 

In maximum margin matrix factorization scheme of collaborative filtering, ratings matrix with multiple discrete values is treated by specially extending hinge loss function to suit multiple levels. We view this process as analogous to extending two-class classifier to a unified multi-class classifier. Alternatively, multi-class classifier can be built by arranging multiple two- class classifiers in a hierarchical manner. We investigate this aspect for collaborative filtering and propose a novel method of constructing a hierarchical bi-level maximum margin matrix factorization to handle matrix completion of ordinal rating matrix. The advantages of the proposed method over other matrix factorization based collaborative filtering methods are given by detailed experimental analysis. We also observe that there could be several possible alternative criteria to formulate the factorization problem of discrete ordinal rating matrix, other than the maximum margin criterion. Taking the cue from the alternative formulation of support vector machines, a novel loss function is derived by considering \textit{proximity} as an alternative criterion instead of margin maximization criterion for matrix factorization framework. We validate our hypothesis by conducting experiments on real and synthetic datasets. 

We extended the concept of matrix factorization for yet another important problem of machine learning namely multi-label classification which deals with the classification of data with multiple labels. We propose a novel piecewise-linear embedding method with a low-rank constraint on parametrization to capture nonlinear intrinsic relationships that exist in the original feature and label space. Extensive comparative studies to validate the effectiveness of the proposed method against the state-of-the-art multi-label learning approaches is given through detailed experimental analysis. We also study the embedding of labels together with the group information with an objective to build an efficient multi-label classifier. We assume the existence of a low-dimensional space onto which the feature  vectors and label vectors can be embedded. We ensure that labels belonging to the same group share the same sparsity pattern in their low-rank representations. We perform comparative analysis which manifests the superiority of our proposed method over state-of-art algorithms for multi-label learning.%

%
%
 \begin{singlespace}
  \tableofcontents
  \thispagestyle{empty}
  
 \listoftables
  \addcontentsline{toc}{chapter}{LIST OF TABLES}
  \listoffigures
  \addcontentsline{toc}{chapter}{LIST OF FIGURES}
  \end{singlespace}
%
\abbreviations

  \noindent
  \begin{tabbing}
  xxxxxxxxxxx \= xxxxxxxxxxxxxxxxxxxxxxxxxxxxxxxxxxxxxxxxxxxxxxxx \kill
   \textbf{CF} \> Collaborative Filtering \\
   \textbf{FE} \>  Feature Space Embedding \\ 
   \textbf{GroPLE} \>  Group Preserving Label Embedding for Multi-Label Classification \\
   \textbf{HMF} \> Hierarchical Matrix Factorization  \\
   \textbf{LE} \>  Label Space Embedding \\
   \textbf{MF} \> Matrix Factorization \\   
   \textbf{MLC} \> Multi-label Classification  \\   
   \textbf{MLC-HMF} \>  Multi-label Classification Using Hierarchical Embedding  \\
   \textbf{MMMF} \> Maximum Margin Matrix Factorization  \\   
   \textbf{PMMMF} \> Proximal Maximum Margin Matrix Factorization  \\
  \end{tabbing}
  
  \pagebreak
  \pagenumbering{arabic}

  \newpage
\thispagestyle{empty}
\chapter{Introduction}
\label{chap:intro}

With the advent of technology and high usage of modern equipment/ devices, large amounts of data are being generated world over. We now stand at the brink of data-driven transformation where the data can be harnessed to derive meaningful insights that can help organizations for better functioning as well as assist the user in his/ her decision making process. Examples include (a) suggesting the right choice of products to a user  or identifying the most appropriate customers for a product (b) assisting an organization in classifying the users for better serviceability etc.

The vast amount of accumulated data is a crucial competitive asset and can be tailored to satisfy an individual's/ organization's needs. However, the crucial challenge here is in the processing of the data. The amount of data is enormous and therefore requires a large amount of time to explore all of them. For example, if one has to purchase an item, he/ she needs to process all the information to select which items meet their needs. To avoid the information overload, we need some assistance in the form of recommendation or classification in our day to day life. For instance, when we are purchasing an electronic gadget, we usually rely on the suggestions of friends who shares similar tastes. Similarly, when we are planning to invest in equity funds, we need some assistance to differentiate between different equity based on their features before we decide to invest. In the recent past, machine learning techniques are most commonly used to understand the nature of data so as to help in reducing the burden on individuals in the decision making process. Machine learning techniques for \emph{Recommendation} and \emph{Classification} has become a prime focus of research to tackle the problem of information overload.

Recommendation (Recommender) Systems are software tools that aim at making informed decisions about the services that a user may like~\cite{sarwar2002incremental,koren2011advances,lu2015recommender}. Given a list of users, items and user-item interactions (ratings), Recommender System predicts the score/ affinity of item $j$ for user $i$ and thereby helps in understanding the user behaviour which in turn can be used to make personalized recommendations. Examples include classic recommendation tasks such as recommending books, movies, music etc., as well as new web-based applications such as predicting preferred news articles, websites etc. Recommender Systems can be broadly classified into two categories namely, content-based filtering and collaborative filtering\cite{lu2015recommender}. In content-based filtering, users and items are represented using a set of features (profile) and an  item is recommended by finding the similarity between the user and the item profile. In the case of collaborative filtering the user-item association is obtained based on the preferences of the user given so far and the preference information of other users. 

Classification is a supervised learning technique which deals with the categorization of a data object into one of the several predefined classes. Majority of the methods for supervised machine learning proceeds from a formal setting in which the data objects (instances) are represented in the form of feature vectors wherein each object is associated with a unique class label from a set of disjoint class labels $L$, $|L| > 1$. Depending on the total number of disjoint classes in $L$, a learning task is categorized as binary classification (when $|L| = 2$) or multi-class classification (when $|L| > 2$)~\cite{read2009classifier,tsoumakas2007random}. In this thesis, we are focusing on a special class of classification problem called multi-label classification. Unlike the traditional classification setting, in multi-label classification problem, each instance can be simultaneously associated with a subset of labels. For example, in image classification, an image can be simultaneously tagged with a subset of labels such as \textit{natural}, \textit{valley}, \textit{mountain} etc. Similarly, in document classification, a document can simultaneously belong to \textit{Computer Science} and \textit{Physics}. The multi-label classification task aims at building a model that can automatically tag a new example with the most relevant subset of class labels.

In this thesis we focus on developing novel techniques for collaborative filtering and multi-label classification. It should be kept in mind that in both recommendation and classification problem the data is actually organized in matrix form. For example, in collaborative filtering, users preferences on items can be represented as a matrix, whose rows represent users, columns represent items, and each element of the matrix represent the preference of a user for an item.  Similarly, in multi-label classification problem, the set of data objects and their corresponding label vectors can be represented as a matrix. The data object can be represented as a row of the feature matrix and the associated label vector can be represented as the corresponding row of label matrix.  The key technical challenge involved in designing new algorithms for recommendation and classification is dependent on how well one can handle the huge and sparse matrices which usually has thousands and millions of rows and which are usually noisy.

Recent years have witnessed extensive applications of low-rank linear factor models for exploiting the complex relationships existing in such data matrices. The goal is to learn a low-dimensional embedding where the data object can be represented with a small number of features. For instance, in the case of collaborative filtering, the idea is to learn low-dimensional latent factors for every user and item. Similarly, in multi-label classification the goal of low-rank factor model learning is to embed the feature and label vector to a low dimensional space so that the intrinsic relationship in the original space can be captured. Matrix factorization (MF) methods have attracted significant attention in the areas of computer vision~\cite{cao2015low,cao2016robust}, pattern recognition~\cite{wang2015subspace,zhang2015ensemble}, image processing~\cite{fevotte2015nonlinear,haeffele2014structured}, information retrieval~\cite{kim2015simultaneous,tang2016supervised} and signal processing~\cite{vu2016combining,wang2014discriminative} for learning low-rank latent factor models. The objective of MF is to learn low-rank latent factor matrices $U$ and $V$ so as to simultaneously approximate the observed entries under some loss measure. Here, the interpretation of factor matrices $U$ and $V$ are application dependent. We present a brief discussion on matrix factorization approach for collaborative filtering and multi-label classification. 

Matrix factorization is just one way of doing collaborative filtering (CF) wherein it is possible to discover the latent features underlying the interactions between two different kinds of entities (user/ item). CF approaches also assume that a user's preference on an item is determined by a small number of factors and how each of those factors applies to the user and the item (low dimensional linear factor model). The intuition behind using matrix factorization is that there should be some latent features that determine how a user rates an item. It is possible that two users would highly rate a movie if it belongs to a particular genre like action/ comedy. Discovering these latent features definitely helps in predicting the rating with respect to a certain user and a certain item because the features associated with the user should match with the features associated with the item. It is also the case that in trying to discover the different features, an assumption is made that the number of features would be smaller than the number of users and the number of items. The idea is that it is not reasonable to assume that each user is associated with a unique feature in which case it would be absurd to make any recommendations because each of these users would not be interested in the items rated by other users.

The discussion above shows that CF can be formulated as a MF problem in the sense that, in a $d$-factor model, given a rating matrix $Y \in \mathbb{R}^{N\times M} $ the idea is to find two low-rank matrices $U\in \mathbb{R}^{N\times d}$ and $V \in \mathbb{R}^{M\times d}$ such that $Y \approx UV^{T}$ where $d$ is a parameter. Each entry $y_{ij} \in \{ 0,1,2,\dots,R\}$ in $Y$, defines the preference given by $i$th user for $j$th item. $y_{ij} = 0$, indicates that the user has not given any preference for $j$th item and $R$ is the total level of ratings. The goal is to predict the preference for all items for which the user's preference is unobserved.  The $i$th row of $U$ represents the latent factor of the  $i$th user and the $j$th row of $V$ represents the latent feature of the $j$th item.  The prediction for $y_{ij}$ can be achieved by the linear combination of the $i$th row of $U$ with the $j$th row of $V$ i.e.
\begin{equation}
 y_{ij} = \sum_{p=1}^{d} U_{ip}V_{jp} =  U_{i}V^{T}_{j}
\end{equation} 
The outcome of matrix factorization method is a low-rank, dense matrix $X$ which is an approximation of the given sparse rating matrix Y. Here X is computed as $ X = UV^{T}$. 

Multi-label classification can be seen as a generalization of \textit{single label} classification where an instance is associated with a unique class label from a set of disjoint labels $L$. Formally, given $N$  training examples in the form of a pair of feature matrix  $X$ and  label matrix $Y$ where each example $x_i \in \mathbb{R}^D, 1\le i \le N$, is a row of $X$ and its associated labels $y_i \in  \{-1,+1\}^L$ is the corresponding row of $Y$, the task of multi-label classification is to learn a parameterization $h : \mathbb{R}^D \rightarrow \{-1,+1\}^L$ that maps each instance to a set of labels. Multi-label classification has applications in many areas such as machine learning ~\cite{babbar2017dismec,yen2016pd,read2009classifier,zhang2006multilabel}, computer vision ~\cite{liu2015low,wang2016cnn,cabral2011matrix,boutell2004learning} and data mining ~\cite{nam2014large,zhou2014activity,tsoumakas2007random,schapire2000boostexter}. Multi-label classification is nowadays applied to massive data sets of considerable size and under such conditions, time- and space-efficient implementations of learning algorithms is of major concern. For example, there are millions of articles available on Wikipedia each tagged with a set of labels and in most of the cases, the author of an article associates very few labels which they know from a broad set of relevant labels.   

To cope with the challenge of exponential-sized output space, exploiting intrinsic information in feature and label spaces has been the major thrust of research in recent years and use of parametrization and embedding have been the prime focus. Researchers have studied several aspects of embedding which include label embedding, feature embedding, dimensionality reduction and feature selection. These approaches differ from one another in their capability to capture other intrinsic properties such as label correlation and local invariance. To glean out the potentially hidden dynamics that exists in the original space, most of the embedding based approaches focus on learning a low-dimensional representation for the original feature (label) vector. There are two strategies of embedding for exploiting inter-label correlation; (1) Feature Space Embedding (FE); and (2) Label Space Embedding (LE). FE aims to design a projection function which can map the instance in the original feature space to an embedded space and then a mapping is learnt from the embedded space to the label space. The second approach is to transform the label vectors to an embedded space, followed by the association between feature vectors and embedded label space for classification purpose.

In recent years, matrix factorization based approach which aims at determining two matrices $U$ and $V$ is frequently used to achieve the FE and LE. In FE, the goal is to transform each $D$-dimensional feature vector (a row of matrix $X$) from the original feature space to a $L$-dimensional label vector (corresponding row in $Y$) and usually the mapping is achieved through a linear transformation matrix $W \in \mathbb{R}^{D \times L}$. The MF based approach assumes that the label matrix $Y$ is of low-rank due to the presence of similar labels and thereby models the inter-label correlation implicitly using low-rank constraints on the transformation matrix $W$. The transformation matrix $W$ is approximated using the product of two latent factor matrices $U$ and $V$. The matrix $U$ acts as a transformation matrix which transforms the data from the original feature space to an embedded space and the matrix $V$ can be interpreted as the decoding matrix from the embedded space to the label space. 

In LE, the goal is to transform each $L$-dimensional label vector (a row of matrix $Y$) from the original label space to a $d$-dimensional embedded vector. Thereafter, a predictive model is trained from the original feature space to the embedded space. With proper decoding process that maps the projected data back to the original label space, the task of multi-label prediction is achieved. The LE approach based on MF aims at approximating the label matrix $Y$ as a product of two matrices $U$ and $V$ with the assumption that the label matrix exhibits a low-rank structure. The $i$th row of matrix $U$ can be viewed as the embedded representation of the label vector associated with the $i$th instance and the matrix $V$ can be viewed as the decoding matrix from the embedded space to the label space. 

\section{Contributions of the Thesis}
The major contributions of the thesis are as follows. In maximum margin matrix factorization scheme (a variant of basic matrix factorization method), ratings matrix with multiple discrete values is treated by specially extending \emph{hinge loss function} to suit multiple levels. We view this process as analogous to extending two-class classifier to a unified multi-class classifier. Alternatively, multi-class classifier can be built by arranging multiple two- class classifiers in a hierarchical manner. We investigate this aspect for collaborative filtering and propose a novel method of constructing a hierarchical bi-level maximum margin matrix factorization to handle matrix completion of ordinal rating matrix~\cite{KUMAR20171}. 

We observe that there could be several possible alternative criteria to formulate the factorization problem of discrete ordinal rating matrix, other than the maximum margin criterion. Taking a cue from the alternative formulation of support vector machines, a novel loss function is derived by considering \textit{proximity} as an alternative criterion instead of margin maximization criterion for matrix factorization framework~\cite{KUMAR201762}.

We extended the concept of matrix factorization for yet another important problem of machine learning namely multi-label classification which deals with the classification of data with multiple labels. We propose a novel piecewise-linear embedding method with a low-rank constraint on the parametrization to capture nonlinear intrinsic relationships that exist in the original feature and label space~\cite{KUMAR2018263}.

We study the embedding of labels together with the group information with an objective to build an efficient multi-label classifier. We assume the existence of a low-dimensional space onto which the feature  vectors and label vectors can be embedded. We ensure that labels belonging to the same group share the same sparsity pattern in their low-rank representations.

\section{Structure of the Thesis}
The thesis is organized as follows. In Chapter $2$, 
we start our discussion with an introductory discussion on matrix factorization and the associated optimization problem formulation. Thereafter we discuss some common loss functions used in matrix factorization to measure the deviation between the observed data and the corresponding approximation. We also discuss several ways of norm regularization which is needed to avoid overfitting in matrix factorization models. In the later part of the chapter we discuss at length the application of matrix of factorization techniques in collaborative filtering and multi-label classification.

Chapter $3$ starts with a discussion on bi-level maximum margin matrix factorization (MMMF) which we subsequently use in our proposed algorithm. We carry out a deep investigation of the well known maximum margin matrix factorization technique for discrete ordinal rating matrix. This investigation led us to propose a novel and efficient algorithm called HMF (\textit{Hierarchical Matrix Factorization}) for constructing a hierarchical bi-level maximum margin matrix factorization method to handle matrix completion of ordinal rating matrix.  The advantages of HMF over other matrix factorization based collaborative filtering methods are given by detailed experimental analysis at the end of the chapter.

Chapter $4$ introduces a novel method termed as PMMMF (\textit{Proximal Maximum Margin Matrix Factorization}) for factorization of matrix with discrete ordinal ratings. Our work is motivated by the notion of Proximal SVMs (PSVMs)~\cite{li2015multia,Fung2001} for binary classification where two \emph{parallel} planes are generated,  one for each class, unlike the standard SVMs~\cite{vapnik2013nature,burges1998tutorial,mangasarian1999generalized}. Taking the cue from here, we make an attempt to introduce a new loss function based on the \textit{proximity} criterion instead of margin maximization criterion in the context of matrix factorization. We validate our hypothesis by conducting experiments on real and synthetic datasets. 

Chapter $5$ extended the concept of matrix factorization for yet another important problem in machine learning namely multi-label classification. We visualize matrix factorization as a kind of low-dimensional embedding of the data which can be practically relevant when a matrix is viewed as a transformation of data from one space to the other. At the beginning of the chapter we discuss briefly about the traditional approach of multi-label classification and establish a bridge between multi-label classification and matrix factorization. We present a novel multi-label classification method, called MLC-HMF (\textit{Multi-label Classification using Hierarchical Embedding}), which learns  piecewise-linear embedding with a low-rank constraint on parametrization to capture nonlinear intrinsic relationships that exist in the original feature and label space. Extensive comparative studies to validate the effectiveness of the proposed method against the state-of-the-art multi-label learning approaches is discussed in the experimental section.

In Chapter $6$, we study the embedding of labels together with group information with an objective to build an efficient multi-label classifier. We assume the existence of a  low-dimensional space onto which the feature vectors and label vectors can be embedded. In order to achieve this, we address three sub-problems namely; (1) Identification of groups of labels; (2) Embedding of label vectors to a low rank-space so that the sparsity characteristic of individual groups remains invariant; and (3) Determining a linear mapping that embeds the feature vectors onto the same set of points, as in stage 2, in the low-rank space. At the end, we perform comparative analysis which manifests the superiority of our proposed method over state-of-art algorithms for multi-label learning. 

We conclude the thesis with a discussion on future directions in Chapter $7$.

\section{Publications of the Thesis}
\begin{enumerate}
\item Vikas Kumar, Arun K Pujari, Sandeep Kumar Sahu, Venkateswara Rao Kagita, Vineet Padmanabhan. "Collaborative Filtering Using Multiple Binary Maximum Margin Matrix Factorizations." Information Sciences 380 (2017): 1-11.
\item Vikas Kumar, Arun K Pujari, Sandeep Kumar Sahu, Venkateswara Rao Kagita, Vineet Padmanabhan. "Proximal Maximum Margin Matrix Factorization for Collaborative Filtering." Pattern Recognition Letters 86 (2017): 62-67.
\item Vikas Kumar, Arun K Pujari, Vineet Padmanabhan, Sandeep Kumar Sahu,\linebreak Venkateswara Rao Kagita. "Multi-label Classification Using Hierarchical Embedding." Expert Systems with Applications 91 (2018): 263-269.
\item Vikas Kumar, Arun K Pujari, Vineet Padmanabhan, Venkateswara Rao Kagita. "Group Preserving Label Embedding for Multi-Label Classification." Pattern\linebreak Recognition,  Under Review.
\end{enumerate}

  \newpage
\thispagestyle{empty}
\setcounter{chapter}{1}
\chapter{Foundational Concepts}
\label{relatedWorkVikas} 
In majority of data-analysis tasks, the datasets are naturally organized in matrix form. For example, in collaborative filtering, users preferences on items can be represented as a matrix, whose rows represent users, columns represent items, and each element of the matrix represent the preference of a user for an item.   In a clustering problem, a row of the matrix represents data object (instance) and the columns represent associated features (attributes). In image inpainting problem, an image can be represented by a matrix where each entry of the matrix corresponds to pixel values. Similarly, in document analysis, a set of document can be represented as a matrix wherein the rows represent terms and the columns represent documents. Each entry represents the frequency of the associated terms in a particular document.  

Recent advances in many data-analysis tasks have been focused on finding a meaningful (simplified) representation of the original data matrix. A simplified representation typically helps in better understanding the structure, relationship within the data or attributes and retrieving the hidden information that  exists in the original data matrix. Matrix Factorization (MF) methods have attracted significant attention for finding the latent (hidden) structure from the original data matrix. Given an original matrix $Y \in \mathbb{R}^{N \times M}$,  MF aims at determining two  matrices $U\in \mathbb{R}^{N\times d}$ and $V\in \mathbb{R}^{M\times d}$ such that $Y \approx UV^{T}$ where the inner dimension $d$ is called the \textit{numerical rank} of the matrix. The numerical rank is much smaller than $M$ and $N$, and hence, factorization allows the matrix to be stored inexpensively. Here, the interpretation of factor matrices $U$ and $V$ are application dependent. For example, in collaborative filtering, given that $Y$ is a $N \times M$ size user-item rating matrix, MF aims at determining the latent factor representation of users and items. The $i$th row of $U$ represents the latent factor of the  $i$th user and the $j$th row of $V$ represents the latent feature of the $j$th item~\cite{koren2009matrix,rennie2005fast,mnih2007probabilistic,koren2008factorization,salakhutdinov2007restricted}. In a clustering problem with $M$ data points (columns of $Y$) and $d$ cluster,  the $j$th column of $U$ can be interpreted as the representative of the $j$th cluster and $i$th column of $V^T$ can be visualized as the reconstruction weights of the $i$th data point from those representatives~\cite{eriksson2012high,xu2003document,liu2013multi,wang2008semi}. 

The MF field is too vast to cover and there are many research directions one can pursue starting from theoretical aspects to application-oriented aspects. In this thesis, we focus our discussion on two special applications of matrix factorization, namely collaborative filtering~\cite{koren2009matrix,rennie2005fast,mnih2007probabilistic,koren2008factorization,salakhutdinov2007restricted} and multi-label classification~\cite{cabral2011matrix,wicker2012multi,balasubramanian2012landmark,jing2015semi}. The purpose of this chapter is to introduce the basic matrix factorization framework and concepts which will be used throughout the thesis (Section~\ref{matrixFactorizationRW}). In Section~\ref{lossFunctionInMF}, we discuss some of the common loss functions used in MF to measure the deviation between the observed data and corresponding approximation. We briefly discuss the different ways to avoid overfitting in MF model in Section~\ref{regularizationInMF}. In Section~\ref{CFViaMF} and~\ref{MLCViaMF}, we explore how matrix factorization can be applied to interesting applications of machine learning namely collaborative filtering and multi-label classification, respectively. 

\section{Matrix Factorization}
\label{matrixFactorizationRW}
Matrix factorization is a key technique employed for many real-world applications wherein the objective is to select $U$ and $V$ among all the matrices that fit the given data and the one with the lowest rank is preferred.  In most applications, the task that needs to be performed is not just to compute any factorization, but also to enforce additional constraints on the factors $U$ and $V$. The matrix factorization problem can be formally defined as follows:
\begin{Definition}[Matrix Factorization]
Given a data matrix $Y\in\mathbb{R}^{N\times M}$, matrix factorization aims at determining two  matrices $U\in \mathbb{R}^{N\times d}$ and $V\in \mathbb{R}^{M\times d}$ such that $Y \approx UV^{T}$ where the inner dimension $d$ is called the \textit{numerical rank} of the matrix. The numerical rank is much smaller than  $M$ and $N$, and hence, factorization allows the matrix to be stored inexpensively. $ X = UV^{T} $ is called as low-rank approximation of $Y$.
\end{Definition}
\vspace{2cm}
A common formulation for this problem is a regularized loss problem which can be given  as follows.
\begin{equation}\label{basic-fact-formulation_RW}
\underset{U, V}{min} \; \ell(Y, U, V) + \lambda R(U, V)
\end{equation}
where $Y$ is the data matrix, $X= UV^T$ is its low-rank approximation, $\ell(\cdot)$ is a loss function that measures how well $X$ approximates $Y$, and $R(\cdot)$ is a regularization function that promotes various desired properties in $X$ (low-rank, sparsity, group-sparsity, etc.). When $ \ell(\cdot)$ and $R(\cdot)$ are convex functions of $X$ (equivalently, of $U$ and $V$)  the above problem can be solved efficiently. The loss function $\ell(Y,U,V)$ can further be decmposed into the sum of pairwise discrepancy between the observed entries and their corresponding prediction, that is, $\ell(Y,U,V) = \sum_{i=1}^{N}{\sum_{j=1}^{M} \ell(y_{ij},U_i, V_j)}$, where $y_{ij}$ is the $ij$th entry of the matrix $Y$ and $U_i$, the $i$th row of $U$, represent the latent feature vector of $i$th user and $V_j$, the $j$th row of $V$, represent the latent feature vector of $j$th item. There have been umpteen number of proposals for factorizing a matrix and these differ, by and large, among themselves in defining the loss function and the regularization function. The structure of the latent factors depends on the types of loss (cost) functions and the constraints imposed on the factor matrices. In the following subsections, we present a brief literature review of the major loss functions and regularizations.

\section{Loss Function}
\label{lossFunctionInMF}
Loss function $\ell: Y \times Y \rightarrow \mathbb{R}$ in MF model is used to measure the discrepancy between the observed entry and the corresponding prediction. Based on the types of observed data, the loss function in MF can be roughly grouped into three categories: (1) Loss function for Binary data; (2) Loss function for Discrete ordinal data; and (3) Loss function for Real-valued data.

\subsection{Binary Loss Function:} For a binary data matrix $Y = [y_{ij}]$ where the entries are restricted to take only two values $y_{ij} \in \{+1, -1\}$, we review some of the important loss functions such as \textit{zero-one loss}, \textit{hinge loss}, \textit{smooth hinge loss}, \textit{modified least square} and \textit{logistic loss}~\cite{rennie2005loss} in the following subsections.

\noindent \textbf{Zero-One Loss :} The zero-one loss is a standard loss function to measure the penalty for an incorrect prediction which takes only two values zero or one. For a given observed entry $y_{ij}$ and the corresponding prediction $U_iV_j^T$, zero-one loss is defined as    
\begin{equation}
 \ell(z) = 
 \begin{cases}
 0 & \text{if $z~\geq~0$;} \\
 1 & \text{otherwise,}
 \end{cases}
\end{equation}
where $z= y_{ij}(U_iV_j^T)$.

\noindent \textbf{Hinge Loss:} There are two major drawbacks with zero-one loss (1) Minimizing objective function involving zero-one loss is difficult to optimize as the function is non-convex; (2) It is insensitive to the magnitude of prediction whereas in general, when the entries are binary, the magnitude of score $y_{ij}(U_iV_j^T)$ represent the confidence in our prediction. The hinge loss is a convex upper bound on zero-one error~\cite{globerson2006nightmare} and sensitive to the magnitude of prediction. In the case of classification with large confidence (margin), hinge loss is the most preferred loss function and is defined as
\begin{equation}
\ell(z) = 
\begin{cases}
$0$ & \text{if z $\geq$ 1;} \\
1 - z & \text{otherwise.}
\end{cases}
\end{equation}
\noindent \textbf{Smooth Hinge Loss:} Hinge loss, $h(z)$ is non-differentiable at $z = 1$ and very sensitive to outliers~\cite{rennie2005fast}. An alternative of hinge loss is proposed in~\cite{rennie2005loss} called smooth hinge and is defined as
\begin{equation}
\ell(z) = 
\begin{cases}
0 & \text{if z $\geq$ 1;} \\
\frac{1}{2}(1 - z)^{2} & \text{if $0 < z < 1$;}\\
\frac{1}{2} - z & \text{otherwise.}
\end{cases}
\end{equation}
Smooth Hinge shares many properties with hinge loss and is insensitive to outliers. A detailed discussion about hinge and smooth hinge loss is given in Chapter $3$.

\noindent \textbf{Modified Square Loss:} In~\cite{zhang2001text}, the hinge function is replaced by a smoother quadratic function to make the derivative smooth. The modified square loss is defined as
\begin{equation}
\ell(z) = 
\begin{cases}
0 & \text{if z $\geq$ 1;} \\
(1 - z)^{2} & \text{otherwise.}\\
\end{cases}
\end{equation}
\noindent \textbf{Logistic Loss:}  The logistic loss function is strictly convex function which enjoys properties similar to that of the hinge loss~\cite{mairal2009supervised}. The logistic loss is defined as
\begin{equation}
 \ell(z) = \log (1+\exp^{-z}).
\end{equation}

\subsection{Discrete Ordinal Loss Function}  For a discrete ordinal matrix $Y = [y_{ij}]$ where entries are no more \emph{like} and \emph{dislike} but can be any value from a discrete range, that is,  $y_{ij} \in \{1, 2, \dots, R\}$, there is need to extend binary class loss function to multi-class loss function. There are two major approaches for extending loss function for binary class classification to multi-class classification. The first approach is to directly solve a multi-class problem by modifying the binary class objective function by adding a constraint to it for every class as suggested in \cite{rennie2005loss} \cite{zhang2013multicategory} . The second approach is to decompose the multi-class classification problem into a series of independent binary class classification problems 
as given in~\cite{byun2002applications}. 

To extend binary loss function to multi-class setting, most of the approaches define a set of threshold values $\theta_1 < \theta_1 < \dots < R-1$ such that the real line is divided into $R$ regions. The region defined by threshold $\theta_{q-1}$ and $\theta_q$ corresponds to rating $q$~\cite{rennie2005loss}. For simplicity of notation, we assume $\theta_0 = -\infty$ and $\theta_R = +\infty$. There are different approaches for constructing a loss function based on a set of threshold values such as immediate-threshold and all-threshold~\cite{rennie2005loss}. For each observed entry and corresponding prediction pair $(y_{ij}, U_iV_j^T)$, the immediate-threshold based approach calculates the loss as the sum of immediate-threshold violation. 
\begin{equation}
 \ell(y_{ij}, U_iV_j^T) = \ell(U_iV_j^T - \theta_{i,y_{ij}-1})~+~\ell(\theta_{i,y_{ij}} - U_iV_j^T) 
\end{equation}
On the other hand, all-threshold loss is calculated as the sum of loss for all threshold which is the cost of crossing multiple rating-boundaries and is defined as
\begin{equation}
 \ell(U_iV_j^T, y_{ij}) = \sum_{r=1}^{y_{ij} - 1} {\ell(U_iV_j^T - \theta_{i,r})} + \sum_{r=y_{ij}}^{R-1} {\ell(\theta_{i,r} - U_iV_j^T)}.
\end{equation}

\subsection{Real-valued Loss Function}
For a real-valued data matrix $Y = [y_{ij}]$ where the entries are in real-valued domain $\mathbb{R}$, we review some of the important loss functions such as square-loss \cite{paatero1997}, KL-divergence \cite{lee2001}, $\beta$-divergence \cite{cichocki2006csiszar} and Itakura-Saito divergence~\cite{fevotte2009nonnegative} loss. These are given  in the following subsections. 

\noindent \textbf{Squared Loss: } The square loss is the most common loss function used for real-valued prediction. The penalty for misclassification is calculated as the square distance between the observed entry and the corresponding prediction. The square loss is defined as
\begin{equation}
\ell(y_{ij},\hat y_{ij}) = (y_{ij} - \hat y_{ij})^2
\end{equation}
where $\hat y_{ij} = U_iV_j^T$.

\noindent \textbf{Kullback-Leibler Divergence Loss:} The KL-divergence loss also know as I-divergence loss is the measure of information loss when $U_iV_j^T$ is used as an approximate to $y_{ij}$.  The KL-divergence loss is defined as
\begin{equation}
\ell(y_{ij},\hat y_{ij}) = y_{ij} \ln \frac{y_{ij}}{\hat y_{ij}} - y_{ij} + \hat y_{ij}. 
\end{equation}

\noindent \textbf{Itakura-Saito Divergence:} Itakura and Saito~\cite{fevotte2009nonnegative} is obtained from the the maximum likelihood (ML) estimation. The loss function is defined as
\begin{equation}
 \ell(y_{ij},\hat y_{ij}) = \frac{y_{ij}}{\hat y_{ij}} -y_{ij} \ln \frac{y_{ij}}{\hat y_{ij}} - 1.
\end{equation}

\noindent \textbf{$\beta$-Divergence Loss:} Cichocki et al.~\cite{cichocki2006csiszar} proposed a generalized family of loss function called $\beta$-divergence defined as
\begin{equation}
\ell(y_{ij},U_i,V_j) = 
\begin{cases}
\frac{y_{ij}^\beta}{\beta(\beta-1)} + \frac{\hat y_{ij}^\beta}{\beta} - \frac{y_{ij} \hat y_{ij}^{\beta-1}}{\beta-1}  & \text{if $\beta \in \mathbb{R} \setminus \{0,1\}$;} \\
y_{ij} \ln \frac{y_{ij}}{\hat y_{ij}} - y_{ij} + \hat y_{ij}  & \text{if $\beta = 1$;}\\
\frac{y_{ij}}{\hat y_{ij}} -y_{ij} \ln \frac{y_{ij}}{\hat y_{ij}} - 1  & \text{if $\beta = 0$,}\\
\end{cases}
\end{equation}
where $\beta \ge 0$ is a generalization parameter. At limitng case $\beta =0$ and $\beta=1$, the $\beta$-divergence corresponds to Itakura-Saito divergence and Kullback-Leibler divergence, respectively. The squared loss can be obtained as a special case at $\beta = 2$.

\section{Regularization}
\label{regularizationInMF}
Most of the machine learning algorithms suffer from the problem of overfitting where the model fits the training data too well but have poor generalization capability for new data.  Thus, a proper technique should be adopted to avoid overfitting of the training data. In MF community, to make the model unbiased i.e., to avoid the model to fit only with a particular dataset, many researchers have tried with different regularization terms along with some loss function~\cite{koren2010collaborative,paterek2007improving,takacs2007major,koren2009matrix}. Regularization is not only used to prevent overfitting but also to achieve different structural representation of the latent factors. Several methods based on norm regularization has been proposed in the literature which includes $\ell_{1}$, $\ell_{2}$ and $\ell_{2,1}$. 

\noindent \textbf{$\ell_{1}$ Norm:} The $\ell_{1}$ norm, also know as sparsity inducing norm, is used to produce sparse latent factors and thus avoids overfitting by retaining only the useful factors. In effect, this implies that most units take values close to zero while only few take significantly non-zero values. For a a given matrix $A$, the $\ell_{1}$ norm is defined as
\begin{equation}
 \|A\|_{1} = \sum_{ij} |a_{ij}|.
\end{equation}

\noindent \textbf{$\ell_{2}$ Norm:} The most popular and widely investigated regularization term used in MF model is $\ell_{2}$ norm which is also known as Frobenius norm. For a given matrix $A$, the $\ell_{2}$  norm is defined as
\begin{equation}
 \|A\|_{F} = \sqrt{\sum_{ij} a^2_{ij}}
\end{equation}
\noindent where $\|\cdot\|_2$ is the $\ell_{2}$  norm of a matrix. Minimizing the objective function containing the $\ell_{2}$  norm  as regularization term gives two benefits 1) It avoids overfitting by  penalizing the large latent factor values. 2) Approximating the target matrix with low-rank factor matrix is a typical non-convex optimization problem and in fact, the $\ell_{2}$ norm has also been suggested as a convex surrogate for the rank in control applications~\cite{fazel2001rank,rennie2005fast}. 

\noindent \textbf{$\ell_{2,1}$ Norm:} The $\ell_{2,1}$ norm also known as group sparsity norm is used to induce a sparse representation at the level of groups. For a given matrix $A$, the $\ell_{2,1}$ norm is defined as
\begin{equation}
 \|A\|_{2,1} = \sum_{i=1}^{n}{\sqrt{\sum_{j=1}^{m}A_{ij}^2}}.
\end{equation}

\section{Collaborative filtering with Matrix Factorization}
\label{CFViaMF}
In collaborative filtering, the goal is to infer user preferences for items based on his/ her previously given preference and a large collection of preferences of other users. Given a partially observed $N \times M$ user-item rating matrix $Y$ with $N$ number of users and $M$ number of items the goal is to predict unobserved preference of users for items. The collaborative filtering problem can be formally defined as follows: 
\begin{Definition}[Collaborative Filtering]
 Let $Y$ be a $N \times M$ size user-item rating matrix and $\Omega$ be the set of observed entries. For each $(i,j)\in \Omega$, the entry $y_{ij} \in \{1,2,\dots,R\}$ defines the preference of $i$th user for $j$th item. For each $(i,j) \notin \Omega$, $y_{ij} = 0$ indicates that preference of $i$th  user  for $j$th item is not available (unsampled entry). Given a partially observed rating matrix $Y \in \mathbb{R}^{N \times M}$, the goal is to predict $y_{ij}$ for $(i,j) \notin \Omega$.  
\end{Definition}

Matrix factorization is a key technique employed for completion of user-item rating matrix wherein the objective is to learn low-rank (or low-norm) latent factors $U$ (for users) and $V$ (for items) so as to simultaneously approximate the observed entries under some loss measure and predict the unobserved entries. There are various ways of doing so. It is shown in~\cite{lee1999learning} that to approximate $Y$, the entries in the factor matrix $U$ and $V$ need to be non-negative so that only additive combination factors are allowed. The basic idea is to learn factor matrices $U$ and $V$ in such a way that, the squared sum distance between the observed entry and corresponding prediction is minimized. The optimization problem is formulated as
\begin{equation}
 \underset{U \ge 0, V \ge 0}{min} J(U, V)= \sum_{(i,j) \in \Omega}{(y_{ij}-U_iV_j^T)^2}.
\end{equation}

\noindent Singular value decomposition (SVD) is used in~\cite{sarwar2002incremental} to learn the factor matrices $U$ and $V$. The key technical challenge when SVD is applied to sparse matrices is that it suffers from severe overfitting. When SVD factorization is applied on sparse data, error function needs to be modified so as to consider only the observed ratings by setting the non-observed entries to zero. This minor modification results in a non-convex optimization problem. Instead of minimizing the rank of a matrix, maximum margin matrix factorization (MMMF)~\cite{srebro2004maximum} proposed by srebro et al. aims at minimizing the Froebenius norms of $U$ and $V$, resulting in convex optimization problems. It is shown that MMMF can be formulated as a semi-definite programming (SDP) problem and solved using standard SDP solvers. However, current SDP solvers can only handle MMMF problems on matrices of dimensionality up to a few hundred.  Hence, a direct gradient based optimization method for MMMF is proposed in \cite{rennie2005fast} to make fast collaborative prediction. The detailed discussion about MMMF is given in Chapter 3. 
To further improve the performance of MMMF, in \cite{decoste2006collaborative}, MMMF is casted using ensemble methods which includes bagging and random weight seeding.  MMMF was further extended in \cite{ weimer2008improving} by introducing offset terms, item dependent regularization and a graph kernel on the recommender graph. In \cite{xu2012nonparametric}, a noparametric Bayesian-style MMMF was proposed that utilizes nonparametric techniques to resolve the unknown number of latent factors in MMMF model \cite{weimer2008improving}\cite{ rennie2005fast}\cite{ decoste2006collaborative}.  A probabilistic interpretation of MMMF was presented in \cite{ xu2013fast} model through data augmentation.

The proposal of MMMF hinges heavily on extended hinge loss function.  Research on different loss functions and their extension to handle multiple classes has not attracted much attention of researchers though there are some important proposals \cite{liu2011hard}. MMMF has become a very popular research topic since its publication and several extensions have been proposed \cite{decoste2006collaborative,weimer2008improving,xu2012nonparametric,xu2013fast}. There has also been  some research on matrix factorization on binary or bi-level preferences \cite{Volkovs2015}. But many view binary preference as a special case of matrix factorization with discrete ratings.  In~\cite{zhong2012contextual} the rating matrix is decomposed hierarchically by grouping similar users and items together, and each sub-matrix is  factorized locally. To the best of our knowledge, there is no research on hierarchical MMMF. 

\section{Multi-label Classification with Matrix Factorization}
\label{MLCViaMF}
 In machine learning and statistics, the classification problem is concerned with the assignment of a class (category) to a data object (instance) from a given set of discrete classes. For example, classifying a document into one of the several known categories such as \textit{sports}, \textit{crime}, \textit{business}, \textit{politics} etc. In a traditional classification problem, data objects are represented in the form of feature vectors, each associated with a unique class label from a set of disjoint class labels $L$, $|L| > 1$. Depending on the total number of disjoint classes in $L$, a learning task is categorized as \textit{binary} classification (when $|L| = 2$) or \textit{multi-class} classification (when $|L| > 2$)~\cite{sorower2010literature}. However, in many real-word classification tasks, data object can be simultaneously  associated with one or more than one class in $L$. For example, a document can simultaneously belong to more than one class such as \textit{politics} and \textit{business}. The objective of multi-label classification (MLC) is to build a classifier that can automatically tag an example with the most relevant subset of labels. This problem can be seen as  a generalization of the \textit{single label} classification where an instance is associated with a unique class label from a set of disjoint labels $L$. The multi-label classification problem can be formally defined as follows: 
 \begin{Definition}[Multi-label Classification] 
 Given $N$  training examples in the form of a pair of feature matrix  $X$ and  label matrix $Y$ where each example $x_i \in \mathbb{R}^D, 1\le i \le N$, is a row of $X$ and its associated labels $y_i \in  \{-1, 1\}^L$ is the corresponding row of $Y$. The $+1$ entry at the $j$th coordinate of vector $y_i$ indicates the presence of label $j$ in data point $x_i$.  The task of multi-label classification is to learn a parametrization $h : \mathbb{R}^D \rightarrow \{-1, 1\}^L$ that maps each instance (or, a feature vector)  to a set of  labels (a label vector). 
 \end{Definition}
 
MLC is a major research area in the machine learning community and finds application in several domains such as computer vision~\cite{cabral2011matrix,boutell2004learning}, data mining~\cite{tsoumakas2007random,schapire2000boostexter} and text classification~\cite{zhang2006multilabel,schapire2000boostexter}. Due to the exponential size of the output space, exploiting intrinsic information in the feature and the label space has been the major thrust of research in recent years and the use of parametrization and embedding techniques have been the prime focus in MLC. The embedding based approach assumes that there exists a low-dimensional space onto which the given set of feature vectors and/ or label vectors can be embedded. The embedding strategies can be grouped into two categories namely; (1) Feature space embedding; and (2) Label space embedding. Feature space embedding aims to design a projection function which can map the instance in the original feature space to the label space. On the other hand, the label space embedding approach transform the label vectors to an embedded space via linear or local non-linear embeddings, followed by the association between feature vectors and embedded label space for classification purpose. With proper decoding process that maps the projected data back to the original label space, the task of multi-label prediction is achieved~\cite{hsu2009multi,rai2015large,tai2012multilabel}. We present a brief review of the FE and LE approaches for multi-label classification. The detailed discussion is given in Chapter~\ref{mlchmfChapter}.

Given $N$  training examples in the form of a pair of feature matrix  $X$ and  label matrix $Y$ where each example $x_i \in \mathbb{R}^D, 1\le i \le N$, is a row of $X$ and its associated labels $y_i \in  \{-1, +1\}^L$ is the corresponding row of $Y$,  the goal of FE is to learn a transformation matrix $W\in \mathbb{R}^{D \times L}$ which maps instances feature space to label space. This approach requires $D \times L$ parameter to model the classification problem, which will be expensive when $D$ and $L$ are large~\cite{yu2014large}. In ~\cite{yu2014large} a generic empirical risk minimization (ERM) framework is used  with low-rank constraint on linear parametrization $W = UV^T$, where $U \in \mathbb{R}^{D \times d} $ and $V \in \mathbb{R}^{L \times d}$ are of rank $d \ll D$. The problem can be restated as follows.
\begin{equation} \label{basic-low-rank-formulation_RL0}
		\underset{U, V}{min} \; \ell(Y, XUV^T) + \frac{\lambda}{2} (\|U\|_F^2 + \|V\|_F^2)\\
\end{equation}
where $\|\cdot\|_{F}$ is Frobenius norm. The formulation in Eq.~(\ref{basic-low-rank-formulation_RL0}) can capture the intrinsic information of both feature and label space. It can also be seen as a joint learning framework in which dimensionality reduction and multi-label classification are performed simultaneously~\cite{ji2009linear,yu2016semisupervised}.

The matrix factorization (MF) based approach for LE aims at determining two matrices $U \in \mathbb{R}^{ N \times d}$ and $V \in \mathbb{R}^{ d \times L}$~\cite{bhatia2015sparse,xu2016robust}. The matrix $U$ can be viewed as the \emph{basis matrix}, while the matrix $V$ can be treated as the \emph{coefficient matrix} and a common formulation  is the following optimization problem. 	
\begin{equation} \label{basic-LE-formulation0}
	\underset{U, V}{min} \; \ell(Y, U, V) + \lambda R(U, V)\\
\end{equation}
where  $\ell(\cdot)$ is a loss function that measures how well $UV$ approximates $Y$, $R(\cdot)$ is a regularization function that promotes various desired properties in $U$ and $V$ (sparsity, group-sparsity, etc.)  and the constant $\lambda \ge 0$ is the regularization parameter which controls the extent of regularization. In~\cite{lin2014multi}, a MF based approach is used to learn the label encoding and decoding matrix simultaneously. The problem is formulated as follows.
\begin{equation}
		\label{faie0}
		\underset{U, V}{min}~\|Y - UV\|_{F}^{2} + \alpha\Psi(X, U)
\end{equation}
where $U \in \mathbb{R}^{ N \times d}$ is the code matrix, $V \in \mathbb{R}^{ d \times L}$ is the decoding matrix, $\Psi(X,U)$ is used to make $U$ feature-aware by considering correlations between $X$ and $U$ as side information and the constant $\alpha \ge 0$ is the trade-off parameter.

  \newpage
\thispagestyle{empty}
\setcounter{chapter}{2}
\chapter{Collaborative Filtering Using Hierarchical Matrix Factorizations} 
\label{hmf_chapter}
In the previous chapter, we discussed the basic matrix factorization model and the associated formulation of matrix factorization as an optimization problem. We have also discussed at length the application of matrix of factorization techniques in collaborative filtering and multi-label classification.  In this chapter,  we elaborate on those basics and propose a matrix factorization based collaborative filtering approach.
\section{Introduction}
\label{hmfIntroduction}
In this chapter, we describe the proposed method, termed as \textit{HMF (Hierarchical Matrix Factorization)}. HMF is a novel method for constructing a hierarchical bi-level maximum margin matrix factorization to handle matrix completion of ordinal rating matrix. The proposed method draws motivation from research on multi-class classification. There are two major approaches of extending two-class classifiers to multi-class classifiers. The first approach explicitly reformulates the classifier, resulting in a unified multiclass optimization problem (\textit{embedded  technique}). The second approach (\textit{combinational technique}) is to decompose a multiclass problem into multiple, independently trained, binary classification problems and to combine them appropriately so as to form a multiclass classifier. In maximum margin matrix factorization (MMMF)~\cite{rennie2005fast}, the authors adopted \textit{embedded} approach by extending bi-level hinge loss to multi-level cases. 

Combinational techniques have been very  popular and  successful as they all exhibit some relative advantages over embedded techniques and this prompts us to question whether some sort of combinational approach can be employed in the context of MMMF. There are several approaches in combinational techniques and these are \textit{One-Against-All} (OAA)~\cite{hsu2002comparison, rifkin2004defense, weston1998multi}, \textit{One-Against-One} (OAO)~\cite{debnath2004decision, hsu2002comparison} etc. HMF falls into the category of OAA approach with a difference. The OAA approach of classification employs one binary classifier for each class against all other classes. In the present context `class' corresponds to the number of ordinal ranks. Interestingly, since the ranks are ordered, in the present case,  OAA strategy is used by taking advantage of the ordering of 'classes'. Unlike the traditional OAA strategy, (which means that one bi-level matrix factorization to be used for rank $q$ as one label (say, $-1$) and all other ranks as the other label (say, $+1$)), here we employ for each rank $q$, all ranks below $q$ as $-1$ and all ranks above $q$ as $+1$.

MMMF and HMF, like any other matrix factorization methods, use latent factor model approach. The latent factor model is concerned with learning from limited observations, latent factor vector $U_i$ for each user and latent factor vector $V_j$ for each  item, such that the dot product of $U_i$ and $V_j$  gives the ranking of user $i$ for item $j$. In MMMF, besides learning the latent factors, the set of thresholds also needs to be learned. It is assumed that there are $R-1$ thresholds $\theta_{i,r}$ for each user $i$  ($R$ is the number of ordinal values or number of ranks).  Thus, the rating of user $i$ for item $j$ is decided by comparing the dot product  of $U_i$ and $V_j$   with each $\theta_{i,r}$.  Thus, the fundamental assumption of MMMF is that the latent factor vectors determine the properties of users and items and the threshold values capture the characteristics of rating. HMF differs from MMMF on this principle. The underlying principle of HMF is that the latent factors of users and items are going to be different, if the users' \textit{likes} or \textit{dislikes} cutoff thresholds are different. The latent factors are different according to situations. For instance, the latent factors when ranks above $q$ are identified as \textit{likes} is different from  situations wherein the ranks above $q+1$ are identified as \textit{likes}.
Thus if we have $R$ ratings then, there will be $R-1$ pairs of latent factors $(U^q, V^q)$. Unlike the process of learning single pair of latent factors, $U$ and $V$, HMF learns several $U$'s and $V$'s in this process without any additional computational overheads. The process is proved to be a more accurate matrix completion process. There has not been any attempt in this direction and we believe that our present study will open up new areas of research in future.

The rest of the chapter is organized as follows. In Section~\ref{bi_level_MMMF}, we describe the bi-level MMMF which we use subsequently in our algorithm.  Section~\ref{F-MMMF} summarizes the well-known existing MMMF method. We introduce our proposed method of Hierarchical Matrix Factorization (HMF) in Section~\ref{hmf_section}. The advantages of HMF over other matrix factorization based collaborative filtering methods are given by detailed experimental analysis in Section~\ref{experiment_section_hmf}. Section~\ref{conclusion_hmf} concludes the chapter.


\section{Bi-level MMMF}
\label{bi_level_MMMF}
In this section we describe Maximum Margin Matrix Factorization for a bi-level rating matrix. The matrix completion of bi-level matrix is concerned with the following problem. 

\noindent
\textbf{Problem P \textsubscript{$\pm{1}$}(Y): }
Let $Y$ be a $N \times M$ partially observed user-item rating matrix and $\Omega$ is the set of observed entries. For each $(i, j) \in \Omega$, the entry $y_{ij} \in  \{-1, +1\}$ defines the preference of $i$th user for $j$th item with $+1$ for \textit{likes} and $-1$ for \textit{dislikes}. For each $(i, j) \notin \Omega$, the entry $y_{ij} = 0$ indicates that the preference of $i$th  user  for $j$th item is not available. Given a partially observed rating matrix $Y\in\mathbb{R}^{N\times M}$, the goal is to infer $y_{ij}$ for $(i, j) \notin \Omega$.

Matrix factorization is one of the major techniques employed for any matrix completion problem. In this line, the above problem can be rephrased using latent factors. Given a partially observed rating matrix $Y\in\mathbb{R}^{N\times M}$ and the observed preference set $\Omega$, matrix factorization aims at determining two low-rank (or, low-norm) matrices $U\in \mathbb{R}^{N\times d}$ and $V\in \mathbb{R}^{M\times d}$ such that $Y \approx UV^{T}$. The row vectors $U_i$, $1 \le i \le N$ and $V_j$, $1 \le j \le M$ are the $d$-dimensional representations of the users and the items, respectively. A common formulation of  P\textsubscript{$\pm{1}$}(Y) is to find $U$ and $V$ as solution of the following optimization problem.
\begin{equation}\label{basic-fact-formulation}
\underset{U, V}{min}~J(U, V) = \sum_{(i,j)\in \Omega} \ell(y_{ij}, U_i, V_j) + \lambda R(U, V)
\end{equation}
where $\ell(\cdot)$ is a loss function that measures how well $(U_i.V_j^T)$ approximates $y_{ij}$, and $R(\cdot)$ is a regularization function. The idea of the above formulation is to alleviate the problem of outliers through a robust loss function and at the same time to avoid overfitting and to make the optimization function smooth with the use of regularization function.

A number of approaches can be used to optimize the objective function \ref{basic-fact-formulation}. Gradient Descent method and its variants start with random $U$ and $V$ and iteratively update $U$ and $V$ using the equations \ref{U_update} and \ref{V_update}, respectively.
\begin{align}
 U_{ip}^{t+1} &= U_{ip}^{t} - c \frac{\partial J}{\partial U_{ip}^t} \label{U_update}\\
 V_{jq}^{t+1} &= V_{jq}^{t} - c \frac{\partial J}{\partial V_{jq}^t} \label{V_update}\
\end{align}
where $c$ is the step length parameter and suffixes $t$ and $(t + 1)$ indicate current values and updated values.

The step-wise description of the process is given as Pseudo-code in Algorithm~\ref{bilevelbasicAlgo}.

\begin{algorithm}[ht!]
\SetAlgoLined
\SetKwData{Left}{left}\SetKwData{This}{this}\SetKwData{Up}{up}
\SetKwFunction{Union}{Union}\SetKwFunction{FindCompress}{FindCompress}
\SetKwInOut{Input}{input}\SetKwInOut{Output}{output}
\Input{Bi-level Rating Matrix: $Y$, Number of Latent Factors: $d$,  Regularization $~~$Parameter: $\lambda$}
\Output{Factor Matrices: $U$ and $V$}
\BlankLine
$t\leftarrow 0$\;
Initialize: $U^t$,  $V^t$\; 

\While{Stopping criteria met} {
    Calculate $\frac{\partial J}{\partial U^t_{ip}}$ and $\frac{\partial J}{\partial V^t_{jq}}$ at $U^t$ and $V^t$, respectively\;
    $t\leftarrow t+1$\;    
    $U_{ip}^{t+1} \leftarrow U_{ip}^{t} - c \frac{\partial J}{\partial U_{ip}^t}$\;
    $V_{jq}^{t+1} \leftarrow  V_{jq}^{t} - c \frac{\partial J}{\partial V_{jq}^t}$\;
}
Let $U$ and $V$ are the factor matrices obtained after convergence\;
\Return $U$ and $V$\;
\caption{A\textsubscript{$\pm{1}$} ($Y$, $d$, $\lambda$)}
\label{bilevelbasicAlgo}
\end{algorithm}

Once $U$ and $V$ are computed by Algorithm~\ref{bilevelbasicAlgo}, the matrix completion process is accomplished from the factor matrices as follows.
\begin{equation}
  \hat y_{ij} = 
  \begin{cases}
  -1, &  \text { if $(i,j) \notin \Omega \wedge  U_iV^{T}_j < \theta$; } \\
  +1, & \text{if $(i,j) \notin \Omega \wedge U_iV^{T}_j \ge \theta$;} \\
  y_{ij}, & \text{if  $(i,j) \in \Omega$,}
  \end{cases}
  \label{bilevelMapping}
\end{equation}
where $\theta$ is the user-specified threshold value. \\

The latent factor $V_j$ of each item $j$ can be viewed as a point in $d$-dimensional space and the latent factor $U_i$ of user $i$ can be viewed as a decision hyperplane in this space. The objective of bi-level matrix factorization is to learn the embeddings of these points and hyperplanes in $\mathbb{R}^d$ such that each hyperplane (corresponding to a user) separates (or, equivalently, classifies) the items by \textit{likes} and \textit{dislikes} of a user. Let us consider the following partially observed bi-level matrix $Y$ (Table \ref{bilevelY}) for illustration. The unobserved entries are indicated by $0$ entries.

\begin{figure}[ht!]
	\centering
	\includegraphics[width=5.5in,height=4.8in]{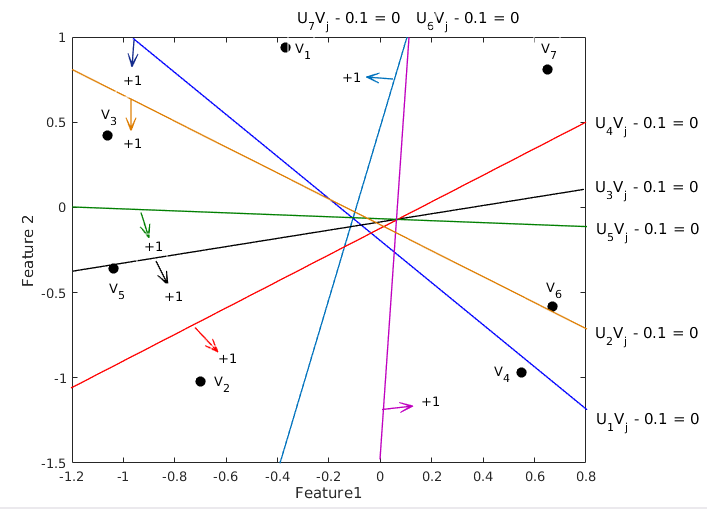}
	\captionof{figure}{Geometrical interpretation of bi-level matrix factorization.}
	\label{bilevelClassification}
\end{figure}

\begin{table}[ht!]
\renewcommand{\arraystretch}{1}
\centering
\caption{Bi-level matrix $Y$}
 \begin{tabular}{|r|r|r|r|r|r|r|}
  \hline
  0&1&0&0&1&0&-1\\
  \hline
  -1&0&1&1&1&0&-1\\
  \hline
  0&1&-1&0&0&1&-1\\
  \hline
  -1&1&-1&1&-1&1&0\\
  \hline
  -1&0&-1&1&1&0&0\\
  \hline
  0&-1&0&1&0&1&1\\
  \hline
  1&1&0&-1&0&0&0\\
  \hline
 \end{tabular}
  \label{bilevelY}  
\end{table}
\noindent
Let us assume that the following $U$ and $V$ are the latent factor matrices with $d=2$. 
\begin{table}[ht!]
\centering
\caption{Latent factor matrices corresponding to $Y$.}
 \begin{minipage}{0.2\textwidth}
 \centering
  \begin{tabular}{|r|r|}
      \hline
      -0.63&-0.50\\ \hline
      -0.69&-0.96\\ \hline
      0.27&-1.09\\ \hline
      0.63&-0.84\\ \hline
      -0.02&-1.03\\ \hline
      1.19&-0.03\\ \hline
      -1.11&0.21\\ \hline
    \end{tabular}
    \captionof{subfigure}{U}
 \end{minipage}%
 \begin{minipage}{0.2\textwidth}
 \centering
  \begin{tabular}{|r|r|}
      \hline
	-0.37&0.94\\ \hline
	-0.70&-1.02 \\ \hline
	-1.06&0.42 \\ \hline
	0.55&-0.97 \\ \hline
	-1.04&-0.36 \\ \hline
	0.67&-0.58 \\ \hline
	0.65&0.81 \\ \hline
  \end{tabular}
  \captionof{subfigure}{V}
 \end{minipage}
  \label{bilevelUandV}  
\end{table}
\vspace{-0.7cm}
The same is depicted graphically in Figure~\ref{bilevelClassification}. $V_{1}, V_{2}, \dots, V_{7}$ are depicted as points. The hyperplanes $U_{1}, U_{2}, \dots, U_{7}$ with threshold $0.1$ are depicted as lines. An arrow is shown to indicate the $positive$ side of each line. In other words, if a point falls this side, the preference is $like~(+1)$ and the other side preference is $dislikes~(-1)$. Entry $y_{34}$ is predicted based on the position of $V_4$ with respect to $U_3$. $V_4$ falls to the \textit{positive} side of line corresponding to $U_3$ and hence, the entry $y_{34}$ is predicted as $+1$.
The latent factor $U$ and $V$ are obtained by a learning process making use of the generic algorithm (Algorithm~\ref{bilevelbasicAlgo}) with the observed entries as the training set. The objective of this learning process is to minimize the loss due to discrepancy between computed entries and the observed entries. In the example (Figure \ref{bilevelClassification}) point $V_3$ and $V_5$ are in the $positive$ side of $U_2$ and $V_7$ in the \textit{negative} side of $U_2$. These points are located at different distance.

There are many matrix factorization algorithms for bi-level matrices which adopts the generic algorithm describe in Algorithm~\ref{bilevelbasicAlgo}. These algorithms are designed based on the specification of the loss function and the regularization function. We adopt here a maximum-margin formulation as the guiding principle.

\begin{figure}[ht!]
	\centering
	  \includegraphics[width=4.8in,height=4in]{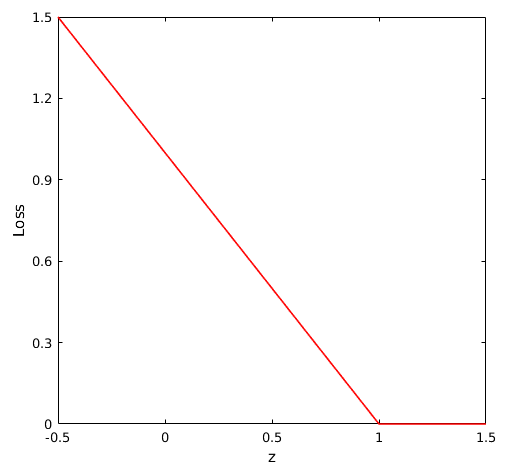}
	\caption{Loss function values for the hinge}
	\label{hingeLossFig}
\end{figure}

Loss functions in matrix factorization models are used to measure the discrepancy between the observed entry and the corresponding prediction. Furthermore, especially when predicting discrete values such as ratings, loss functions other then sum-squared loss are often more appropriate~\cite{rennie2005fast}. The trade-off between generalization loss and empirical loss has been prevailing since the advent of support vector machine (SVM). Maximum margin approach aims at providing higher generalization ability and avoiding overfitting. In this context, hinge loss function is the most preferred loss function and is defined as follows.
\begin{equation}
\label{hingeLossFun}
h(z) = 
\begin{cases}
0, & \text{if z $\geq$ 1}; \\
1 - z, & \text{otherwise},
\end{cases}        
\end{equation}
where $z = y_{ij}(U_{i}V_{j}^{T})$. The hinge loss is illustrated in Figure~\ref{hingeLossFig}. 
\vspace{1cm}

The following real-valued prediction matrix is obtained from the latent factor matrices $U$ and $V$ (Table~\ref{bilevelUandV}) corresponding to the matrix Y (Table~\ref{bilevelY}).
\begin{table}[ht!]
\renewcommand{\arraystretch}{1}
\caption{Real-valued prediction corresponding to bi-level matrix Y.}
\centering
 \begin{tabular}{|r|r|r|r|r|r|r|}
 \hline
  -0.24&0.95&0.46&0.14&0.84&-0.13&-0.81\\
  \hline
  -0.65&1.46&0.33&0.55&1.06&0.09&-1.23\\
  \hline
  -1.12&0.92&-0.74&1.21&0.11&0.81&-0.71\\
  \hline
  -1.02&0.42&-1.02&1.16&-0.35&0.91&-0.27\\
  \hline
  -0.96&1.06&-0.41&0.99&0.39&0.58&-0.85\\
  \hline
  -0.47&-0.80&-1.27&0.68&-1.23&0.81&0.75\\
  \hline
  0.61&0.56&1.26&-0.81&1.08&-0.87&-0.55\\
  \hline
 \end{tabular}
 \label{bilevelUV}
\end{table}

\noindent
The hinge loss function values corresponding to the observed entries in $Y$ (Table~\ref{bilevelY}) and the real-valued prediction (Table~\ref{bilevelUV}) is shown in Table~\ref{HingeLossValue}.
\begin{table}[ht!]
\renewcommand{\arraystretch}{1}
\caption{Hinge loss corresponding to observed entries in Y.}
\centering
 \begin{tabular}{|c|c|c|c|c|c|c|}
 \hline
  0&0.05&0&0&0.16&0&0\\
  \hline
  0.35&0&0.67&0.45&0&0&0.35\\
  \hline
  0&0.08&0.26&0&0&0.19&0\\
  \hline
  0&0.58&0&0&0.65&0.09&0\\
  \hline
  0.04&0&0.59&0.01&0.61&0&0.04\\
  \hline
  0&0.20&0&0.32&0&0.19&0\\
  \hline
  0.39&0.44&0&0.19&0&0&0.39\\
  \hline
 \end{tabular}
 \label{HingeLossValue}
\end{table}

\begin{figure}[ht!]
	\centering
	  \includegraphics[width=4.6in,height=4in]{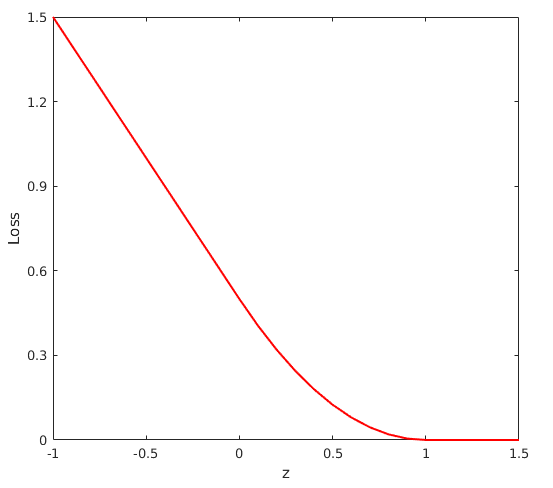}
	\caption{Loss function values for the smooth hinge}
	\label{smoothHingeLossFig}
\end{figure}

Hinge loss, $h(z)$ is non-differentiable at $z = 1$ and is very sensitive to outliers as mentioned in~\cite{rennie2005fast}. Therefore an alternative called \emph{smooth hinge loss} is proposed in~\cite{rennie2005loss} and can be defined as,
\begin{equation}
 h(z) = 
  \begin{cases}
  0, & \text{if z $\geq$ 1}; \\
  \frac{1}{2}(1 - z)^{2}, & \text{if $0 < z < 1$};\\
  \frac{1}{2} - z, & \text{otherwise}.
  \end{cases}
  \label{smoothHinge} \\
\end{equation}
The smooth hinge loss is illustrated in Figure~\ref{smoothHingeLossFig}. 

The smooth hinge loss function values corresponding to the observed entries in $Y$ (Table~\ref{bilevelY}) and the real-valued prediction (Table~\ref{bilevelUV}) is shown in Table~\ref{SmoothHingeLossValue}.
\begin{table}[ht!]
\caption{Smooth hinge loss corresponding to observed entries in Y.}
\centering
 \begin{tabular}{|c|c|c|c|c|c|c|}
 \hline
  0&0&0&0&0.01&0&0\\
  \hline
  0.06&0&0.23&0.10&0&0&0.06\\
  \hline
  0&0&0.03&0&0&0.02&0\\
  \hline
  0&0.17&0&0&0.21&0&0\\
  \hline
  0&0&0.17&0&0.19&0&0\\
  \hline
  0&0.02&0&0.05&0&0.02&0\\
  \hline
  0.08&0.10&0&0.02&0&0&0.08\\
  \hline
 \end{tabular}
 \label{SmoothHingeLossValue}
\end{table}\\
Figure \ref{hingeLossFig} and \ref{smoothHingeLossFig} show the loss function values for the hinge and smooth hinge loss, respectively. It can be seen that the smooth hinge loss shares important similarities to the hinge loss and has a smooth transition between a zero slope and a constant negative slope. Table~\ref{HingeLossValue} and ~\ref{SmoothHingeLossValue} show the hinge loss and smooth hinge loss function values corresponding to the observed entries in $Y$ (Table~\ref{bilevelY}). It can also be seen that the smooth hinge is less sensitive to the outliers as compared to the hinge loss function. For example, the rating given by user $4$ for item $2$ is positive and the same reflect in the embedding (Figure~\ref{bilevelClassification}). The loss incurred by hinge and smooth hinge are $0.58$ and $0.17$, respectively. Even though the point is embedded with margin the hinge loss function gives more reward to the model for the increase in objective value.

We reformulate P\textsubscript{$\pm{1}$}(Y) problem for the bi-level rating matrix as the following optimization problem. \vspace{-0.5cm}
\begin{equation}\label{binary-hinge-objective}
\underset{U, V}{min} J(U, V)= \underset{(i,j) \in \Omega} {\Sigma} h \big(y_{ij}(U_{i}V_{j}^{T})\big)  + \frac{\lambda}{2} (\|U\|^{2}_{F} + \|V\|^{2}_{F})
\end{equation}
where $\|. \|_{F}$ is the Frobenius norm which is the same as defined in Chapter $2$,  $\lambda > 0$ is the regularization parameter and $h(\cdot)$ is the smooth hinge loss function as defined previously.
 
The gradients of the variables to be optimized are determined as follows. The gradient with respect to each element of $U$ is calculated as
\begin{align}
\frac{\partial J}{\partial U_{ip}} &= \sum_{j|(i,j) \in \Omega}^{}  \frac{\partial h\big( y_{ij}(U_{i}V_{j}^{T})\big)}{\partial U_{ip}} + \frac{\lambda}{2}\bigg(\frac{\partial \|U\|^{2}_{F}}{\partial U_{ip}} +  \frac{\partial \|V\|^{2}_{F}}{\partial U_{ip}}\bigg) \nonumber \\
 &= \sum_{j|(i,j) \in \Omega}^{} y_{ij} h'\big( y_{ij}(U_{i}V_{j}^{T})\big) \frac{\partial (U_{i}V_{j}^{T}) }{\partial U_{ip}} + \lambda U_{ip} \nonumber \\
 &= \sum_{j|(i,j) \in \Omega}^{} y_{ij} h'\big( y_{ij}(U_{i}V_{j}^{T})\big)V_{jp}+ \lambda U_{ip}
\end{align}

\noindent
Similarly, the gradient with respect to each element of $V$ is calculated as follows.
\begin{align}
\frac{\partial J}{\partial V_{jq}} 
 &= \sum_{i|(i,j) \in \Omega}^{} y_{ij} h'( y_{ij}(U_{i}V_{j}^{T}))U_{iq}+ \lambda V_{jq}
\end{align}
where $h'(z)$ is defined as follows.
\begin{equation} 
  h'(z) = 
  \begin{cases}
  0, & \text{if z $\geq$ 1}; \\
  z-1, & \text{if $0 < z < 1$};\\
  -1, & \text{otherwise}.
  \end{cases}
  \label{smoothHingederivation}
\end{equation}

Algorithm \ref{bilevelmmmfAlgo} outlines the main flow of the bi-level maximum margin matrix factorization (BMMMF).
\begin{algorithm}[ht!]
\SetAlgoLined
\SetKwData{Left}{left}\SetKwData{This}{this}\SetKwData{Up}{up}
\SetKwFunction{Union}{Union}\SetKwFunction{FindCompress}{FindCompress}
\SetKwInOut{Input}{input}\SetKwInOut{Output}{output}
\Input{Bi-level Rating Matrix: $Y$, Number of Latent Factors: $d$,  Regularization $~~$Parameter: $\lambda$}
\Output{Factor Matrices: $U$ and $V$}
\BlankLine
$t\leftarrow 0$\;
Initialize: $U^t$,  $V^t$\; 
\While{Stopping criteria met} {
    $\frac{\partial J}{\partial U^t_{ip}} \leftarrow \sum_{j|(i,j) \in \Omega}^{} y_{ij} h'( y_{ij}(U_{i}^{t}V_{j}^{t^T}))V_{jp}^{t}+ \lambda U_{ip}^{t}$ \;    
    $\frac{\partial J}{\partial V^t_{jq}} \leftarrow \sum_{i|(i,j) \in \Omega}^{} y_{ij} h'( y_{ij}(U_{i}^{t}V_{j}^{t^T}))U_{iq}^{t}+ \lambda V_{jq}^{t}$\;
    $t\leftarrow t+1$\;    
    $U_{ip}^{t+1} \leftarrow U_{ip}^{t} - c \frac{\partial J}{\partial U_{ip}^t}$\;
    $V_{jq}^{t+1} \leftarrow  V_{jq}^{t} - c \frac{\partial J}{\partial V_{jq}^t}$\;
}
Let $U$ and $V$ are the factor matrices obtained after convergence\;
\Return $U$ and $V$\;
\caption{BMMMF ($Y$, $d$, $\lambda$)}
\label{bilevelmmmfAlgo}
\end{algorithm}

\begin{figure}[ht!]
	\noindent
	\centering
	  \includegraphics[width=4.5in,height=4in]{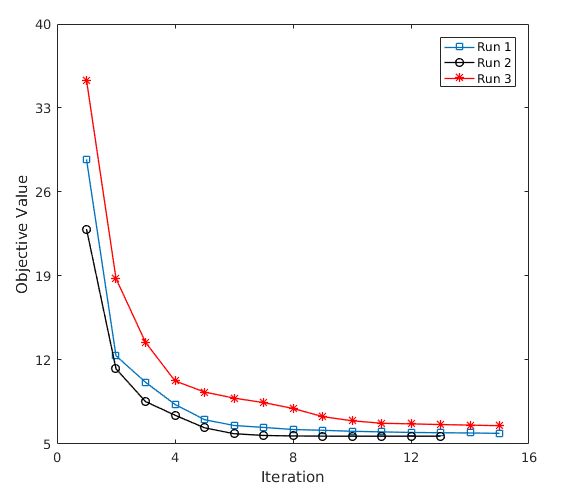}
	\caption{Convergence graph of BMMMF with different initial points}
	\label{bilevelObjectiveValue}
\end{figure}

We also show the behaviour of $J$ (Equation~\ref{binary-hinge-objective}). We plot the value of $J$ obtained in every iteration for the same $Y$ (Table~\ref{bilevelY}) starting from different initial points. It can be seen from Figure~\ref{bilevelObjectiveValue} that $J$ is having asymptotic convergence.

\section{Multi-level MMMF} \label{F-MMMF}
As discussed in the previous chapter, MMMF \cite{srebro2004maximum} and subsequently, Fast MMMF~\cite{rennie2005fast}  are proposed primarily for collaborative filtering with ordinal rating matrix when user-preferences are not in the form of like/ dislike but  are values in a discrete range. The matrix completion of ordinal rating matrix is concerned with the following problem. 

\noindent
\textbf{Problem P\textsubscript{ord}(Y): }
Let $Y$ be a $N \times M$ partially observed user-item rating matrix and $\Omega$ is the set of observed entries. For each $(i, j) \in \Omega$, the entry $y_{ij} \in \{1, 2, \dots, R\}$ defines the preference of the $i$th user for the $j$th item. For each $(i, j) \notin \Omega$, $y_{ij} = 0$ indicates that the preference of $i$th  user  for $j$th item is not available. Given a partially observed rating matrix $Y\in\mathbb{R}^{N\times M}$, the goal is to predict $y_{ij}$ for $(i, j) \notin \Omega$.

\noindent
\textbf{Need for multiple thresholds: }
Unlike P\textsubscript{$\pm{1}$}(Y), P\textsubscript{ord}(Y) has domain of $y_{ij}$ with more than two values. When the domain has two values, P\textsubscript{ord}(Y) is equivalent to P\textsubscript{$\pm{1}$}(Y). Continuing our discussion on geometrical interpretation of P\textsubscript{$\pm{1}$}(Y), the \textit{likes $(+1)$} and \textit{dislikes $(-1)$} of user $i$ are separated with both sides of the hyperplane defined by $U_i$. The same concept when extended to P\textsubscript{ord}(Y), it is necessary to define threshold values $\{\theta_1, \dots, \theta_{R-1}\}$ such that the region between two parallel hyperplanes defined by the same $U_i$ with different threshold $\theta_{q-1}$ and $\theta_{q}$ corresponds to the rating $q$. Thus, in P\textsubscript{ord}(Y), 
in addition to learning the latent factor matrices $U$ and $V$ it is also needed to learn the thresholds $\theta$'s.
There may be a debate on the number of $\theta$'s needed, but following the original proposal of MMMF~\cite{rennie2005fast}, we follow $R-1$ thresholds for each user and hence there are $n(R-1)$ thresholds. 

There are different approaches for constructing a loss function based on a set of thresholds such as immediate-threshold and all-threshold~\cite{rennie2005loss}. For each observed entry and the corresponding prediction pair $(y_{ij}, U_iV_j^T)$, the immediate-threshold based approach calculates the loss as sum of immediate-threshold violation which is $\ell(U_iV_j^T - \theta_{i,y_{ij}-1})~+~\ell(\theta_{i,y_{ij}} - U_iV_j^T)$. On the other hand, all-threshold loss is calculated as the sum of loss for all thresholds which is the cost of crossing multiple rating-boundaries and is defined as follows.
\begin{equation}
 \ell(U_iV_j^T, y_{ij}) = \sum_{r=1}^{y_{ij} - 1} {\ell(U_iV_j^T - \theta_{i,r})} + \sum_{r=y_{ij}}^{R-1} {\ell(\theta_{i,r} - U_iV_j^T)}
\end{equation}
\noindent
Using the $R-1$ thresholds $\theta$'s, MMMF extended the hinge loss function meant for binary classification to ordinal settings. The difference between immediate-threshold and all-threshold hinge is illustrated with the help of the following example. Let us consider the partially observed ordinal rating matrix $Y$ with $R=5$, the learnt factor matrices $U$, $V$ and the set of thresholds $\theta$'s for each user (Table \ref{ordinalY_U_V_theta}).
\begin{table}[ht!]
\renewcommand{\arraystretch}{1.2}
\centering
\caption{Runnning example.}
\begin{adjustbox}{max width=\textwidth}
 \begin{minipage}{0.50\textwidth}
    \centering
    \begin{tabular}{|r|r|r|r|r|r|r|r|r|r|}
      \hline
	1&2&4&3&5&2&1&4&3&0 \\  \hline
	0&4&5&3&2&4&1&1&3&2 \\  \hline
	4&3&5&5&2&0&2&3&1&4 \\  \hline
	5&0&1&2&3&1&4&3&5&2 \\  \hline
	2&5&3&4&2&4&0&3&1&1 \\  \hline
	0&1&5&3&1&5&3&2&4&4 \\  \hline       
	4&2&5&1&4&3&2&5&0&3 \\  \hline 
	2&2&3&1&3&4&5&0&5&4 \\  \hline
	5&4&5&2&3&1&4&0&1&2 \\  \hline   
	3&0&4&5&1&2&1&5&3&4 \\  \hline
    \end{tabular}
    \captionof{subfigure}{Y}
  \end{minipage}%
  \begin{minipage}{0.22\textwidth}
 \centering
  \begin{tabular}{|r|r|}
      \hline
      0.22&-0.12 \\  \hline
      0.90& 0.05 \\  \hline
      1.45& 0.31 \\  \hline
     -1.52& 0.04 \\  \hline
      0.94& 1.06 \\  \hline
      1.09&-1.45 \\  \hline
      0.10& 0.14 \\  \hline
     -0.60&-1.65 \\  \hline
     -0.24& 0.71 \\  \hline
      0.48& 0.59 \\  \hline
    \end{tabular}
    \captionof{subfigure}{U}
 \end{minipage}%
 \begin{minipage}{0.22\textwidth}
 \centering
  \begin{tabular}{|r|r|}
      \hline
	-0.79& 1.01 \\  \hline
	 0.10& 1.21 \\  \hline
	 1.51&-0.32 \\  \hline
	 0.76& 0.63 \\  \hline
	-0.53& 0.24 \\  \hline
	 1.45&-0.79 \\  \hline
	-0.98&-0.74 \\  \hline
	-0.17& 0.72 \\  \hline
	-0.72&-1.32 \\  \hline
	 0.39&-0.63 \\  \hline
  \end{tabular}
  \captionof{subfigure}{V}
 \end{minipage}%
 \begin{minipage}{0.40\textwidth}
 \centering
  \begin{tabular}{|r|r|r|r|}
      \hline
	-0.61&	-0.18&	0.51&	1.21 \\  \hline
	-0.74&	 0.09&	0.69&	1.41 \\  \hline
	-1.42&	-0.65&	0.28&	0.91 \\  \hline
	-1.35&	-0.25&	0.74&	0.96 \\  \hline
	-0.50&	 0.25&	0.77&	1.44 \\  \hline
	-1.08&	-0.60&	0.56&	1.51 \\  \hline
	-1.10&	-0.24&	0.34&	0.62 \\  \hline
	-1.41&	-0.62&	0.13&	0.99 \\  \hline
	-0.73&	-0.24&	0.02&	0.83 \\  \hline
	-0.73&	-0.51&	0.09&	0.90 \\  \hline
  \end{tabular}
  \captionof{subfigure}{$\theta$'s}
 \end{minipage}
 \end{adjustbox}
 \label{ordinalY_U_V_theta}
\end{table}\\
\noindent
The following real-valued prediction matrix is obtained from the above latent factor matrices $U$ and $V$ corresponding to the matrix $Y$.
\begin{table}[ht!]
\renewcommand{\arraystretch}{1.2}
\centering
\caption{Real-valued prediction corresponding to ordinal rating matrix $Y$.}
\begin{adjustbox}{max width=\textwidth}
   \begin{minipage}{1.2\textwidth}
    \centering
    \begin{tabular}{|r|r|r|r|r|r|r|r|r|r|}
      \hline
	-0.30&	-0.12&	 0.37&	 0.09&	-0.15&	 0.41&	-0.13&	-0.12&	 0.00&	 0.16 \\  \hline
	-0.66&	 0.15&	 1.34&	 0.72&	-0.47&	 1.27&	-0.92&	-0.12&	-0.71&	 0.32 \\  \hline
	-0.83&	 0.52&	 2.09&	 1.30&	-0.69&	 1.86&	-1.65&	-0.02&	-1.45&	 0.37 \\  \hline
	 1.24&	-0.10&	-2.31&	-1.13&	 0.82&	-2.24&	 1.46&	 0.29&	 1.04&	-0.62 \\  \hline
	 0.33&	 1.38&	 1.08&	 1.38&	-0.24&	 0.53&	-1.71&	 0.60&	-2.08&	-0.30 \\  \hline
	-2.33&	-1.65&	 2.11&	-0.09&	-0.93&	 2.73&	 0.00&	-1.23&	 1.13&	 1.34 \\  \hline
	 0.06&	 0.18&	 0.11&	 0.16&	-0.02&	 0.03&	-0.20&	 0.08&	-0.26&	-0.05 \\  \hline
	-1.19&	-2.06&	-0.38&	-1.50&	-0.08&	 0.43&	 1.81&	-1.09&	 2.61&	 0.81 \\  \hline
	 0.91&	 0.84&	-0.59&	 0.26&	 0.30&	-0.91&	-0.29&	 0.55&	-0.76&	-0.54 \\  \hline
	 0.22&	 0.76&	 0.54&	 0.74&	-0.11&	 0.23&	-0.91&	 0.34&	-1.12&	-0.18 \\  \hline
    \end{tabular}
  \end{minipage}%
  \end{adjustbox}
\end{table}\\
\noindent
One can see that the entry $y_{13}$ is observed as $4$ and the corresponding real-valued prediction is $0.37$. When immediate-threshold hinge is the loss measure, the overall loss is calculated as follows.
\begin{align*}
 h(y_{13}, U_1V_3^T) &= h(U_1V_3^T - \theta_{i,3}) + h(\theta_{i,4} - U_1V_3^T)\\
		     &= h(0.37 - 0.51) + h(1.21 - 0.37)\\
		     &= h(-0.14) +  h(0.84)\\
		     &= 0.65
\end{align*}

\noindent
where $h(\cdot)$ is the smooth hinge loss function as defined previously. For the same example, the overall loss with all-threshold hinge function is calculated as follows.
\begin{align*}
 h(y_{13}, U_1V_3^T) &= h(U_1V_3^T - \theta_{i,1}) + h(U_1V_3^T - \theta_{i,2}) + h(U_1V_3^T - \theta_{i,3}) + h(\theta_{i,4} - U_1V_3^T)\\
		     &= h(0.37+ 0.61) + h(0.37 + 0.18)) +h(0.37 - 0.51) + h(1.21 - 0.37)\\
		     &= h(0.98) + h(0.55) +  h(-0.14) +  h(0.84)\\
		     &= 0.75
\end{align*}
Continuing our discussion on geometrical interpretation, immediate-threshold loss tries to embed the point $V_3$ into the region defined by the parallel hyperplanes $(U_1, \theta_{i,3})$ and $(U_1, \theta_{1,4})$ which basically mean that $U_1V_3^T - \theta_{1,3} > 0$ and $U_1V_3^T - \theta_{1,4} < 0$. The all-threshold hinge function not only tries to embed the points rated as $r$ into the region defined by $(U_i, \theta_{r-1})$ and $(U_i, \theta_r)$  but also consider the position of the points with respect to other hyperplanes. It is also desirable that every point $V_j$ rated by user $i$ should satisfy the condition  $U_iV_j^T - \theta_{i,r-1} > 0$ for $r < y_{ij}$ and $U_iV_j^T - \theta_{i,r} < 0$ for $r \ge y_{ij}$.

In MMMF~\cite{rennie2005fast}, each hyperplane acts as a maximum-margin separator which is ensured by considering smooth hinge as the loss function (all-threshold hinge). The resulting optimization problem for P\textsubscript{ord}(Y) is
\begin{equation}
\underset{U, V}{min}\;J(U,V,\theta) \sum_{(i,j) \in \Omega} { \bigg( \sum_{r = 1}^{y_{ij}-1}{h (U_{i}V_{j}^{T} - \theta_{i,r})} + \sum_{r = y_{ij}}^{R-1}{h (\theta_{i,r} - U_{i}V_{j}^{T})}\bigg)} + \frac{\lambda}{2}(\|U\|^{2}_{F} + \|V\|^{2}_{F})
\end{equation}
where $\|. \|_{F}$ is the Frobenius norm,  $\lambda > 0$ is  the regularization parameter, $\Omega$ is the set of observed entries,  $h(z)$ is the smooth hinge loss as defined previously and $\theta_{i,r}$ is the threshold for rank $r$ of user $i$. The equation given above can be rewritten as follows.
\begin{equation}
\label{mmmfOptimizationProblem}
\underset{U, V}{min}\;J(U,V,\theta) \; \sum_{r = 1}^{R-1}{\sum_{(i,j) \in \Omega}h \big(T_{ij}^{r}(\theta_{i,r} - U_{i}V_{j}^{T}))} + \frac{\lambda}{2}(\|U\|^{2}_{F} + \|V\|^{2}_{F})
\end{equation}

\noindent
where $T$ is defined as

\begin{equation*}
T_{ij}^r = 
\begin{cases}
+1, & \text{if r $\geq$ $y_{ij}$;} \\
-1, & \text{if r $ < $ $y_{ij}$.}
\end{cases}
\end{equation*}

The gradients of the variables to be optimized are determined as follows. The gradient with respect to each element of $U$ is calculated as follows.
\begin{align}
\frac{\partial J}{\partial U_{ip}} &= \sum_{r = 1}^{R-1}{ \sum_{j|(i,j) \in \Omega}{ \frac{\partial h \big(T_{ij}^{r}(\theta_{i,r} - U_{i}V_{j}^{T})\big) }{\partial U_{ip}} } } + \frac{\lambda}{2}\bigg(\frac{\partial \|U\|^{2}_{F}}{\partial U_{ip}} +  \frac{\partial \|V\|^{2}_{F}}{\partial U_{ip}}\bigg) \nonumber \\
 &=\sum_{r = 1}^{R-1}{ \sum_{j|(i,j) \in \Omega}{ T_{ij}^{r} h' \big(T_{ij}^{r}(\theta_{i,r} - U_{i}V_{j}^{T})\big) \frac{\partial (\theta_{i,r} - U_{i}V_{j}^{T})}{\partial U_{ip}} } } + \lambda U_{ip} \nonumber \\
 &= \lambda U_{ip} - \sum_{r = 1}^{R-1}{ \sum_{j|(i,j) \in \Omega}{ T_{ij}^{r} h' \big(T_{ij}^{r}(\theta_{i,r} - U_{i}V_{j}^{T})\big)V_{jp}  } }
\end{align}

\noindent
where $h'(\cdot)$ is the same as defined previously. Similarly, the gradients with respect to each element of $V$ is calculated as
\begin{align}
\frac{\partial J}{\partial V_{jq}} 
 &= \lambda V_{jq} - \sum_{r = 1}^{R-1}{ \sum_{i|(i,j) \in \Omega}{ T_{ij}^{r} h' \big(T_{ij}^{r}(\theta_{i,r} - U_{i}V_{j}^{T})\big)U_{iq}  } } 
\end{align}

\noindent
and the gradient with respect to each $\theta_{i,r}$ is determined as follows.
\begin{align}
\frac{\partial J}{\partial \theta_{i,r}} &= \sum_{j|(i,j) \in \Omega}{ \frac{\partial h \big(T_{ij}^{r}(\theta_{i,r} - U_{i}V_{j}^{T})\big) }{\partial \theta_{i,r}} }  + \frac{\lambda}{2}\bigg(\frac{\partial \|U\|^{2}_{F}}{\partial \theta_{i,r}} +  \frac{\partial \|V\|^{2}_{F}}{\partial \theta_{i,r}}\bigg) \nonumber \\
 &= \sum_{j|(i,j) \in \Omega}{ T_{ij}^{r} h' \big(T_{ij}^{r}(\theta_{i,r} - U_{i}V_{j}^{T})\big) \frac{\partial (\theta_{i,r} - U_{i}V_{j}^{T})}{\partial \theta_{i,r}} }   \nonumber \\
 &= \sum_{j|(i,j) \in \Omega}{ T_{ij}^{r} h' \big(T_{ij}^{r}(\theta_{i,r} - U_{i}V_{j}^{T})\big)}
\end{align}
Once $U$, $V$ and $\theta$'s are computed, the matrix completion process is accomplished as follows.
\begin{equation*}
\hat y_{ij} = 
\begin{cases}
r, & \text{if $(i,j) \notin \Omega \wedge (\theta_{i,r} \le x_{ij} \le \theta_{i,r+1}) \wedge (0 \le r \le R-1)$}; \\
y_{ij}, & \text{if $(i,j) \in \Omega$},
\end{cases}
\end{equation*}
where, $\hat y_{ij}$ is the prediction for item $j$ by user $i$. For simplicity of notation, we assume $\theta_{i,0} = -\infty$ and $\theta_{i,R} = +\infty$ for each user $i$.


\begin{figure*}
	\centering
	\includegraphics[width=5.8in,height=4.1in]{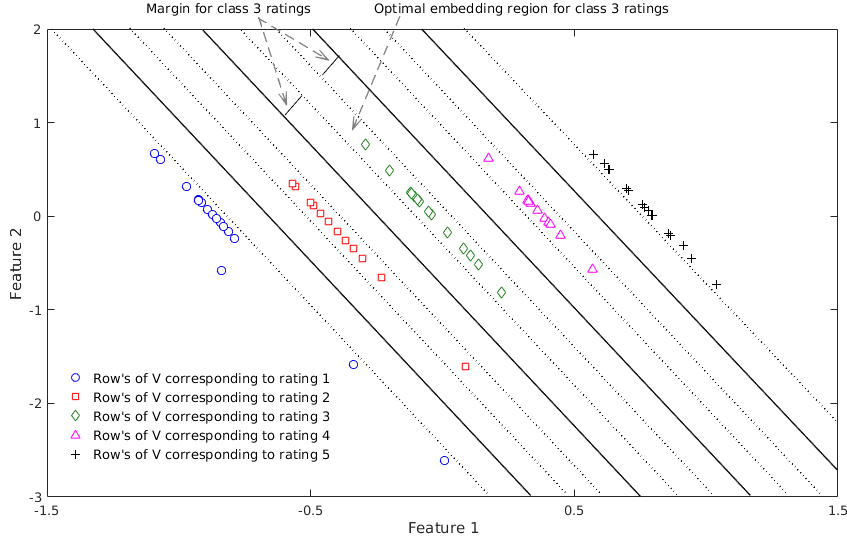}
	\captionof{figure}{Classification by  MMMF for the $i$th user}
	\label{mmmfClassification}
\end{figure*}

\begin{figure*}
 \centering
 \fbox{\includegraphics[width=5.6in,height=3.4in]{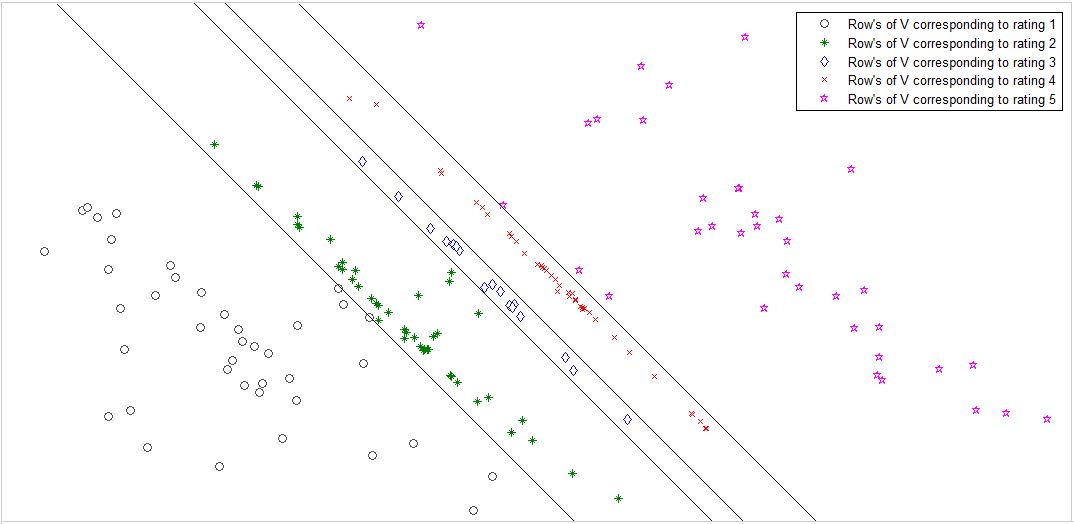}}
 \caption{Classification by MMMF for $i$th user}
\label{mmmfClassificationUnequalMargin}
 \end{figure*}

The objective of MMMF is to learn the embedding of items as points, users as hyperplanes and ratings as thresholds such that the points fall as correctly as possible into regions with the margin of ratings defined by the decision hyperplane $(U_j, \theta_{j,r})$. We illustrate this concept by taking a synthetic dataset of size $5 \times 1000$ with $10\%$ of observed entries. The number of ordinal ratings is $5$ and $d=2$. Figure~\ref{mmmfClassification} gives the decision hyperplane for a user and embedding of points corresponding to the items. Since $R=5$, there are $4$ hyperplanes that subdivide the entire space into $5$ regions corresponding to $5$ ratings. The data so chosen is balanced in the sense that the number of items for each rating is uniform. From the figure, it is clear that the embedding is learnt perfectly and the points corresponding to the items fall in the region of respective ratings with the margin. The margin is also shown in the figure by drawing parallel lines artifically. Thus, one can see that the embedding is achieved with true sense of maximum-margin. However, this is not so for all cases. It is observed through several experiments that the margin for the parallel lines is not the same and these hyperplanes separate classes with unequal margin with bias toward classes with sparse training samples. When the data is balanced, as in the previous case (Figure~\ref{mmmfClassification}), the unequal margin phenomenon is not evident. In another synthetic example of size $5 \times1000$ with $20\%$ of observed entries, $R=5$, $d=2$ and where the data is not balanced the inequalities in the margin size is evident. This is depicted in Figure~\ref{mmmfClassificationUnequalMargin}. In this example the observed entries for rating $2, 3$ and $4$ are substantially less than those for ratings $1$ and $5$. The observed entries for rating $3$ is substantially less than the observed entries for ratings $2$ and $4$. When the points are embedded in two dimensional plane the parallel separating line corresponding to different thresholds for a given user have different margins and this is depicted in Figure~\ref{mmmfClassificationUnequalMargin}. One can see that the hyperplanes tend to come closer towards the ratings with sparse samples. The hyperplane separating regions $4$ and region $5$ is not giving sufficient margin for rating $4$ but overfits rating $5$. The hyperplane separating regions $3$ and $4$ have unequal margin on both sides. So as a conclusion, it is observed that the proposed MMMF~\cite{rennie2005fast} is not truly handling margin maximization for predictions of ratings.
  
\begin{figure}[ht!]
	\centering
	  \includegraphics[width=4.5in,height=4.2in]{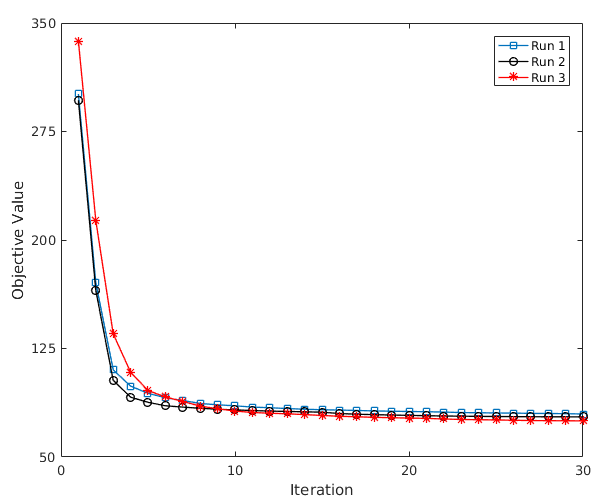}
	\caption{Convergence graph of  MMMF with different initial points}
	\label{mmmfObjectiveValue}
\end{figure}

We also show the behaviour of $J$ defined in Equation~\ref{mmmfOptimizationProblem}. We plot the value of $J$ obtained in every iteration for the same $Y$ (Table~\ref{ordinalY_U_V_theta}) starting from different initial points. It can be seen from Figure~\ref{mmmfObjectiveValue} that $J$ is having asymptotic convergence.
 
For ordinal ratings with discrete values such as $1$-$5$ stars, it is necessary to think of a sort of multiclass separation such that items are classified based on ratings which take on values from a finite discrete set. MMMF \cite{rennie2005fast} follows the principle of single machine extension and  handles ordinal rating matrix factorization by proposing a multi-class extension of the hinge loss function.

\section{HMF- The Proposed Method}
\label{hmf_section}
In this section, we describe the proposed method, termed as \textit{HMF (Hierarchical Matrix Factorization)}. As discussed in Section~\ref{hmfIntroduction}, the proposed method draws motivation from research on multi-class classification. There are two major approaches for addressing multi-class classification problems namely \textit{embedded techniques}~\cite{weston1999support, lee2004multicategory, crammer2001algorithmic, blondel2013block} and \textit{combinational technique}~\cite{hsu2002comparison, rifkin2004defense, bahlmann2002online}. The first approach explicitly formulate the multi-class problem directly. The second approach (\textit{combinational technique}) is to decompose the multi-class problem to multiple independent binary classification problems. In (MMMF)~\cite{rennie2005fast}, an \textit{embedded} approach is proposed to address the P\textsubscript{ord}(Y) problem. The authors has extended hinge loss defined for binary-level to multi-level cases. As discussed in Section~\ref{F-MMMF}, the extension of hinge-loss to multi-level does not preserve the property of maximum margin and also impose the parallel constraints on decision hyperplane. Our goal is to examine a solution to the P\textsubscript{ord}(Y) problem by constructing a hierarchy of bi-level MMMF using the principle of combinational approach. We believe that by following the principle of combinational approach the maximum margin advantage of bi-level matrix factorization can be retained and the parallel constraints on the decision hyperplane can also be relaxed.

As discussed previously, HMF is a stage-wise matrix completion technique that makes use of several bi-level MMMFs in a hierarchical fashion. At every stage $q$ of HMF, the original problem  P\textsubscript{ord}(Y) is converted to P\textsubscript{$\pm{1}$}(Y) and solved using the method described in Section~\ref{bi_level_MMMF}. The output at stage $q$ is the learnt latent factor matrices $U^q$ and $V^q$. The so computed latent factors $U^q$ and $V^q$ are used to predict the entries of the matrix of a specific ordinal value $q$. In other words, at Stage $q$ , HMF attempts to complete only those unobserved entries of the rating matrix where the rating $q$ is predicted. Thus for a $R$ -level ordinal rating matrix, we use $R-1$ stages. The method of conversion from P\textsubscript{ord}(Y) to P\textsubscript{$\pm{1}$}(Y) at stage $q$ is discussed below.

\noindent
\textbf{Conversion of  P\textsubscript{ord}(Y) to P\textsubscript{$\pm{1}$}(Y): }
At every stage $q$ of HMF, a BMMMF is employed and for this purpose, the  P\textsubscript{ord}(Y) is converted to P\textsubscript{$\pm{1}$}(Y) problem as follows.
\begin{equation}
 y_{ij}^{q} =
\begin{cases}
-1, & \text{if $(i,j) \in \Omega \wedge y_{ij} \leq q$};\\
+1, & \text{if $(i,j) \in \Omega \wedge y_{ij} > q$};\\
0, & \text{if $(i,j) \notin \Omega$.}
\end{cases}
\label{binaryConversion}
\end{equation}
This conversion states that any rating above $q$ is treated as $likes$ and below or equal to $q$ is treated as $dislike$.

The outcome of BMMMF at stage $q$ are the factor matrices $U^{q}$ and $V^{q}$ approximating $Y^q$. Let $\hat Y^q$ be the bi-level prediction matrix obtained from the factor matrices  $U^{q}$ and $V^{q}$ and 
$\Omega^q$ is the index set corresponding to the $-1$ entries in $\hat Y^q$. For simplicity of notation, we assume $\Omega^0$ is empty. The entries in ordinal rating prediction matrix $\hat Y$ are predicted  as $q$ based on the following rule.
\begin{equation}
\hat y_{ij} 
= 
\begin{cases} 
q & \text{if $(i,j) \notin {(\Omega^0 \cup  \Omega^1 \dots \cup \Omega^{q-1})} \wedge (\Omega^q = -1 )$}\\
y_{ij} & \text{if $(i,j) \in \Omega$ }
\end{cases}
\label{binaryConversion1}
\end{equation}
At the last stage, stage $R-1$, after employing the above rule to predict entries with values $R-1$, we complete the matrix by assigning $R$ for all remaining unobserved entries.

Algorithm \ref{hmfAlgo} outlines the main flow of the proposed method HMF for the ordinal rating matrix completion problem.
\begin{algorithm}[ht!]
\SetAlgoLined
\SetKwData{Left}{left}\SetKwData{This}{this}\SetKwData{Up}{up}
\SetKwFunction{Union}{Union}\SetKwFunction{FindCompress}{FindCompress}
\SetKwInOut{Input}{input}\SetKwInOut{Output}{output}
\Input{Ordinal Rating Matrix: $Y$, Maximum Rating: $R$, Number of Latent $~~$Factors: $d$, Threshold: $\theta$, Regularization Parameters: $\{\lambda_1, \dots, \lambda_{R-1}\}$} 
\Output{$\hat Y$}
\BlankLine
Initialize: $\hat Y \leftarrow 0$\;
\For{$q =~1~to~R-1$}{
$Y^q \leftarrow 0$\;
\For{$(i,j) \in \Omega$}{
\uIf{ $y_{ij} \le q$}{
 $y_{ij}^q = -1$\;}
\Else{
 $y_{ij}^q = +1$\;}
}
($U^q$, $V^q$) $\leftarrow$ BMMMF ($Y^q$, $d$, $\lambda_q$)\;
\For{ever user-item pair $(i,j)$}{
\uIf{ $U_i^qV_j^{q^T} < \theta$}{
 $\hat y_{ij}^q$ = -1\;}
\Else{
 $\hat y_{ij}^q$ = +1\;}
}
\For{ever user-item pair $(i,j)$}{
\If{$\big( (i,j) \notin \Omega \wedge \hat y_{ij}^q = -1   \big) \vee y_{ij} = q$}{$\hat y_{ij} \leftarrow q$}
}
}
$\hat y_{ij} \leftarrow R$, ~if $\hat y_{ij} = 0$, $\forall (i,j) \notin \Omega$ \;
\Return $\hat Y$\;
\caption{HMF ($Y$, $R$, $d$, $\theta$, $\lambda$)}
\label{hmfAlgo}
\end{algorithm}

We illustrate the working of HMF using a toy example. Let us consider the following partially observed ordinal rating matrix $Y$ of size $5 \times 7$ with $5$ levels.
\begin{table}[ht!]
\renewcommand{\arraystretch}{1}
 \centering
 \caption{Ordinal rating matrix Y}
 \begin{tabular}{|c|c|c|c|c|c|c|}
  \hline
  3&0&0&5&2&0&0 \\ \hline
  5&4&0&1&5&3&4 \\ \hline
  1&0&4&0&3&1&0 \\ \hline
  5&4&0&0&0&0&1 \\ \hline
  0&3&2&0&5&2&0 \\ \hline  
 \end{tabular}
 \label{ordinalYhmf}
\end{table}

\noindent
For notational convenience, we use $ij$ to denote $(i,j)$. The set of observed entries $\Omega = \{11, 14, 15, 21, 22, 24, 25, 26, 27, 31, 33, 35, 36, 41, 42, 47, 52, 53, 55, 56 \}$.  For each of the $4$ stages, the original $5$-level rating matrix $Y$ is converted to a bi-level ($\pm1$) matrix. Taking $d=2$, we employed bi-level MMMF to get factor matrix pairs $(U^q, V^q)$ for each stage $q$ independently. At stage $1$, the bi-level matrix $Y^1$ obtained from $Y$, the learnt latent factor matrices $U^1$ and $V^1$ and the corresponding bi-level prediction $\hat Y^1$ is shown in Table~\ref{bilevelStage1output}.

\begin{table}[ht!]
\renewcommand{\arraystretch}{1}
\centering
\caption{Runnning example.}
\begin{adjustbox}{max width=\textwidth}
 \begin{minipage}{0.38\textwidth}
    \centering
    \vspace*{-1cm}
    \begin{tabular}{|r|r|r|r|r|r|r|}
      \hline
      1& 0& 0& 1& 1& 0& 0 \\ \hline
      1& 1& 0&-1& 1& 1& 1 \\ \hline
     -1& 0& 1& 0& 1&-1& 0 \\ \hline
      1& 1& 0& 0& 0& 0&-1 \\ \hline
      0& 1& 1& 0& 1& 1& 0 \\ \hline  
    \end{tabular}
    \captionof{subfigure}{$Y^1$}
  \end{minipage}%
  \begin{minipage}{0.20\textwidth}
 \centering
 \vspace*{-1cm}
  \begin{tabular}{|r|r|}
      \hline
      -0.48&-0.54 \\ \hline
       0.12&-1.09 \\ \hline
       0.98&-0.13 \\ \hline
      -0.77&-0.29 \\ \hline
      -0.01&-0.94 \\ \hline
    \end{tabular}
    \captionof{subfigure}{$U^1$}
 \end{minipage}%
 \begin{minipage}{0.20\textwidth}
 \centering
  \begin{tabular}{|r|r|}
       \hline
      -0.70&-0.63 \\ \hline
      -0.36&-0.72 \\ \hline
       0.47&-0.53 \\ \hline
      -0.51& 0.25 \\ \hline
       0.28&-0.80 \\ \hline
      -0.52&-0.66 \\ \hline
       0.58&-0.37 \\ \hline
  \end{tabular}
  \captionof{subfigure}{$V^1$}
 \end{minipage}%
 \begin{minipage}{0.38\textwidth}
    \centering
    \vspace*{-1cm}
    \begin{tabular}{|r|r|r|r|r|r|r|}
      \hline
      1& 1& 1& 1& 1& 1&-1 \\ \hline
      1& 1& 1&-1& 1& 1& 1 \\ \hline 
     -1&-1& 1&-1& 1&-1& 1 \\ \hline
      1& 1&-1& 1& 1& 1&-1 \\ \hline
      1& 1& 1&-1& 1& 1& 1 \\ \hline
    \end{tabular}
    \captionof{subfigure}{$\hat Y^1$}
  \end{minipage}
 \end{adjustbox}
 \label{bilevelStage1output}
\end{table}%
\vspace{-2.5em}

\noindent
At stage $1$, the set of entries $\Omega^1 = \{17, 24, 31, 32, 34, 36, 43, 47, 54\}$ are predicted as $-1$. The set of entries $\Omega^1 \setminus \Omega $ i.e., $\{17, 32, 34, 43, 54\}$ and the set of entries where $1$ are observed now contains rating $1$ in the partially complete prediction matrix $\hat Y$ (Table~\ref{Y1cap}). 
\begin{table}[ht!]
\centering
\caption{Partially complete matrix after stage 1}
 \begin{tabular}{|r|r|r|r|r|r|r|}
      \hline
      0&0&0&0&0&0&1 \\ \hline
      0&0&0&1&0&0&0 \\ \hline
      1&1&0&1&0&1&0 \\ \hline
      0&0&1&0&0&0&1 \\ \hline
      0&0&0&1&0&0&0 \\ \hline
 \end{tabular}
 \label{Y1cap}
\end{table}

\noindent Similarly, at stage $2$, the original matrix is mapped to bi-level matrix $Y^2$. The bi-level MMMF is employed on $Y^2$ to get the factor matrices $U^2$ and $V^2$. The factor matrices and the corresponding bi-level prediction $\hat Y^2$ is shown in the Table~\ref{bilevelStage2output}.
\begin{table}[ht!]
\centering
\caption{Runnning example.}
\begin{adjustbox}{max width=\textwidth}
 \begin{minipage}{0.38\textwidth}
    \centering
    \vspace*{-1cm}
    \begin{tabular}{|r|r|r|r|r|r|r|}
      \hline
      1&0&0&1&-1&0&0 \\ \hline
      1&1&0&-1&1&1&1 \\ \hline
      -1&0&1&0&1&-1&0 \\ \hline
      1&1&0&0&0&0&-1 \\ \hline
      0&1&-1&0&1&-1&0 \\ \hline   
    \end{tabular}
    \captionof{subfigure}{$Y^2$}
  \end{minipage}%
  \begin{minipage}{0.2\textwidth}
 \centering
 \vspace*{-1cm}
  \begin{tabular}{|r|r|}
      \hline
      -0.69&-0.47 \\ \hline
      -0.43&1.06 \\ \hline
      0.92&0.01 \\ \hline
      -0.79&0.05 \\ \hline
      0.24&0.86 \\ \hline
    \end{tabular}
    \captionof{subfigure}{$U^2$}
 \end{minipage}%
 \begin{minipage}{0.2\textwidth}
 \centering
  \begin{tabular}{|r|r|}
       \hline
      -0.89&0.21 \\ \hline
      -0.43&0.64 \\ \hline
      0.42&-0.54 \\ \hline
      -0.21&-0.67 \\ \hline
      0.53&0.76 \\ \hline
      -0.76&0.01 \\ \hline
      0.35&0.56 \\ \hline
  \end{tabular}
  \captionof{subfigure}{$V^2$}
 \end{minipage}%
  \begin{minipage}{0.38\textwidth}
    \centering
    \vspace*{-1cm}
    \begin{tabular}{|r|r|r|r|r|r|r|}
      \hline
      1&-1&-1&1&-1&1&-1 \\ \hline
      1&1&-1&-1&1&1&1 \\ \hline
      -1&-1&1&-1&1&-1&1 \\ \hline
      1&1&-1&1&-1&1&-1 \\ \hline
      -1&1&-1&-1&1&-1&1 \\ \hline
    \end{tabular}
    \captionof{subfigure}{$\hat Y^2$}
  \end{minipage}%
 \end{adjustbox}
 \label{bilevelStage2output}
\end{table}
\vspace{-1.5em}

\noindent As discussed previously, at any stage $q$, the candidate set of entries for prediction will exclude the set of observed entries and the set of entries predicted before stage $q$. At stage $2$, the set of entries $\Omega^2 = \{12, 13, 15, 17, 23, 24, 31, 32, 34, 36, 43, 45, 47, 51, 53,\linebreak 54, 56\}$ are predicted as $-1$. The set of entries $\Omega^2 \setminus \{\Omega^1 \cup \Omega\}$ i.e., $\{12, 13, 23, 45, 51\} $  along with the entries where $2$ are observed now contains rating $2$ in the partially complete prediction matrix $\hat Y$ (Table~\ref{Y2cap}).

\begin{table}[ht!]
\renewcommand{\arraystretch}{1}
\centering
\caption{Partially complete matrix after stage 2}
 \begin{tabular}{|r|r|r|r|r|r|r|}
      \hline
      0&2&2&0&2&0&1 \\ \hline
      0&0&2&1&0&0&0 \\ \hline
      1&1&0&1&0&1&0 \\ \hline
      0&0&1&0&2&0&1 \\ \hline
      2&0&2&1&0&2&0 \\ \hline      
 \end{tabular}
 \label{Y2cap}
\end{table}

\noindent
The bi-level mapping $Y^3$, the latent factor matrices $U^3$, $V^3$ and the corresponding bi-level prediction $\hat Y^3$ obtained at stage $3$ is shown in able~\ref{bilevelStage3output}.
\begin{table}[ht!]
\renewcommand{\arraystretch}{1}
\centering
\caption{Runnning example.}
\begin{adjustbox}{max width=\textwidth}
 \begin{minipage}{0.38\textwidth}
    \centering
    \vspace*{-1cm}
    \begin{tabular}{|r|r|r|r|r|r|r|}
      \hline
      -1& 0& 0& 1&-1& 0& 0 \\ \hline
       1& 1& 0&-1& 1&-1& 1 \\ \hline
      -1& 0& 1& 0&-1&-1& 0 \\ \hline
       1& 1& 0& 0& 0& 0&-1 \\ \hline
       0&-1&-1& 0& 1&-1& 0 \\ \hline   
    \end{tabular}
    \captionof{subfigure}{$Y^3$}
  \end{minipage}%
  \begin{minipage}{0.2\textwidth}
 \centering
 \vspace*{-1cm}
  \begin{tabular}{|r|r|}
      \hline
      -0.42& 0.70 \\ \hline
       0.26&-1.05 \\ \hline
      -0.80& 0.44 \\ \hline
       0.84& 0.08 \\ \hline
      -0.05&-0.81 \\ \hline
    \end{tabular}
    \captionof{subfigure}{$U^3$}
 \end{minipage}%
 \begin{minipage}{0.2\textwidth}
 \centering
  \begin{tabular}{|r|r|}
       \hline
       0.72&-0.54 \\ \hline
       0.80&-0.12 \\ \hline
      -0.18& 0.68 \\ \hline
      -0.26& 0.62 \\ \hline
       0.26&-0.83 \\ \hline
       0.59& 0.52 \\ \hline
      -0.39&-0.57 \\ \hline
  \end{tabular}
  \captionof{subfigure}{$V^3$}
 \end{minipage}%
  \begin{minipage}{0.38\textwidth}
    \centering
    \vspace*{-1cm}
    \begin{tabular}{|r|r|r|r|r|r|r|}
      \hline
    -1&    -1&     1&     1&    -1&     1&    -1 \\ \hline
     1&     1&    -1&    -1&     1&    -1&     1 \\ \hline
    -1&    -1&     1&     1&    -1&    -1&     1 \\ \hline
     1&     1&    -1&    -1&     1&     1&    -1 \\ \hline
     1&     1&    -1&    -1&     1&    -1&     1 \\ \hline
    \end{tabular}
    \captionof{subfigure}{$\hat Y^3$}
  \end{minipage}%
 \end{adjustbox}
 \label{bilevelStage3output}
\end{table}
\vspace{-1.5em}

\noindent
It can be seen from Table~\ref{bilevelStage3output} that the entries set $\Omega^3 = \{11, 12, 14, 15, 23, 25, 31, 32,\\ 35, 36, 43, 44, 46, 53, 54, 56\}$ are predicted as $-1$. The set of entries $\Omega^3 \setminus \{\Omega^1 \cup \Omega^2 \cup \Omega\}$ i.e., $\{44\}$ and the set of entries where $3$ are observed now is fixed to $3$ in $\hat Y$ (Table~\ref{Y3cap}).
\begin{table}[ht!]
\renewcommand{\arraystretch}{1}
\centering
\caption{Partially complete matrix after stage 3}
 \begin{tabular}{|r|r|r|r|r|r|r|}
      \hline
      3&2&2&0&2&0&1 \\ \hline
      0&0&2&1&0&3&0 \\ \hline
      1&1&0&1&3&1&0 \\ \hline
      0&0&1&3&2&0&1 \\ \hline
      2&3&2&1&0&2&0 \\ \hline      
 \end{tabular}
 \label{Y3cap}
\end{table}

\pagebreak
\noindent
Similarly, at stage $4$, the observed matrix $Y$ is mapped to bi-level matrix $Y^4$. Table~\ref{bilevelStage4output} shows the bi-level matrix $Y^4$, the factor matrix $U^4$, $V^4$  and the corresponding bi-level prediction matrix $\hat Y^4$ obtained at stage $4$.
\begin{table}[ht!]
\renewcommand{\arraystretch}{1.05}
\centering
\caption{Runnning example.}
\begin{adjustbox}{max width=\textwidth}
 \begin{minipage}{0.38\textwidth}
    \centering
    \vspace*{-1cm}
    \begin{tabular}{|r|r|r|r|r|r|r|}
      \hline
      -1& 0& 0& 1&-1& 0& 0 \\ \hline
       1&-1& 0&-1& 1&-1&-1 \\ \hline
      -1& 0&-1& 0&-1&-1& 0 \\ \hline
       1&-1& 0& 0& 0& 0&-1 \\ \hline
       0&-1&-1& 0& 1&-1& 0 \\ \hline   
    \end{tabular}
    \captionof{subfigure}{$Y^4$}
  \end{minipage}%
  \begin{minipage}{0.2\textwidth}
 \centering
 \vspace*{-1cm}
  \begin{tabular}{|r|r|}
      \hline
       0.76& 0.28 \\ \hline
      -0.76&-0.73 \\ \hline
       0.81&-0.58 \\ \hline
      -0.71&-0.43 \\ \hline
      -0.43&-0.85 \\ \hline
    \end{tabular}
    \captionof{subfigure}{$U^4$}
 \end{minipage}%
 \begin{minipage}{0.2\textwidth}
 \centering
  \begin{tabular}{|r|r|}
       \hline
      -0.89&-0.15 \\ \hline
       0.54& 0.55 \\ \hline
      -0.17& 0.69 \\ \hline
       0.58& 0.35 \\ \hline
      -0.86&-0.27 \\ \hline
       0.01& 0.83 \\ \hline
       0.54& 0.40 \\ \hline
  \end{tabular}
  \captionof{subfigure}{$V^4$}
 \end{minipage}%
  \begin{minipage}{0.38\textwidth}
    \centering
    \vspace*{-1cm}
    \begin{tabular}{|r|r|r|r|r|r|r|}
      \hline
    -1&     1&     1&     1&    -1&     1&     1 \\ \hline
     1&    -1&    -1&    -1&     1&    -1&    -1 \\ \hline
    -1&     1&    -1&     1&    -1&    -1&     1 \\ \hline
     1&    -1&    -1&    -1&     1&    -1&    -1 \\ \hline
     1&    -1&    -1&    -1&     1&    -1&    -1 \\ \hline
    \end{tabular}
    \captionof{subfigure}{$\hat Y^4$}
  \end{minipage}%
 \end{adjustbox}
 \label{bilevelStage4output}
\end{table}
\vspace{-1.5em}

\noindent
The set of entries $\Omega^4 = \{11, 15, 22, 23, 24, 26, 27, 31, 33, 35, 36, 42, 43, 44, 46, 47, 52, 53\\, 54, 56, 57\}$ are predicted as $-$ at stage $4$. The set of entries $\Omega^4 \setminus \{ \Omega^1 \cup \Omega^2 \cup \Omega^3 \cup \Omega\}$ i.e., $\{46, 57\}$ and the set of entries where $4$ are observed now contains rating $4$ in the partially complete prediction matrix $\hat Y$ (Table~\ref{Y4cap}).
\begin{table}[ht!]
\renewcommand{\arraystretch}{1}
\centering
\caption{Partially complete matrix after stage 4}
 \begin{tabular}{|r|r|r|r|r|r|r|}
      \hline
      3&2&2&0&2&0&1 \\ \hline
      0&4&2&1&0&3&4 \\ \hline
      1&1&4&1&3&1&0 \\ \hline
      0&4&1&3&2&4&1 \\ \hline
      2&3&2&1&0&2&4 \\ \hline      
 \end{tabular}
 \label{Y4cap}
\end{table}

\noindent
After completion of stage $4$, the set of entries where $0$ is present in $\hat Y^4$ are filled with rating $5$. The resulting complete matrix is shown in Table~\ref{Y5cap}.
\begin{table}[ht!]
\renewcommand{\arraystretch}{1}
\centering
\caption{Complete matrix computed by HMF}
 \begin{tabular}{|r|r|r|r|r|r|r|}
      \hline
      3&2&2&5&2&5&1 \\ \hline
      5&4&2&1&5&3&4 \\ \hline
      1&1&4&1&3&1&5 \\ \hline
      5&4&1&3&2&4&1 \\ \hline
      2&3&2&1&5&2&4 \\ \hline      
 \end{tabular}
 \label{Y5cap}
\end{table}

\begin{figure}[ht!]
\centering
\begin{adjustbox}{max size=\textwidth}
 \begin{minipage}{1\textwidth}
    \includegraphics[width=\textwidth,height=8.3in]{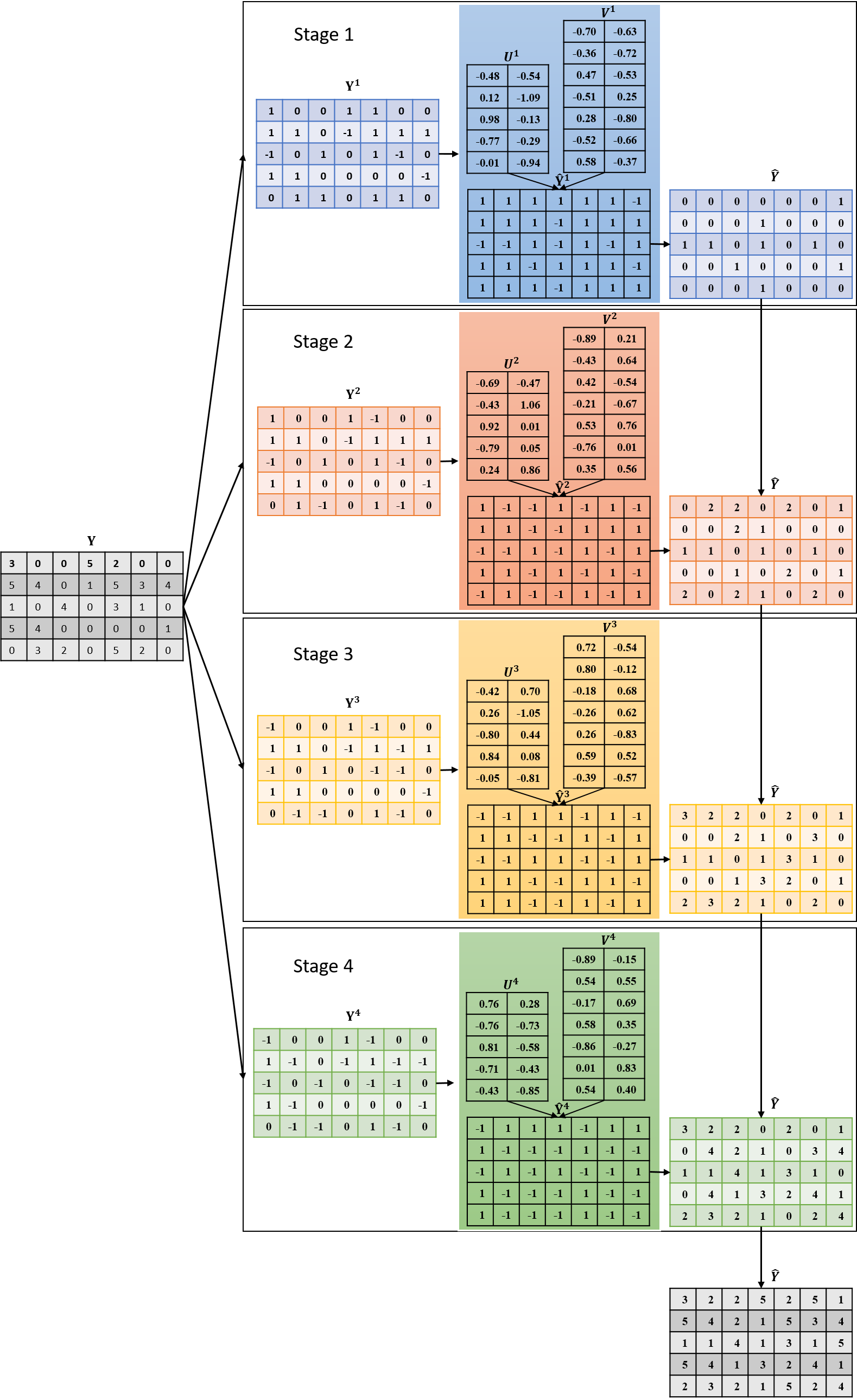}
    \caption{Hierarchical matrix factorization}
    \label{hmf-tree}
    \end{minipage}
 \end{adjustbox}
\end{figure}

For the example given in Table~\ref{ordinalYhmf},  the complete process of HMF is summarized in Figure~\ref{hmf-tree}. 

\section{Parallelization of HMF}
The process described in Section~\ref{hmf_section} is sequential as the candidate entries for stage $q$ are those which are completed till Stage $q-1$. A minor modification makes the process suitable for distributed computing. Since every pair of $(U, V)$ is used to predict only a non-overlapping subset of elements of $Y$, the computing need required at each stage $q$ can be ported to a parallel or a distributed environment. In other words, the factorization for each $Y^q$ can be accomplished in parallel on a multiprocessor system.
\begin{figure}[ht!]
	\centering
	\includegraphics[width=5in,height=3.1in]{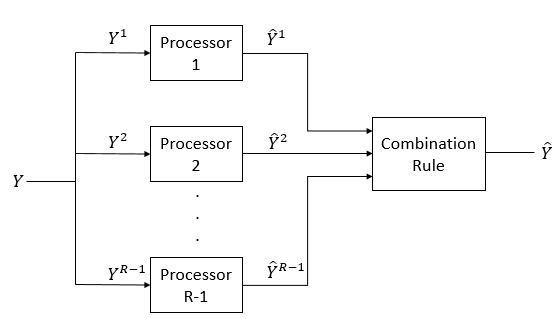}
	\captionof{figure}{Block diagram of a parallel architecture for HMF}
	\label{blockDiagramPHMF}
\end{figure}

Figure~\ref{blockDiagramPHMF} shows the block diagram of a parallel architecture for HMF. Each processor solves a P\textsubscript{$\pm{1}$}(Y) problem defined at stage $q$. After that, the output of each processor i.e., $\hat Y^1, \hat Y^2, \dots, \hat Y^{R-1}$ is  combined to obtain the complete prediction matrix $\hat Y$. Algorithm~\ref{phmfAlgo} outlines the main flow of the parallel architecture for HMF.
\begin{algorithm}[ht!]
\SetAlgoLined
\SetKwData{Left}{left}\SetKwData{This}{this}\SetKwData{Up}{up}
\SetKwFunction{Union}{Union}\SetKwFunction{FindCompress}{FindCompress}
\SetKwInOut{Input}{input}\SetKwInOut{Output}{output}
\Input{Ordinal Rating Matrix: $Y$, Maximum Rating: $R$, Number of Latent $~~$Factors: $d$, Threshold: $\theta$, Regularization Parameters: $\{\lambda_1, \dots, \lambda_{R-1}\}$} 
\Output{$\hat Y$}
\BlankLine
Initialize: $\hat Y \leftarrow 0$\;
\For{each processor q in parallel}{ 
$Y^q \leftarrow 0$\;
\For{$(i,j) \in \Omega$}{
\uIf{ $y_{ij} \le q$}{
 $y_{ij}^q = -1$\;}
\Else{
 $y_{ij}^q = +1$\;}
}
$\hat Y^q  \leftarrow 0$\;
($U^q$, $V^q$) $\leftarrow$ BMMMF ($Y^q$, $d$, $\lambda_q$)\;
\For{ever user-item pair $(i,j)$}{
\uIf{ $U_i^qV_j^{q^T} < \theta$}{
 $\hat y_{ij}^q$ = -1\;}
\Else{
 $\hat y_{ij}^q$ = +1\;}
 }
 }
\For{$q =~1~to~R-1$}{
\For{ever user-item pair $(i,j)$}{
\If{$\big( (i,j) \notin \Omega \wedge \hat y_{ij}^q = -1   \big) \vee y_{ij} = q$}{$\hat y_{ij} \leftarrow q$}
}
}
\Return $\hat Y$\;
\caption{PHMF ($Y$, $R$, $d$, $\theta$, $\lambda$)}
\label{phmfAlgo}
\end{algorithm}

\section{Experimental Analysis}
\label{experiment_section_hmf}
In this section we analyze the performance of HMF by taking into account factors such as accuracy and efficiency. In the following sections,  we describe the experimental setup including the data sets and relevant statistics, the experimental protocols, the competing algorithms, the evaluation metrics, the parameter setting and following this, we discuss the experimental results.

\begin{subsection}{Data Sets}
\label{hmfDatasets}
For a comprehensive performance evaluation, we conducted experiments on both \textit{real} and \textit{synthetic} data sets. We used three benchmark datasets for our experiments, namely \textit{MovieLens 100K}, \textit{MovieLens 1M} and \textit{EachMovie}, which are the standard data sets used in the matrix factorization community. All of these datasets can be downloaded from \textit{grouplens}\footnote{http://grouplens.org/datasets/ \label{datasetLabel}}. The \textit{MovieLens 100K} data set consists of $100,000$ ratings given by $943$ users for $1682$ movies. The \textit{MovieLens 1M} dataset consists of $1,000,209$ ratings, given by $6,040$ users for $3,952$ movies, out of which $3,706$ are actually rated and every user has at least $20$ ratings. There are five possible rating values, $\{1, 2 \dots , 5\}$ in \textit{MovieLens 100K} and \textit{MovieLens 1M}. The \textit{EachMovie} dataset consists of $2,811,983$ ratings, given by $72,916$ users for $1,628$ movies, out of which $1,623$ are actually rated and $36,656$ users has given at least $20$ ratings. There are six possible rating values, $\{0, 0.2,\dots , 1\}$ and we mapped them to \{1, 2 \dots, 6\}. The detailed characteristics of these data sets are summarized in Table~\ref{datasetsCharacteristicsHMF}.
\begin{table}[h!]
	\renewcommand{\arraystretch}{1.2}
	\centering
	\caption{ Description of the experimental datasets}
	\begin{tabular}{lllllll}
		\toprule
		Data set&\#Users&\#Items&\#Ratings&Sparsity&Rating-scale&Filtering \\
		\hline
		MovieLens 100K&943&1682&100,000&93.7\%&5&20\\
		MovieLens 1M&6040&3900&1,000,209&95.7\%&5&20\\
		EachMovie&72,916&1628&2,811,983&97.6\%&6&20\\
		\bottomrule
	\end{tabular}
	\label{datasetsCharacteristicsHMF}
\end{table}

\noindent \textbf{Synthetic Data Generation:}
We generated the synthetic dataset of size $1000 \times 1000$ with ratings in the range $1$-$5$. Synthetic data is generated by a novel method. The motivation for using \textit{synthetic} dataset in the experiments is to see whether a method can successfully retrieve the factor matrices when the input matrix is known to be the exact product of the latent factor matrices. Since the rating matrix is discrete and cannot be the exact product of two real-valued matrices, we make use of a novel iterative technique to generate the rating matrices.  Let $Y_0$ be a random rating matrix of size $N \times M$ and  $d \le min(n,m)$. We start with a random $N \times d$ factor matrix $U$. Then the latent factor matrix $V$ is 	obtained as $V = (\inv{(U^{T}U)}U^{T}Y_{0})^T$. We then compute $[UV^{T}]$ as $Y_{1}$, where $[\cdot]$ denotes rounding to the nearest integer\footnote{In case if nearest integer is exceeding (or below) the range of rating we set it to maximum (minimum) rating.}~. $Y_{1}$, so computed, becomes new $Y_0$ and $U$ is computed as   $U =  Y_{0}\inv{(VV^{T})}V$. This process is repeated iteratively, by alternatively updating $U$ and $V$, until the matrices stbilize. The current $Y_0$ becomes the $synthetic$ rating matrix.
\end{subsection}

\begin{subsection}{Experimental Protocols} Our experimental set up is based on two different and popular experimental protocols that have been proposed in the literature for evaluating the empirical performance of collaborative filtering methods i.e., weak and strong generalization~\cite{marlin2003modeling}. 
\begin{itemize}
	\item Weak generalization: Testing based on weak generalization protocol  is based on the 
	notion that the \emph{test set} is formed by randomly holding out one rating 
	from each user's rating set and the rest of the known ratings are 
	considered as part of the \emph{training set}. A prediction model is trained using the data in the 
	training set, and its performance is evaluated over the test set. Therefore weak generalization 
	is a one stage process wherein one can measure the ability of a model to generalize to 
	other items rated by the \emph{same users} used for training the model. 
	
	\item Strong generalization: 
	It is a two stage process in which, initially, a subset of users is randomly selected and completely removed from the training set so as 
	to form the test set. We can call this test set \textbf{G}. The initial prediction model (\textbf{M}) is trained (learned) with all available ratings from the bulk of the users in the training set (all users not in the test set \textbf{G}) . 
	In the second stage (testing phase) one rating  is randomly selected from each user in the set \textbf{G} to form the \textit{held out} set. The remaining ratings for each user in the set \textbf{G} can be used to tweak the prediction model (\textbf{M}) during strong testing. The model is evaluated by predicting the held out ratings. The main idea behind strong generalization is to build a prediction model for a large set of initial users that can be generalized later for a small set of novel users (\textbf{G}).	
\end{itemize}
\end{subsection}

\begin{subsection}{Comparing Algorithms} 
We consider the following nine well-known state-of-the-art algorithms for comparison:
\begin{itemize}
 \item \textbf{URP}~\cite{marlin2003modeling}: The user rating profile (URP) model is the generative version of the vector aspect model for rating profiles. The URP model sample the latent features for each user from a Dirichlet prior and  then perform variational inference on the user profile. Finally, compute the distribution over rating values for a particular unrated item given a user profile.
 \item \textbf{Attitude}~\cite{marlin2003modeling}: The attitude model representsthe latent space description of a user as a vector of attitude expression levels. Given an attitude vector of user, the probability of rating value for each item is a product of attitude expression levels and preference parameters.
 \item \textbf{MMMF}~\cite{rennie2005fast}: The \textit{hinge loss} function defined for bi-level is extended to multi-level for collaborative filtering with ordinal rating matrix when user-preferences are not in the form of like/ dislike but values in a discrete range. 
 \item \textbf{E-MMMF}~\cite{decoste2006collaborative}: To overcome the problem of local minima and impact of outliers (e.g. abnormal or even malicious raters) and other noise in MMMF, E-MMMF investigates diffrent ensemble methods such as \textit{multiple random weight seeds} and \textit{baggings} with MMMF.
 \item \textbf{PMF}~\cite{mnih2007probabilistic}: Salakhutdinov et al.~\cite{mnih2007probabilistic} proposed a probabilistic framework for matrix factorization  where the factor variables are assumed to be marginally independent while rating variables are assumed to be conditionally independent given the factor variables.
 \item \textbf{BPMF}~\cite{salakhutdinov2008bayesian}: Given the ratings, inferring the posterior distribution over the factors  is intractable in PMF. BPMF presented a fully Bayesian treatment of Probabilistic Matrix Factorization by placing hyperpriors over the hyperparameters. BPMF uses Markov chain Monte Carlo (MCMC) method for approximate inference in this model. 
 \item \textbf{GP-LVM}~\cite{lawrence2009non}: Lawrence et al. \cite{lawrence2009non}, developed a non-linear extension to PMF by generating latent components via Gaussian process latent variable models. 
 \item \textbf{iPMMMF and iBPMMMF}~\cite{xu2012nonparametric}: In \cite{xu2012nonparametric}, a noparametric Bayesian-style MMMF was proposed that utilizes nonparametric techniques to resolve the unknown number of latent factors in MMMF model.
 \item \textbf{Gibbs MMMF and iPMMMF}~\cite{xu2013fast}: A probabilistic interpretation of MMMF model through data augmentation is presented in~\cite{xu2013fast}.
\end{itemize}
\end{subsection}

\begin{subsection}{Evaluation Metrics}  
In order to evaluate the performance of the proposed method, we use two popular evaluation metrices: \textit{Normalized Mean Absolute Error (NMAE)}~\cite{marlin2003modeling} and \textit{Frobenius-norm RelativeError (FRE)}~\cite{ling2012decentralized}. NMAE is defined as Mean Absolute Error (MAE) divided by $1.6$ in the case of MovieLens data set and for EachMovie data set it is MAE divide by $1.944$~\cite{marlin2003modeling}. Given $Y$ a $N \times M$ partially observed user-item rating matrix, $\Omega$ the observed entries set and $\hat Y$ be a recovered matrix, the MAE and FRE are defined as follows. 
\begin{align*}
MAE &= \frac{\sum_{(i,j) \in \Omega}{|y_{ij} - \hat y_{ij}|}}{|\Omega|} & FRE &= \sqrt{\frac{\sum_{(i,j) \in \Omega}{(y_{ij} - \hat y_{ij})^2}}{\sum_{(i,j) \in \Omega}{y_{ij}^{2}}}} 
\end{align*}
\end{subsection}
\vspace{-10pt}
\begin{subsection}{Parameter Setting} 
We first demonstrate the effect of regularization parameter $\lambda$. Determining $\lambda$ value is one of the critical aspect for learning latent factors optimized over a loss function. 
To select the best regularization value for each dataset, we follow the same approach as defined in \cite{decoste2006collaborative}. 
\begin{figure}[ht!]
  \centering
  \includegraphics[width=4in,height=3.5in]{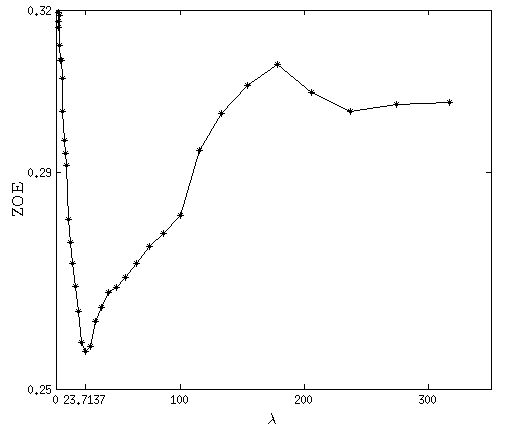}
  \caption{Validation score for different values of $\lambda$ on \textit{MovieLens} data set}
  \label{movieLensLambda}
\end{figure}
\begin{figure}[ht!]
\centering
  \includegraphics[width=4in,height=3.5in]{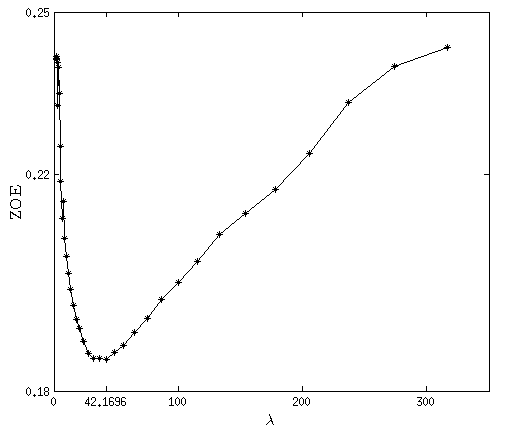}
  \caption{Validation score for different values of $\lambda$ on \textit{EachMovie} data set}
  \label{EachMovieLambda}
\end{figure}
We held out one rating from every user to form the validation set which is later used to evaluate the model performance. We repeat the selection process (train/ validation set) $3$ times for every candidate $\lambda$ and calculate Zero-One Error (ZOE) on validation set. Finally, we select $\lambda$ corresponding to the smallest ZOE. The candidate values of $\lambda$ are $\{10^{\frac{i}{16}}\}, \forall i \in \{ 4, 5, \dots , 40\}$. We repeat the same procedure for every level and select the best $\lambda$ for the experiments reported in Table~\ref{tab:error}. Figures~\ref{movieLensLambda} and~\ref{EachMovieLambda} depict the variation of ZOE for different values of $\lambda$ for two data sets for Stage $3$.  We carry out this exercise for each stage.  
\end{subsection}

\begin{subsection}{Results and Discussion} 
\noindent \textbf{Comparative Analysis:} Table~\ref{tab:error} gives the comparative analysis of the proposed HMF against state-of-art algorithms on two different data sets with two different experimental setups. We use the same $d$ value ($d=100$) as reported in all the comparative methods to attain fair evaluation. We observe that our algorithm exhibits better accuracy in case of \textit{EachMovie} data set than any of the algorithms in both strong and weak generalization. In the case of \textit{MovieLens 1M} data set, our method outperforms other methods in weak generalization and is better than all but GP-LVM for strong generalization. The results reported here are average values of three runs.
\begin{table}[ht!]
	\renewcommand{\arraystretch}{1.4}
	\centering
	\caption{Average and standard deviation of NMAE of different models}
	\label{tab:error}
	\resizebox{\textwidth}{!}{
	\begin{tabular}{|l|*5{|c|c||c|c|c|}}
		\hline
		& \multicolumn{2}{c||}{MovieLens} & \multicolumn{2}{c|}{EachMovie}\\
		\cline{2-5}
		Algorithms&weak&strong&weak&strong\\
		\hline
		URP 			& .4341 $\pm$ .0023	& .4444  $\pm$ .0032	& .4422 $\pm$ .0008	& .4557 $\pm$ .0008 \\ \hline
		Attitude		& .4320 $\pm$.0055	& .4375 $\pm$ .0028	& .4520 $\pm$ .0016	& .4550 $\pm$ .0023 \\ \hline
		MMMF 			& .4156 $\pm$ .0037	& .4203 $\pm$ .0138	& .4397 $\pm$ .0006	& .4341 $\pm$ .0025 \\ \hline
		E-MMMF 			& .4029 $\pm$ .0027 	& .4071 $\pm$ .0093	& .4287 $\pm$ .0020	& .4295 $\pm$ .0030 \\ \hline
		PMF 			& .4332 $\pm$ .0033	& .4413 $\pm$ .0074	& .4466 $\pm$ .0016	& .4579 $\pm$ .0016 \\  \hline
		BPMF 			& .4235 $\pm$ .0023	& .4450 $\pm$ .0085	& .4352 $\pm$ .0014	& .4445 $\pm$ .0005 \\ \hline
		GP-LVM 			& .4026 $\pm$ .0020	& \textbf{.3994 $\pm$ .0145}	& .4179 $\pm$ .0018	& .4134 $\pm$ .0049 \\ \hline
		iPMMMF \& iBPMMMF	& .4031 $\pm$ .0030	& .4089 $\pm$ .0146	& .4211 $\pm$ .0019	& .4224 $\pm$ .0051 \\ \hline
		Gibbs MMMF \& iPMMMF	& .4037 $\pm$ .0005	& .4040 $\pm$ .0055	& .4134 $\pm$ .0017	& .4142 $\pm$ .0059 \\ \hline
		HMF			& \textbf{.4019 $\pm$ .0044}	& .4032 $\pm$ .0022	& \textbf{.4118 $\pm$ .0019}	& \textbf{.4095 $\pm$ .0044} \\ \hline
	\end{tabular}
}
\end{table}

As stated earlier, the main motivation of our algorithm is to examine \textit{combinational} approach as an alternative to otherwise well-known \textit{embedded} principle of MMMF.  There is substantial improvement of performance (in terms of  NMAE) of HMF from MMMF. This corroborates our claim that the proposed framework which is based on a \textit{combinational} approach has advantages over factorizations based on \textit{embedded} approach (like  MMMF).  

\noindent \textbf{Number of Latent Factors:} Selecting an optimal (minimum) number of latent factors ($d$) with less compromise on accuracy is important for any matrix factorization technique. In the second set of experiments we analyse the impact of $d$ on the proposed method versus  MMMF. We use \textit{MovieLens 100K} dataset for this study. We randomly select 80\% data for training and rest 20\% for testing. For every \textit{d}, we tune the regularization parameter in both  MMMF and HMF so that the NMAE on training set is in the range $[0.06, 0.08]$. We calculate the NMAE on the test set. The average NMAE of three runs are reported in Figure \ref{result on changing k}, where the abscissa depicts $d$ in decreasing order. 
\begin{figure}[ht!]
	\centering
	\includegraphics[width=4in,height=3.7in]{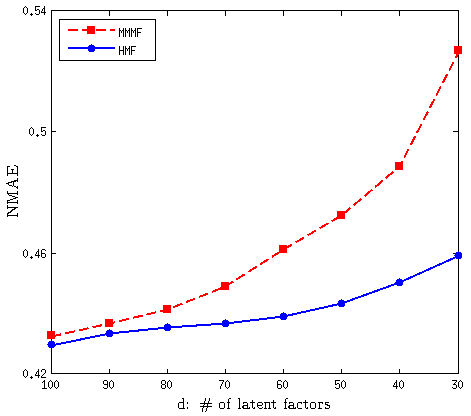}
	\caption{Testing error on changing $d$ value}
	\label{result on changing k}
\end{figure}
It can be seen from Figure~\ref{result on changing k} that when $d$ decreases the performance of  MMMF  drastically deteriorates whereas HMF maintains a balanced performance. This observation indicates that the proposed method can be used for lower ranks of the latent factor matrices without compromising the accuracy of prediction. 

\begin{figure}[ht!]
	\centering
	\includegraphics[width=4.3in,height=3.7in]{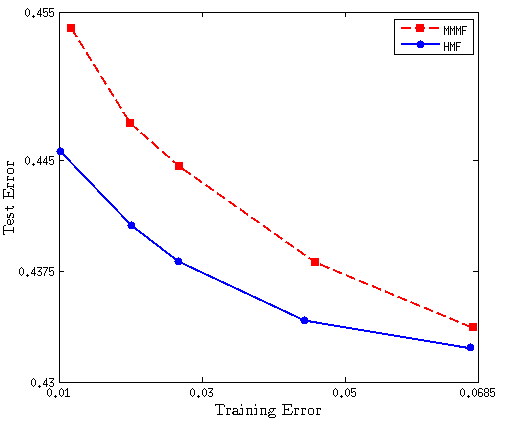}
	\caption{Training error vs Test error}
	\label{TrnErr Vs. TstErr}
\end{figure}
\noindent \textbf{Generalization Analysis:} Traditionally, the concept of maximum margin was introduced with an objective of achieving good generalization error in contrast to  empirical error. It is well-known that both cannot be achieved together and hence there is a trade-off between test-error and training-error. We analyse the interaction between test error and the training error for HMF and  MMMF on  \textit{MovieLens 100K} data set. We fix the value of $d$ to $100$ and plot test-error against training-error (Figure~\ref{TrnErr Vs. TstErr}). It is observed that the trade-off of  MMMF is much higher than that of the proposed method.  When the training-error is reduced, test-error for  MMMF becomes higher than that of our method. Similarly, for low test-error, training error of  MMMF is higher than that of our 
method.

\noindent \textbf{Computation Time:}  Although HMF uses several stages of matrix factorization, it requires less computational time than MMMF. The optimization problem of MMMF has $Nd + Md + N(R-1)$ variables whereas each stage of HMF is an optimization problem of $Nd + Md$ variables. Any Gradient Descent based algorithm requires updating these variables iteratively and hence the number of variables has a major influence on the computational time.  We carry out experiments to compare the computational efficiency of HMF with  MMMF for different values of $\textit{d}$ with fixed  $\lambda$. For MMMF we use the same $\lambda$ as reported in~\cite{decoste2006collaborative}. Aggregated time of HMF for all $R-1$ stages requires substantially less computational time than that of  MMMF for \textit{MovieLens 1M} (Figure ~\ref{runningTimeMovieLens}) and \textit{EachMovie} (Figure ~\ref{runningTimeEachMovie})  datasets. It can be concluded that the proposed framework, HMF is both accurate and computationally efficient. 
\begin{figure}[ht!]
	\centering
	\includegraphics[width=4.3in,height=3.65in]{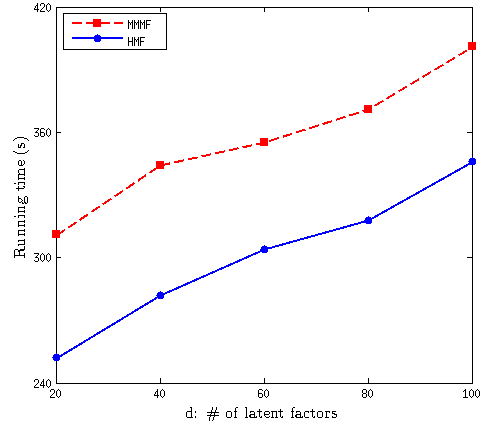}
	\caption{Running time on MovieLens dataset}
	\label{runningTimeMovieLens}
\end{figure}
\begin{figure}[ht!]
	\centering
	\includegraphics[width=4.3in,height=3.65in]{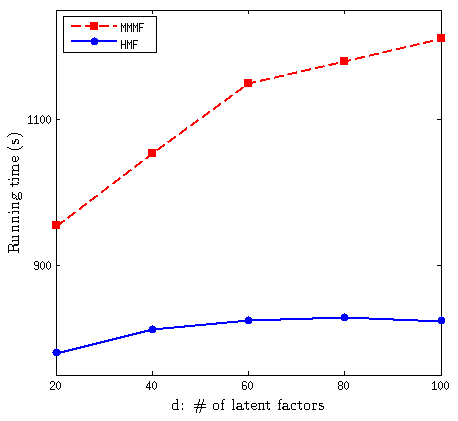}
	\caption{Running time on EachMovie dataset}
	\label{runningTimeEachMovie}
\end{figure}

\begin{figure*}[ht!]
	\begin{tabular}{l}		
		\begin{minipage}{1\textwidth}
			\centering
			\includegraphics[width=5.5in,height=4.1in]{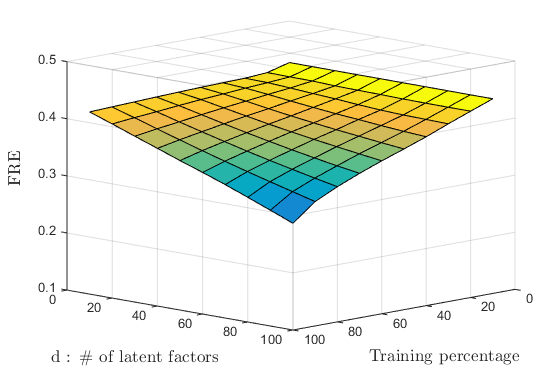}
			\captionof{subfigure}{(a) MMMF}
			\label{mmmfObsPer}
		\end{minipage}
		\\ \\
		\begin{minipage}{1\textwidth}
			\centering
			\includegraphics[width=5.5in,height=4.1in]{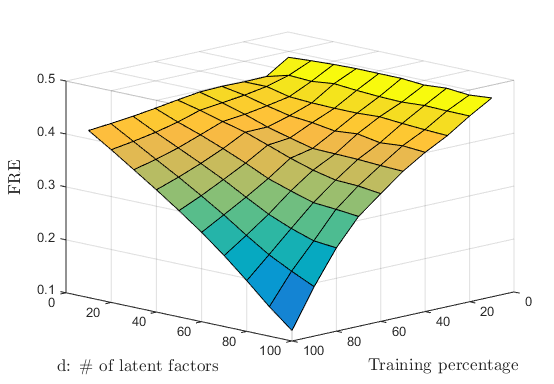}
			\captionof{subfigure}{(b) HMF}
			\label{hmfObsPer}
		\end{minipage}
	\end{tabular}
\caption{Recovery of original rating matrix by MMMF and HMF}
\label{hmfVsmmmfObsPer}
\end{figure*}

\noindent \textbf{Matrix Recovery:} In this experiment we analyzed the effect of sample percentage and the number of latent factors that are necessary for acceptable recovery of the original matrix $Y$. A \textit{synthetic} matrix $Y$ of size $500 \times 500$ is generated using the procedure given in Section~\ref{hmfDatasets}.  We kept the observed entries percentage to $100$ and the number of ordinal ratings to five ($1-5$). Given $Y$, we train HMF and MMMF models with different percentage of sampled entries ($\Omega$) and for different values of the latent dimension $d$. For each ($\Omega, d$)-pair, the regularization parameter $\lambda$ in both HMF and MMMF are tuned from the candidate set $\{10^{\frac{i}{16}}\}, \forall i \in \{ 4, 5, \dots , 40\}$. We repeat the selection process (train/ test set) three times for every ($\Omega, d$)-pair and calculate the average FRE. The result is reported in Figure \ref{hmfVsmmmfObsPer}.  It can be seen from Figure~\ref{hmfVsmmmfObsPer} that there is an increase in the performance of HMF as the observed entries percentage is increasing and for observed entries $\ge 40\%$, HMF is performing better than MMMF. 
\end{subsection}

\section{Conclusions}
\label{conclusion_hmf}
There have been large number of publications on MMMF but these profess the principle of \textit{embedded technique}. The present chapter for the first time investigates matrix factorization of ordinal matrix in a \textit{combinational technique} that uses multiple bi-level matrix factorization. We show several advantages of this approach. Our experimental results show that NMAE for our framework is better than any of these approaches including probabilistic approaches, except in the case of GP-LVM (MovieLens 1M) for strong generalization protocol.  The proposed method exhibits better trade-off of generalization vs. empirical errors and it yields latent factors of lower rank. In the era of big data, when the data is voluminous the proposed method can easily be ported to a parallel or a distributed environment.  In a multiprocessor system, the factorization for each $q$ can be accomplished in parallel. 

  \newpage
\thispagestyle{empty}
\chapter{Proximal Maximum Margin Matrix Factorization for Collaborative Filtering}
\label{pmmmfChapter}
In the previous chapter, we discussed the underlying principle of maximum margin matrix factorization (MMMF) and the motivation for finding an alternative solution. This  led us to a new formulation of matrix factorization, namely, hierarchical matrix factorization which was discussed in the subsequent parts of Chapter $3$. In the present chapter, we propose another alternative for matrix factorization techniques based on maximum margin and propose a novel method called proximal maximum margin matrix factorization (PMMMF). The background information for this chapter is the same as discussed in  Section~\ref{F-MMMF}.
\section{Introduction}
 In MMMF~\cite{rennie2005fast}, as shown in Section~\ref{bi_level_MMMF}, the smooth hinge loss function is defined to ensure that the embedding of points and hyperplanes maintain specific margin to minimize generalization error. In the context of support vector machines (SVM) there have been proposals of alternative formulations different from margin maximization. There are proposals where the hyperplanes are non-parallel for different classes as proposed in twin SVM~\cite{JayadevaKC07}. Similarly, proximal SVM~\cite{Fung2001} attempts to classify based on proximity of the data points to the decision hyperplane. Taking the que from such alternative formulation of SVMs this chapter explores the possibility of \textit{proximal} as an alternative criterion instead of \textit{margin maximization} criterion. In this sense, the idea is to look for embedding of row of $U$ and row of $V$ such that if an item $j$ is rated as $r$ by user $i$ then the objective of the embedding is to ensure that the point $V_j$ is closest to the $r$th parallel hyperplane of $U_i$. This criteria necessitates redefining the loss function. A new loss function is derived in this chapter and it is observed in this process that the threshold values can be computed in closed form avoiding any optimization process. The proposed loss function ensures that the embedded point is in the proximity of the desired hyperplanes and maintain a distance from other hyperplanes. In order to have the parallel hyperplanes corresponding to different ratings uniformally spaced, we have used a hinge loss function. Thus, our loss function is a combination of proximal loss and hinge loss function. There has not been any attempt to find alternative matrix factorization based on maximum margin for ordinal ratings. The present formulation is one of the first alternative formulation to MMMF.

The rest of the chapter is organized as follows.  We introduce our proposed method, termed as Proximal Maximum Margin Matrix Factorization (PMMMF), in Section~\ref{PMMMF- The Proposed Method}. Experimental analysis of the proposed method is reported in Section~\ref{pmmmfExperiment}. We conclude the chapter with a discussion on future directions in Section~\ref{pmmmfConclusion}.

\section{PMMMF- The Proposed Method}
\label{PMMMF- The Proposed Method}
In this section, a novel method of matrix factorization for discrete valued ratings is proposed. Given a partially observed rating matrix $Y$ with $R$ ratings and with $\Omega$ as the observed set, the aim is to determine two factor matrices $U$ and $V$ and a threshold set $\theta$ such that the predicted value $x_{ij} = U_iV_j^T$ is related to rating $y_{ij}$ by proximity of $x_{ij}$ to the corresponding threshold $\theta_{y_{ij}}$. We use, similar to MMMF, $R$ number of thresholds for each user. But, unlike MMMF, the thresholds are computed in closed form and not considered as variables during the optimization step. Thus, the objective is to minimize  
\begin{equation}
\label{proximal-eq}
J = \sum_{r=1}^R{\sum_{(i,j) \in \Omega \wedge (y_{ij} = r) 
}}(U_{i}V_{j}^T-\theta_{i,r})^2 + \frac{\lambda}{2}(\|U\|_{F}^{2} + 
\|V\|_{F}^{2}).
\end{equation}
At the minimizing point we have $\frac{\partial J}{\partial \theta_{i,r}} = 0$ which implies that
\begin{equation*}
\theta_{i,r}^* =  \frac{\sum_{(i,j) \in \Omega \wedge (y_{ij} = r) 
	}U_{i}V_{j}^{T}}{|\Omega(i,r)|}
\end{equation*}
where $\Omega(i,r)$ is the index set of items rated as $r$ by user $i$.

We shall use the above expression to determine the threshold values for fixed  $U$ and $V$. When $\Omega(i,r)$ is empty, we take the corresponding threshold $\theta_{i,r}^*$ as undefined. This assumption is reasonable because, in the absence of any training example of rank $r$ by $i$, the corresponding threshold cannot be learned. Without loss of generality, we assume that $\Omega(i,r)$ is never empty and there is at least one item of rank $r$ by user $i$. Our optimization iteration alternates between two steps . In one step, the factor matrices $U$ and $V$ are updated and then using the updated $U$ and $V$ the optimal value of $\theta^*$ is computed. We use the newly computed $\theta^*$ in the next iteration. 

There can be several possible $U$s and $V$s minimizing the objective function in Eq (\ref{proximal-eq}). In order to ensure that the hyperplanes are widely placed, a hinge loss component is added and the new objective is to minimize 
\begin{equation}
\label{pmmmf}
J = \sum_{r=1}^R{\sum_{(i,j) \in \Omega \wedge (y_{ij} = r) }}(D_{ijr})^2 +  \sum_{r=1}^R{\sum_{(i,j) \in \Omega \wedge (y_{ij} \ne r) }h(T_{ij}^{r}(D_{ijr})}) +   \frac{\lambda}{2}(\|U\|_{F}^{2} + \|V\|_{F}^{2})
\end{equation}
where $ h(z)$ is the same as defined in Eq (\ref{smoothHinge}) , $D_{ijr} =  (U_iV_j^T - \theta_{i,r}^*)$ and   $T_{ij}^r$ is defined as 
\begin{equation}
 T_{ij}^r = 
\begin{cases}
+1 & \text{if r $<$ $Y_{ij}$;} \\
-1 & \text{if r $ > $ $Y_{ij}$.} \\
\end{cases}
\end{equation}
The gradients are determined as follows.
\begin{multline}
\frac{\partial J}{\partial U_{ip}} = \lambda U_{ip} + \sum_{r=1}^{R}{\sum_{(i,j) \in \Omega \wedge (y_{ij} = r)}{2(D_{ijr} )( V_{jp} - \bar V_{ipr})}}\\
+ \sum_{r=1}^{R}{\sum_{(i,j) \in \Omega \wedge (y_{ij} \ne r) }{T_{ij}^{r}h'(T_{ij}^{r}(D_{ijr}))( V_{jp} - \bar V_{ipr})}} \hspace{25pt}
\end{multline}
\begin{multline}
\frac{\partial J}{\partial V_{jq}} = \lambda V_{jq} + \sum_{r=1}^{R}{\sum_{(i,j) \in \Omega \wedge (y_{ij} = r)}{2(D_{ijr} )U_{iq}}} +  \sum_{r=1}^{R}{\sum_{(i,j) \in \Omega \wedge (y_{ij} \ne r)}{T_{ij}^{r}h'(T_{ij}^{r}D_{ijr} ))U_{iq}}} \\ - \sum_{r=1}^{R}{\sum_{(i,j) \in \Omega \wedge (y_{ij} = r)}{\frac{\sum_{(i,t) \in \Omega \wedge y_{it} \ne r}{T_{it}^{r}h'(T_{it}^{r}(D_{itr}))}}{|\Omega(i,r)|}U_{iq}}}
\end{multline}
where $ \bar V_{ipr}$ is defined as 
\begin{equation*}
\bar V_{ipr}= \frac{\sum_{(i,t) \in \Omega (i,r)}{ V_{tp}}}{|\Omega(i,r)|}.
\end{equation*}
Finally, the latent factor matrices $U$ and $V$ are iteratively updated using the following rule.
\begin{align}
 U_{ip}^{t+1} &= U_{ip}^{t} - c \frac{\partial J}{\partial U_{ip}^{t}} \label{U_updatePMMMF}\\
 V_{jq}^{t+1} &= V_{jq}^{t} - c \frac{\partial J}{\partial V_{jq}^{t}} \label{V_updatePMMMF}\
\end{align}

When predicting ordinal ratings, the introduced thresholds $\theta$ are very important since they underpin the large-margin principle of maximum-margin matrix factorization models. The matrix completion process is accomplished from the factor matrices $U$ and $V$ by the following rule.
\begin{equation*}
\hat Y_{ij} = 
\begin{cases}
r &~ \text{if $(\theta_{i,r}^* + \frac{n_{i,r}}{n_{i,r} + n_{i,r+1}}|\theta^{*}_{i,r+1} -  \theta^{*}_{i,r}| < U_iV_j^T)  \wedge$} \\
~ &~ \text{$(0\le r \le R) \wedge (i,j) \notin \Omega$;}\\
y_{ij} & \text{ if $(i,j) \in \Omega $,}
\end{cases}
\end{equation*}
where $n_{i,r}$ is the number of items rated as r by the $i^{th}$ user.  For simplicity of notation we assume  $\theta^{*}_{i,0} = -\infty$, $\theta^{*}_{i,R+1} = +\infty$, $n_{i,0} = 0$ and $n_{i,R+1} = 0$.

\begin{figure*}
	\centering
	\includegraphics[width=5.8in,height=4.5in]{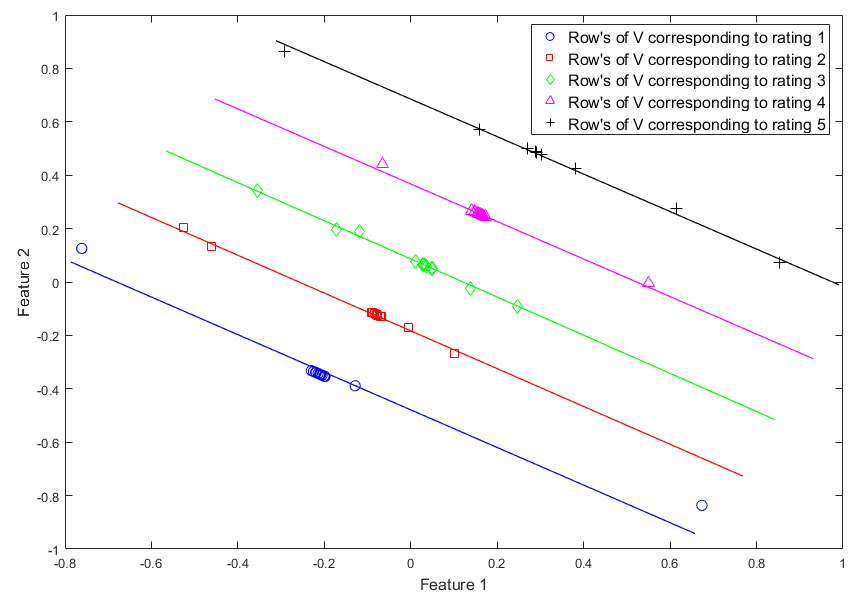}
	\captionof{figure}{Classification by  PMMMF for the $i$th user}
	\label{pmmmfClassification}
\end{figure*}

The objective of PMMMF is to learn the embedding of items as points, users as hyperplanes and ratings as thresholds such that the embedding of points rated as $r$ by user $i$  is in the proximity of the decision hyperplane defined by $(U_{i}, \theta_{i,r})$. While embedding, PMMMF ensure that the hyperplanes are widely placed so that a large margin is created between two classes. In other words, the hyperplane defined by $(U_{i}, \theta_{i,r})$ is no longer a bounding plane as in MMMF, but can be seen as \textit{proximal} plane, around which the items rated similar are clustered and placed as far as possible from other hyperplanes.  We illustrate the concept of PMMMF by taking a synthetic data of size $5 \times 1000$ with $10\%$ of observed entries. The number of ordinal rating is $5$ and $d = 2$. Figure~\ref{pmmmfClassification} gives the decision hyperplane for a user and embedding of points corresponding to items. Since $R = 5$, there are $5$ hyperplanes corresponding to $5$ ratings. From Figure~\ref{pmmmfClassification}, it is clear that the embedding is learnt perfectly and the points rated similar fall in the \textit{proximity} of the respective decsion hyperplane.

\noindent \textbf{Complexity Analysis:} We analyze the computational complexity of the proposed method. In every iteration of gradient descent, PMMMF requires computation of $\theta^{*}$, $\bar{V}$ and $UV^{T}$ with computation cost $MN$, $MN$ and $ MNd$, respectively. The computation cost required for calculation of gradients $\frac{\partial J}{\partial U}$ and $\frac{\partial J}{\partial V}$ are  $2MNdR$ and $ 3MNdR$, respectively. Hence, the overall computation  required in every gradient iteration is $( 5MNdR + MNd + 2MN)$, that is, $\mathcal{O}(MNdR)$.

The optimization problem of PMMMF requires updation of $Nd + Md$ variables unlike MMMF where the optimization problem requires updation of  $Nd + Md + N(R - 1)$ variables  in every iteration of gradient descent method. But on the other hand the updation step in PMMMF takes more time because of repeated computation of the \emph{mean} for obtaining $\theta^{*}$. Though both the methods are of $\mathcal{O}(MNdR)$. The dominating component in PMMMF and MMMF are  $5MNdR$ and $2MNdR$, respectively. In that sense, the computation cost of PMMMF is higher than that of MMMF. Reducing the computation cost of PMMMF is part of our future work.

\section{Experiments}
\label{pmmmfExperiment}
In the following sections, we describe the experimental setup including the data sets, the evaluation metrics, the competing algorithms and following this, we discuss the experimental results.

\begin{subsection}{Data Sets}
We carried out experiments on \textit{real} and \textit{synthetic} data sets. We use \emph{MovieLens $100K$} as the benchmark data set which is standard in the matrix factorization community. The detailed characteristics of \emph{MovieLens-100K} is given in Subsection~\ref{hmfDatasets}. The \emph{MovieLens-100K} data set consists of $100,000$ ratings ($1-5$) given by $943$ users for $1682$ movies. Out of $943$ users, $693$ users have used all possible ratings to express their preferences. We generated the synthetic dataset of size $1000 \times 1000$ with ratings in the range $1-5$. Detailed procedure related to generating the synthetic data is given in Subsection~\ref{hmfDatasets}. 
\end{subsection}

\begin{subsection}{Evaluation Metrics} 
In order to evaluate the performance of the proposed method, we used two popular evaluation metrices: \textit{Mean Absolute Error (MAE)}~\cite{marlin2003modeling} and \textit{Root Mean Square Error (RMSE)}~\cite{ling2012decentralized}. Given $Y$ a $N \times M$ partially observed user-item rating matrix, $\Omega$ be the observed entries set and $\hat Y$ be a recovered matrix, the MAE and RMSE are defined as follows. 
\begin{align}
MAE &= \frac{\sum_{(i,j) \in \Omega}{|y_{ij} - \hat y_{ij}|}}{|\Omega|}   &  RMSE &= \sqrt{\frac{\sum_{(i,j) \in \Omega}{(y_{ij} - \hat y_{ij})^2}}{|\Omega|}} \nonumber
\end{align}
\end{subsection}

\begin{subsection}{Comparing Algorithms}
We consider the following two well-known state-of-the-art algorithms for comparison:
\begin{itemize}
      \item \textbf{ALS}~\cite{koren2009matrix}:  In regularized least squares based matrix factorization (RMF), the goal is to minimize the squared sum distance between the observed entry and the corresponding prediction while overfitting is avoided through a regularized model. One of the several approaches to solve the RMF problem is the alternating minimization algorithms (ALS) where the latent factor matrices are updated in alternate steps.       
      \item \textbf{MMMF}~\cite{rennie2005fast}: The \textit{hinge loss} function defined for bi-level is extended to multi-level for collaborative filtering with ordinal rating matrix when user-preferences are not in the form of like/ dislike but are values in a discrete range. 
\end{itemize}

\noindent All methods are implemented in MATLAB with single computational thread on 4-core $3.40$GHz Intel i7 CPU with $4$GB RAM.  As reported in~\cite{rennie2005fast}, factor numbers higher than $50$ yield similar performances. Hence, they choose $d = 100$ as a compromise between model capacity and computational complexity. Therefore, we also set the latent factor $d$ to be $100$ for all methods to attain fair evaluation. In our experiments, we randomly selected $80\%$ of the observed ratings for training and used the remaining 20\% as the test set. We report the average of the three prediction accuracies.
\end{subsection}

\begin{subsection}{Experimental Results}
\noindent \textbf{Generalization Analysis:} The concept of maximum margin was introduced with an objective of achieving good generalization error in contrast to empirical error. In our first experiment, we analysed the  trade-off between  generalization-error and empirical-error on the \emph{synthetic} dataset by varying the number of observed entries. We vary the regularization parameter $\lambda$  in the range $\{10^{\frac{i}{16}}\}, \forall i \in \{ 1, 5, \dots , 25\}$, for all the methods so that the MAE on training set decreases. Thereafter, we plot the generalization-error against the empirical-error. In all the graphs of Figure \ref{fig:trnVstst}, the curve corresponding to PMMMF is lower than that of the other methods and the sufficient  gap indicates that the rate of increase in empirical-error vs generalization-error is slower. In other words, our proposed approach, PMMMF, exhibits better trade-off than that of MMMF and ALS. 

\begin{figure}[ht!]
\begin{adjustbox}{max width=\textwidth}
	\begin{tabular}{ll}	
		\begin{minipage}{0.5\textwidth}
			\includegraphics[width=3in,height=2.5in]{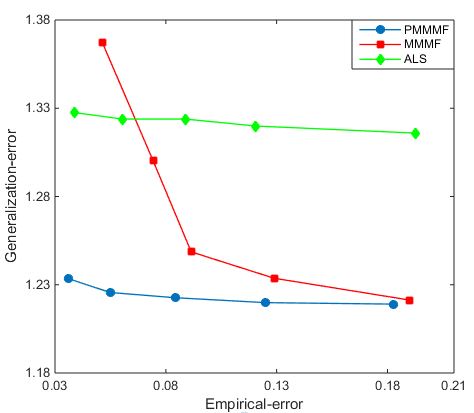}
			\captionof{subfigure}{(a) Observed entries  5\% }
			\label{fig:sub1}
		\end{minipage}
		& 
		\begin{minipage}{0.5\textwidth}
			\includegraphics[width=3in,height=2.5in]{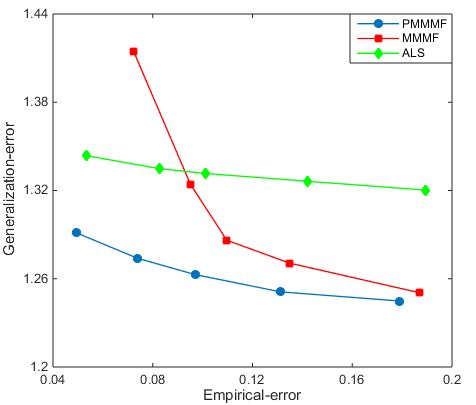}
			\captionof{subfigure}{(b) Observed entries  10\% }
			\label{fig:sub2}
		\end{minipage}
		\\\\ 
		\begin{minipage}{0.5\textwidth}
			\includegraphics[width=3in,height=2.5in]{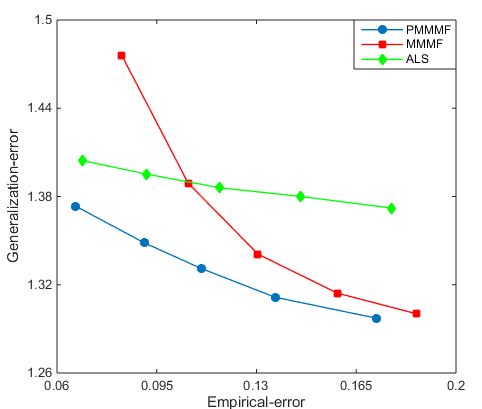}
			\captionof{subfigure}{(c) Observed entries  15\% }
			\label{fig:sub3}
		\end{minipage}
		&
		\begin{minipage}{0.5\textwidth}
			\includegraphics[width=3in,height=2.5in]{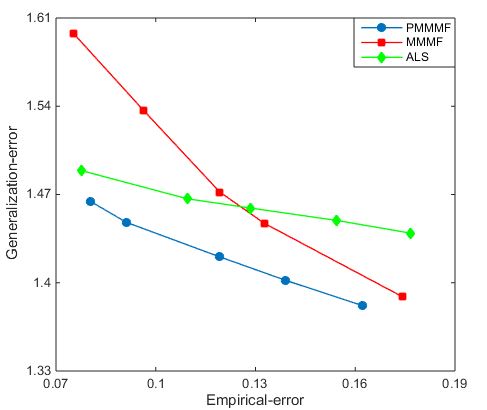}
			\captionof{subfigure}{(d) Observed entries  20\% }
			\label{fig:sub4}
		\end{minipage}
		\\\\
		\begin{minipage}{0.5\textwidth}
			\includegraphics[width=3in,height=2.5in]{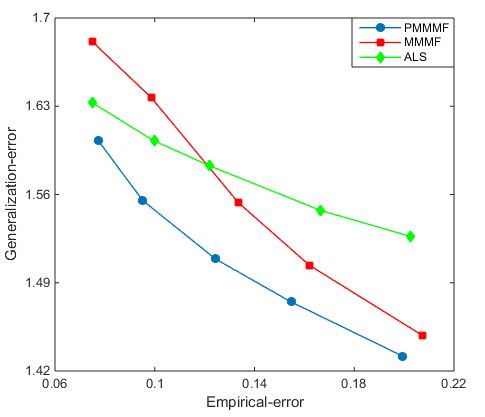}
			\captionof{subfigure}{(e) Observed entries  25\% }
			\label{fig:sub5}
		\end{minipage}
		&
		\begin{minipage}{0.5\textwidth}
			\includegraphics[width=3in,height=2.5in]{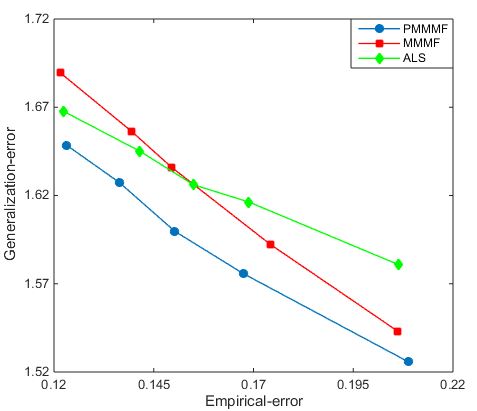}
			\captionof{subfigure}{(f) Observed entries  30\% }
			\label{fig:sub6}
		\end{minipage}
	\end{tabular}
\end{adjustbox}
	\caption{Figure (a), (b), (c), (d), (e) and (f) shows the trade-off between Generalization-error Vs. Empirical-error on synthetic dataset of size $1000 \times 1000$ with $5\%$, $10\%$, $15\%$, $20\%$, $25\%$ and $30\%$ observed entries respectively.}
	\label{fig:trnVstst}
	
\end{figure}

\noindent \textbf{Comparative Analysis:} In the second set of experiments, we compare the accuracy of the proposed PMMMF against MMMF and ALS on \textit{MovieLens 100K} datasets. We tune the regularization parameter $\lambda$ in all the methods so that MAE on the training set is in the range $[0.07, 0.09]$. From the results reported in Table~\ref{tab-error}, it can be seen that our proposed method, PMMMF, exhibits better accuracy than that of MMMF and ALS.
\begin{table}[ht!]
	\renewcommand{\arraystretch}{1.2}
	\centering
	\caption{Average and standard deviation of MAE and RMSE of different models}
	\begin{tabular}{|l|*2{|c|c|}}
		\hline 
		Algorithms&MAE&RMSE\\
		\hline
		MMMF&$0.7355~\pm~ 0.0070 $ &$1.0426~ \pm~ 0.0093$ \\
		\hline
		ALS&$0.7899~\pm~ 0.0039 $ &$1.0838~ \pm~ 0.0055$ \\
		\hline
		PMMMF&$0.7138~ \pm~ 0.0045$&$1.0178~  \pm ~0.0041$  \\
		\hline
	\end{tabular}	
	\label{tab-error}
\end{table}
\end{subsection}

\section{Conclusions and Discussion}
\label{pmmmfConclusion}
This chapter introduced a novel concept of matrix factorization for multi-level ordinal rating matrix. Other than MMMF~\cite{rennie2005fast}, there has not been any attempt for factorizing such matrices. In MMMF, the factorization is achieved by the latent factors which separate different ratings in terms of margins. Recently, proximal SVMs are proposed as an alternative to margin-based classifiers. Proximal SVMs are shown to be superior to traditional SVMs on many counts. Taking the cue from here, we make an attempt in this chapter to introduce \emph{proximity} criterion in place of margin maximization criterion in the context of matrix factorization. The process is non-trivial and novel. Such improvisation yields superior performance. We hope that this will open up new vistas of research in collaborative filtering and matrix completion.   

Like MMMF, PMMMF also assumes that the entries of the rating matrix are ordered. One can think of a more general problem of matrix factorization for matrices having discrete entries which are not ordered. We realize that there are important applications in computer vision and social networking of completion of such matrices. We propose to explore this problem in the future.

 \newpage
\thispagestyle{empty}
\setcounter{chapter}{4}
\chapter{Multi-label Classification Using Hierarchical Embedding}
\label{mlchmfChapter}
In the previous chapters, we discussed about matrix factorization techniques for collaborative filtering. Essentially four techniques where examined therein, namely;  (1) Binary rating matrix  completion using bi-level matrix factorization; (2) Discrete ordinal rating matrix completion using maximum margin matrix factorization; (3) Hierarchical bi-level matrix factorization to handle matrix completion of discrete ordinal rating matrix and; (4) Proximal matrix factorization for rating matrix completion. In all the four methods one common concept that is discussed is the matrix factorization of the rating matrix. The underlying hypothesis is that, matrix factorization is a process of separation of the characteristics of users and items as factor matrices. In a general setting, one can view matrix factorization as a principle of separating or extracting the component latent factors. Another way of visualizing matrix factorization is as a kind of low-dimensional embedding of the data. This is practically relevant when a matrix is viewed as a transformation of data from one space to the other. Matrix factorization is essentially a composition of two transformations, out of which one is for embedding. Hence, matrix factorization can be used in several scenarios where we seek to capture the underlying low-dimensional structure of the data~\cite{zhang2017robust, zhang2013low,de2012least,zhang2015compression}. In this chapter, we explore the application of matrix factorization for yet another important problem of machine learning namely \emph{multi-label classification}. 
\section{Introduction}
\label{Introduction_mlc_hmf}
 In a traditional classification problem, data objects (instances) are represented in the form of feature vectors wherein each object is associated with a unique class label from a set of disjoint class labels $L$, $|L| > 1$. Depending on the total number of disjoint classes in $L$, a learning task is categorized as \textit{binary} classification (when $|L| = 2$) or \textit{multi-class} classification (when $|L| > 2$)~\cite{sorower2010literature}. An example of binary classification problem is the email spam filtering problem where the goal is to classify an email into spam or non-spam whereas the classification of email into one of the predefined classes such as \textit{primary}, \textit{social}, \textit{promotions}, \textit{updates}, \textit{forum} etc., falls into the category of multi-class classification. However, in many real-world classification tasks, the data object can simultaneously belong to one or more classes in $L$. For example, in text categorization, an article can belong to several predefined classes such as \textit{politics}, \textit{sport}, \textit{software}, \textit{business}, \textit{entertainment}; In protein function prediction, a protein can be associated with a set of functional roles such as \textit{metabolism}, \textit{energy}, \textit{cell fate}, \textit{storage protein}, \textit{localization}; In web mining, a web page can be classified as \textit{news}, \textit{academic}, \textit{e-commerce}, \textit{blog}, \textit{forum} etc.; Similarly, in image classification, an image can be annotated with several classes such as \textit{sea}, \textit{sky}, \textit{tree}, \textit{mountain}, \textit{valley}, and so on.

 Multi-label learning is concerned with the classification of data with multiple class labels. The objective of multi-label classification is to build a classifier that can automatically tag an example with the most relevant subset of labels.  This problem can be seen as a generalization of \textit{single label} classification where an instance is associated with a unique class label from a set of disjoint labels $L$.  
 The majority of the methods for supervised machine learning proceeds from a formal setting in which data objects (instances) are represented in the form of feature vectors. Thus, an instance $x$ is represented as $D$ dimensional real-valued feature vector $(x_1, \ldots, x_D) \in \mathbb{R}^D$. In multi-label classification, each training example $x_i, 1\le j \le N$ is associated with a label vector $y_i \in  \{-1, 1\}^L$. The $+1$ entry at the $j$th coordinate of vector $y_i$ indicates the presence of label $j$ in data point $x_i$.  Given the pair, a feature matrix $X \in \mathbb{R}^{N \times D}$ and a label matrix $Y \in \{-1,1\}^{N \times L}$, the task of multi-label classification is to learn a parameterization $h : \mathbb{R}^D \rightarrow \{-1, 1\}^L$ that maps each instance (or, a feature vector)  to a set of  labels (a label vector). The multi-label classification problem can be formally defined as follows: 
 \begin{Definition}[Multi-label Classification] 
 Given $N$  training examples in the form of a pair of feature matrix  $X$ and  label matrix $Y$ where each example $x_i \in \mathbb{R}^D, 1\le i \le N$, is a row of $X$ and its associated labels $y_i \in  \{-1, 1\}^L$ is the corresponding row of $Y$. The $+1$ entry at the $j$th coordinate of vector $y_i$ indicates the presence of label $j$ in data point $x_i$.  The task of multi-label classification is to learn a parametrization $h : \mathbb{R}^D \rightarrow \{-1, 1\}^L$ that maps each instance (or, a feature vector)  to a set of  labels (a label vector). 
 \end{Definition}
 
 Multi-label classification has applications in many areas, such as machine learning~\cite{read2009classifier,zhang2006multilabel}, computer vision~\cite{cabral2011matrix,boutell2004learning}, and data mining~\cite{tsoumakas2007random,schapire2000boostexter}. Existing methods of multi-label classification can be broadly divided into two categories~\cite{sorower2010literature,zhang2014review} - methods based on \emph{problem transformation} and methods based on \emph{algorithm adaptation}. Former approach transforms the multi-label classification problem into single label classification problems so that existing single-label classification algorithms can be applied. During the last decade, a number of problem transformation techniques are proposed in the literature such as Binary Relevance (BR)~\cite{boutell2004learning}, Calibrated Label Ranking~\cite{furnkranz2008multilabel}, Classifier Chains~\cite{read2009classifier}, Random $k$-labelsets~\cite{tsoumakas2007random} to name a few. On the other hand, methods based on algorithm adaption extend or adapt the learning techniques to deal with multi-label data directly. Representative algorithms include AdaBoost.MH and AdaBoost.MR~\cite{schapire2000boostexter} which are two simple extensions of AdaBoost, ML-DT~\cite{clareknowledge} adapting decision tree techniques,  lazy learning techniques such as ML-kNN~\cite{zhang2007ml} and BR-kNN~\cite{spyromitros2008empirical} to name a few.
 
 In case of multi-label classification, one of the major scalability issues arises when we have extremely large feature space and label space. Most of the conventional algorithms of multi-label classification fail in such situations. To cope with the challenge of exponential-sized output space, modeling inter-label correlations  has been the major thrust of research in the area of multi-label classification in recent years~\cite{li2016joint,huang2012multi,bi2014multilabel} and for this, use of  parametrization and embedding have been the prime focus~\cite{li2016joint,huang2012multi,cabral2011matrix,yu2014large,huang2015learning}. There are two strategies of embedding for exploiting inter-label correlation: (1) Feature Space Embedding (FE); and (2) Label Space Embedding (LE). The first is to learn a projection function which can transform the data from original feature-space to an embedded space. In~\cite{huang2015learning,huang2016learning}, a  direct transformation from original feature-space to label-space is suggested with the assumption that each class label is associated with a sparse label specific feature. In~\cite{yu2014large,cabral2011matrix}, the inter-label correlation is modelled implicitly using low-rank constraint on the transformation matrix. The debate is going on as to whether it is the low-rank embedding or the label-specific sparse transformation that models the label correlation accurately. The LE approach transforms the label vectors to an embedded space, followed by the association between feature vectors and embedded label space for classification purpose. With proper decoding process that maps the projected data back to the original label space, the task of multi-label prediction is achieved~\cite{yu2014large, tai2012multilabel, lin2014multi}. It can be seen that both the approaches are essentially a process of parametrization to overcome the complexity of multi-label classification and most often it is proposed to adopt linear parametrization. Some researchers~\cite{kimura2016simultaneous,li2015multi} suggest a natural extension of their proposal of linear parametrization to nonlinear cases but no detailed study is undertaken in this direction. Moreover, all these approaches do not yield results beyond a particular level of accuracy for problems with large data and large number of labels.
 
 Experimental and theoretical study of the recent approaches for multi-label classification reveals many important aspects of the problem. It is clear that a single linear embedding $h$ may not take us very far in finding accurate multi-label classification. There are several reasons for this: the diversity of the training set, the correlation among labels, the feature-label relationship, and most importantly, the learning algorithm to determine the mapping $h$. Normally, $h$ is determined by a process of nonlinear optimization. A research question that naturally arises is whether there can be a parametrization which is piecewise-linear.
 
 
 In this chapter, we investigate this aspect and propose a novel method that generates optimal embeddings for subsets of training examples.  The proposed method is novel in the sense that it judiciously selects a subset of training examples for training and then it assigns a suitable subset of the training set to an embedding. Using multiple embeddings and their assigned training sets, a new instance is classified and we show that the proposed method outperforms all major algorithms on all major benchmark datasets.   
	
The rest of the chapter is organized as follows.  In Section \ref{mlc_hmf_review},
we briefly review the earlier research on multi-label learning. Section~\ref{problemFormulation_mlc_hmf} discuss about embedding approach for multi-label classification. The outline of the proposed method is described in Section~\ref{outline_mlc_hmf}. We introduce our proposed method, termed as MLC-HMF in Section~\ref{proposedMethod_mlc_hmf}. Experimental analysis of the proposed method is reported in Section~\ref{experimentalSection_mlc_hmf}. Finally, Section~\ref{conclusion_mlc_hmf} concludes and indicates several issues for future work.
 
 \section{Multi-label Classification Approaches}
 \label{mlc_hmf_review}
  During the past decade, the problem of multi-label classification has been widely dealt with and different methods have been proposed which can be broadly classified into two categories~\cite{tsoumakas2006multi} - methods based on problem transformation and method based on algorithm adaptation. In this section, we present a brief literature review about these two approaches for multi-label classification.  
 
 \begin{subsection}{Problem Transformation Approach}
 A common approach to multi-label classification problem is to decompose the problem into one or more single-label (binary or multi-class) problems. There are several problem transformation techniques proposed in the literature which includes Binary Relevance~\cite{boutell2004learning}, Calibrated Label ranking~\cite{furnkranz2008multilabel} and Classifier Chains~\cite{read2009classifier}. Here we review these fundamental methods.
 
 \noindent \textbf{Binary Relevance (BR)}~\cite{boutell2004learning}:  BR is a popular problem transformation approach that decomposes the multi-label classification with $L$ labels into $L$ independent binary class classification problem. The BR approach learns $L$ independent binary classifiers $h_l : \mathbb{R}^D \rightarrow \{\pm{1}\}$, $1\le l \le L$, where each binary classifier corresponds to a label in $L$. For each binary classifier $h_l$, the binary training-set $D_l$ is obtained by transforming the original data set $D$ using the following rule. 
 \begin{equation}
 \label{BRTrainingData}
  D_l = \{(x_i,\varphi(y_i,l))| 1\le i \le N\}
 \end{equation}
where 
 \begin{equation}
 \varphi(y_i,l) = \begin{cases}
  +1, &  \text {if $l$th label is in $y_i$} \\
  -1, & \text{otherwise}
  \end{cases}
 \end{equation} 
Once these data sets are ready, some binary classification algorithm is utilized to train the binary classifier $h_l$ corresponding to the prediction of binary association of the $l$th label. For any unseen instance $x_{new} \in \mathbb{R}^D$, the label vector $y_{new}$ is predicted by combining the output of the $L$ classifiers i.e., the union of labels predicted positively by $L$ binary classifiers.  
 
 \begin{table}[ht!]
\centering
\caption{Multi-label data set.}
 \centering
  \begin{tabular}{|r||r|r|r|r||r|r|r|}
      \hline 
      \multirow{2}{*}{\#}&\multicolumn{4}{c||}{Features}&\multicolumn{3}{c|}{Labels}\\ \cline{2-8}
      &\multicolumn{1}{c|}{$a_1$}&\multicolumn{1}{c|}{$a_2$}&\multicolumn{1}{c|}{$a_3$}&\multicolumn{1}{c||}{$a_4$}&$l_1$&$l_2$&$l_3$\\ \hline 
     $x_1$&-0.71&-0.33& 0.40& 0.80& 1 &-1&-1\\ \hline 
     $x_2$& 0.01&-0.57&-0.34& 0.91&-1 & 1& 1\\ \hline 
     $x_3$&-0.08&-0.33& 0.11&-0.31& 1 & 1& 1\\ \hline 
     $x_4$& 0.10&-0.20& 0.33& 0.92& 1 &-1& 1\\ \hline 
     $x_5$& 0.08&0.09&- 0.41& 0.60&-1 & 1&-1\\ \hline 
     \multicolumn{5}{c}{$X$}&\multicolumn{3}{c}{$Y$}   
  \end{tabular}
 \label{mlcDataset}  
\end{table}
 Let us consider the multi-label data set in Table~\ref{mlcDataset}  with five instances $x_1, \dots, x_5$ and three target labels denoted as $l_1,l_2$ and $l_3$. Each instance is  represented by a set of features $a_1, \dots, a_4$. We will refer the feature vector of the $k$th instance by $x_k$ in the subsequent discussion. 
 
\noindent The BR method transforms the original multi-label training-set in Table~\ref{mlcDataset} into three binary training-sets, one for each label. The transformed data set $D_l$ for each classifier $h_l$ is shown in Table~\ref{BRdataset}. After that, some binary classification algorithm is utilized to train the binary classifier $h_l$ on data set $D_l$.
\begin{table}[ht!]
\centering
\caption{Training-set $D_l$ for each classifier $h_l$.}
 \begin{minipage}{0.22\textwidth}
 \centering
 \begin{tabular}{|r||r|}
     \hline 
     \# & $l_1$ \\ \hline 
     $x_1$& 1\\ \hline 
     $x_2$&-1 \\ \hline 
     $x_3$& 1\\ \hline 
     $x_4$& 1 \\ \hline 
     $x_5$&-1 \\ \hline 
  \end{tabular}
    \captionof{subfigure}{$D_1$}
 \end{minipage}%
 \begin{minipage}{0.22\textwidth}
 \centering
  \begin{tabular}{|r||r|}
      \hline 
      \# & $l_2$ \\ \hline 
     $x_1$&-1\\ \hline 
     $x_2$& 1 \\ \hline 
     $x_3$& 1\\ \hline 
     $x_4$&-1 \\ \hline 
     $x_5$& 1 \\ \hline 
  \end{tabular}
  \captionof{subfigure}{$D_2$}
 \end{minipage}%
 \begin{minipage}{0.22\textwidth}
 \centering
  \begin{tabular}{|r||r|}
      \hline 
      \# & $l_3$ \\ \hline 
     $x_1$&-1\\ \hline 
     $x_2$& 1 \\ \hline 
     $x_3$& 1\\ \hline 
     $x_4$& 1 \\ \hline 
     $x_5$&-1 \\ \hline 
  \end{tabular}
  \captionof{subfigure}{$D_3$}
 \end{minipage} 
  \label{BRdataset}  
\end{table} 
\vspace{-1cm}

\noindent The unseen instance $x_{new}$ is then fed into all the binary classifiers and the corresponding label vector prediction $y_{new}$ is obtained by aggregating the output of the individual classifier. Table~\ref{BRprediction} illustrates the classification process using BR approach on unseen instance $x_{new}$. 
\begin{table}[ht!]
\centering
\caption{Binary Relevance example.}
 \centering
  \begin{tabular}{|c||c|}
      \hline 
      Classifiers&Prediction \\ \hline
      $h_1: x_{new} \rightarrow \{l_1, \neg l_1\} \rightarrow \{+1,-1\}$&$+1$ \\ \hline
      $h_2: x_{new} \rightarrow \{l_2, \neg l_2\} \rightarrow \{+1,-1\}$&$+1$ \\ \hline
      $h_3: x_{new} \rightarrow \{l_3, \neg l_3\} \rightarrow \{+1,-1\}$&$-1$ \\ \hline
  \end{tabular}
 \label{BRprediction}  
\end{table}

\noindent Finally, the prediction of each binary classifier $h_1$, $h_2$ and $h_3$ is combined to obtain the label vector prediction $y_{new} = \{+1,-1,-1\}$.

 \noindent \textbf{Calibrated Label ranking (CLR)}~\cite{furnkranz2008multilabel}:  The CLR method transforms the original data sets into $\frac{L(L-1)}{2}$ binary label data set, one for each label pair $(j,k)$, $1 \le j < k \le L$. Each transformed data set $D_{jk}$ retains instances which are associated with only one label from the label pair $(j,k)$. The data set $D_{jk}$ is obtained using the following rule. 
 \begin{equation}
  D_{jk} = \{(x_i,\varphi(y_i,j,k))| \varphi(y_i,j) \ne \varphi(y_i,k), 1\le i \le N\}
 \end{equation}
 where 
 \begin{equation}
 \varphi(y_i,j,k) = \begin{cases}
  +1, &  \text {if $\varphi(y_i,j) = +1$ and $\varphi(y_i,k) = -1$ } \\
  -1, & \text{if $\varphi(y_i,j) = -1$ and $\varphi(y_i,k) = +1$ }
  \end{cases}
 \end{equation}
 
A binary classifier $h_{jk} : \mathbb{R}^D \rightarrow \{\pm{1}\}$ is then trained on each data set $D_{jk}$. Given an unseen instance $x_{new}$, firstly all $\frac{L(L-1)}{2}$  binary classifiers are invoked and the ranking of labels is obtained by counting the votes $x_{new}$ receives for each label. Finally, the label vector $y_{new}$ is then obtained from the ranked label list using some thresholding function.

For example in Table~\ref{mlcDataset}, the CLR approach transforms the original multi-label data set into  $\frac{3(3-1)}{2}$ i.e., $3$ binary label data sets, one for each label pair. The transformed data sets $D_{jk}$, $1\le j < k \le 3$ for each label pair is shown in Table~\ref{CLRdataset}.  
\begin{table}[ht!]
\centering
\caption{Training-set $D_{jk}$ for each classifier $h_{jk}$.}
 \begin{minipage}{0.22\textwidth}
 \centering
 \begin{tabular}{|r||r|}
     \hline 
     \# & $l_{12}$ \\ \hline 
     $x_1$& 1\\ \hline 
     $x_2$&-1 \\ \hline 
     $x_4$& 1 \\ \hline 
     $x_5$&-1 \\ \hline 
  \end{tabular}
    \captionof{subfigure}{$D_{12}$}
 \end{minipage}%
 \begin{minipage}{0.22\textwidth}
  \vspace{-1cm}
 \centering
  \begin{tabular}{|r||r|}
     \hline 
     \# & $l_{13}$ \\ \hline 
     $x_1$& 1\\ \hline 
     $x_2$&-1 \\ \hline 
  \end{tabular}
  \captionof{subfigure}{$D_{13}$}
 \end{minipage}%
 \begin{minipage}{0.22\textwidth}
 \vspace{-1cm}
 \centering
  \begin{tabular}{|r||r|}
     \hline 
     \# & $l_{23}$\\ \hline 
     $x_4$& 1 \\ \hline 
     $x_5$&-1 \\ \hline 
  \end{tabular}
  \captionof{subfigure}{$D_{23}$}
 \end{minipage} 
  \label{CLRdataset}  
\end{table} 
\vspace{-1cm}

\noindent After that, a binary classifier $h_{jk}$ is learnt on the transformed data $D_{jk}$, $1\le j < k \le 3$.  Given an unseen instance $x_{new}$, let it be the case that the classifiers $l_{12}$, $l_{13}$ and $l_{23}$ has predicted $l_1$, $l_2$ and $l_2$, respectively. After counting the votes $x_{new}$ receives from each classifier, $h_{jk}$ for each label, the inferred label ranking is  $l_2 > l_1 > l_3$.

\noindent \textbf{Classifier Chains (CC)}~\cite{read2009classifier}:  The BR approach trains $L$ independent binary classifiers, one for each label. However, its performance can be poor when strong inter-label correlations exist i.e., information of one label is helpful in inferring the information about related label. The CC approach model the inter-label correlations by transforming the multi-label classification problem into a chain of binary classification problems.  Each binary classifier in the chain pass the label information to the subsequent classifiers in the chain. The chain involves same number of binary classifiers as in BR but extends the feature space of each binary classifier  with all prior binary relevance predictions in the chain. Given $L$ labels $\{1,2,\dots,L\}$, let $\pi: \{1, 2, \dots, L\} \rightarrow  \{1, 2, \dots, L\}$ be the permutation function which specifies the ordering of binary classifier in the chain. Each classifier $h_{\pi(j)}$ is responsible for training and prediction of the binary association for $j$th label in the ordered chain. The binary training-set $D_{\pi(1)}$ for the first binary classifier $h_{\pi(1)}$ in the chain is obtained using Eq. (\ref{BRTrainingData}). For subsequent classifiers $h_{\pi(j)}$ in the chain, the original training-set gets appended using the following rule. 
\begin{equation}
 D_j = \{([x_i, pre(h_{\pi(1)}),\dots, pre(h_{\pi(j-1)})],\varphi(y_i,{\pi(j)}))| 1\le i \le N \}
\end{equation}
where $pre(h_{\pi(l)})$ is the prediction of $l$th binary classifier for $x_i$ and $[x_i, pre(h_{\pi(1)}),\dots, \allowbreak pre(h_{\pi(j-1)})]$ concatenates vector $x_i$ with prior binary predictions $pre(h_{\pi(1)}),\dots, \linebreak  pre(h_{\pi(j-1)})$. For any unseen instance $x_{new} \in \mathbb{R}^D$, the classification process begins at $h_{\pi(1)}$ and propagates along the chain $h_{\pi(2)}$, \dots, $h_{\pi(L)}$. The label vector $y_{new}$ is predicted by combining the output of each classifier $h_{\pi(l)}$ in the chain. 
\begin{table}[ht!]
\centering
\caption{Training-set $D_{\pi(l)}$ for each classifier $h_{\pi(l)}$.}
 \begin{minipage}{0.2\textwidth}
 \centering
 \begin{tabular}{|r||r|}
     \hline 
     \multicolumn{1}{|c||}{\#} & \multirow{2}{*}{\scriptsize $pred(h_1)$} \\ \cline{1-1}
     \multicolumn{1}{|c||}{\scriptsize $[x]$}& \\ \hline 
     $x_1$& 1\\ \hline 
     $x_2$&-1 \\ \hline 
     $x_3$&-1\\ \hline 
     $x_4$& 1 \\ \hline 
     $x_5$&-1 \\ \hline 
  \end{tabular}
    \captionof{subfigure}{$D_1$}
 \end{minipage}%
 \begin{minipage}{0.35\textwidth}
 \centering
  \begin{tabular}{|r|r||r|}
      \hline 
     \multicolumn{2}{|c||}{\#} & \multirow{2}{*}{\scriptsize $pred(h_2)$} \\ \cline{1-2} 
     \multicolumn{2}{|c||}{\scriptsize $[x,pred(h_1)]$}& \\ \hline 
     $x_1$& 1&-1\\ \hline 
     $x_2$&-1& 1 \\ \hline 
     $x_3$&-1& 1\\ \hline 
     $x_4$& 1& 1 \\ \hline 
     $x_5$&-1&-1 \\ \hline 
  \end{tabular}
  \captionof{subfigure}{$D_2$}
 \end{minipage}%
 \begin{minipage}{0.35\textwidth}
 \centering
  \begin{tabular}{|r|r|r||r|}
      \hline 
     \multicolumn{3}{|c||}{\#} & \multirow{2}{*}{\scriptsize $pred(h_3)$} \\ \cline{1-3} 
     \multicolumn{3}{|c||}{\scriptsize $[x,pred(h_1),pred(h_2)]$}& \\ \hline 
     $x_1$& 1&-1&-1\\ \hline 
     $x_2$&-1& 1& 1 \\ \hline 
     $x_3$&-1& 1& 1\\ \hline 
     $x_4$& 1& 1& 1 \\ \hline 
     $x_5$&-1&-1& 1 \\ \hline 
  \end{tabular}
  \captionof{subfigure}{$D_3$}
 \end{minipage}%
  \label{CCdataset}  
\end{table} 
\vspace{-1cm}

For the example in Table~\ref{mlcDataset}, let the training order of the classifier obtained by permutation function $\pi$ is $h_1 \rightarrow h_2 \rightarrow h_3$. The CC approach first trains a binary classifier $h_1$ for label $1$; For label $2$, the feature-space of  binary classifier $h_2$ is extended by the prediction of classifier $h_1$; Similarly, for label $3$, the feature-space of the binary classifier $h_3$ is extended by including the predictions of classifiers $h_1$ and $h_2$ (Table~\ref{CCdataset}).

\noindent The new instance $x_{new}$ (let, $x_{new} = [-0.05, -0.1, 0.3, 0.5]$) is then fed into all classifiers from $h_1$ to $h_3$ and the prediction is obtained by aggregating the output of the individual classifier. Table~\ref{CCprediction} illustrates the classification process of CC approach on unseen instance $x_{new}$.
\begin{table}[ht!]
\centering
\caption{Classifier Chain example.}
 \centering
  \begin{tabular}{|l||c|}
      \hline 
      \multicolumn{1}{|c||}{Classifiers} &Prediction \\ \hline
      $h_1: [-0.05, -0.1, 0.3, 0.5] ~~~~~~~~~\rightarrow \{l_1, \neg l_1\} \rightarrow \{+1,-1\}$&$+1$ \\ \hline
      $h_2: [-0.05, -0.1, 0.3, 0.5, \textbf{1}]~~~~~ \rightarrow \{l_2, \neg l_2\} \rightarrow \{+1,-1\}$&$-1$ \\ \hline
      $h_3: [-0.05, -0.1, 0.3, 0.5, \textbf{1}, \textbf{-1}] \rightarrow \{l_3, \neg l_3\} \rightarrow \{+1,-1\}$&$-1$ \\ \hline
  \end{tabular}
 \label{CCprediction}  
\end{table}

\noindent Finally, the prediction of each binary classifier $h_1$, $h_2$ and $h_3$ is combined to obtain the label vector prediction $y_{new} = \{+1,-1,-1\}$.


\noindent \textbf{Random \textit{k}-Labelsets (RA\textit{k}EL)}~\cite{tsoumakas2007random}: The RA\textit{k}EL approach is based on the label powerset (LP) method which cast the multi-label classification problem into multi-class classification problem. In LP approach, the multi-label data is first transformed into a multi-class data by considering each unique subset of labels that exist in the data set as one class. The transformation is achieved using some \textit{injective} function mapping $\sigma: 2^y \rightarrow \mathbb{N}$ from the power set of labels to a natural number. Once the multi-class data set is ready, some multi-class classification algorithm is utilized to train the classifier $h : x \in \mathbb{R}^D \rightarrow \Gamma$, where $\Gamma$ is the set of distinct classes in the transformed data. The major drawbacks with LP approach are: a) Computational complexity incurred due to the large number of unique labelsets. Multi-label classification problem with $L$ labels will have $(min(N, 2^L))$ possible unique label powerset; b) Class imbalance problem, as a large number of labelsets would be associated with very few training instances; and c) Confined to predict labelsets observed in the training-set.

RA\textit{k}EL extends the concept of LP by constructing an ensemble of LP classifiers~\cite{tsoumakas2011random,tsoumakas2007random}. Each LP classifier is trained using a small randomly selected \textit{k}-labelsets from the original labels which result in computationally inexpensive, predictive complete and more balanced multi-class training-set. For example, if we break a data set with $101$ labels into $3$-labelset with the assumption that two labelset will not overlap, then in the worst case scenario $2^3$ binary classifiers need to be trained for each $3$-labelset and $2^3 \times 27$ binary classifiers for overall data set whereas for full LP requires $2^{101}$ binary classifiers. Given the labelset size $k$, let $L^k$ denote the collection of all distinct \textit{k}-labelsets in $L$. The $p$th \textit{k}-labelsets is denoted as $L^k(p) \subseteq L$ where $|L^k(p)| = k$ and 
$1 \le p \le \binom{L}{k}$. For each \textit{k}-labelsets  $L^k(p)$, a multi-class training-set is constructed as
\begin{equation}
 D_{L^k(p)} = \{x_i, \sigma: (y_i \cap L^k(p))  \rightarrow \mathbb{N}| 1 \le i \le m \}. 
\end{equation}
\noindent After that, some multi-class classification algorithm is utilized to train the classifier $h_{L^k(p)}: x \in  D_{L^k(p)} \rightarrow \Gamma$, where $\Gamma$ is the set of distinct classes in $D_{L^k(p)}$. Given the size of labelset $k$ and the number of LP classifiers $n$, RA\textit{k}EL creates an ensemble of $n$ LP classifiers. For each LP classifier $k$-labelsets $L^k(p_r)$, $1 \le r \le n$, is randomly selected and for each, a multi-class classifier $h_{L^k(p_r)}$ is then trained. The unseen instance $x_{new}$ is then fed into all the LP classifiers. After that, RA\textit{k}EL calculates the average vote received for each label and the final output label vector $y_{new}$ is predicted by setting a threshold $\tau$ on the average vote. For the example in Table~\ref{BRdataset}, the classification process of RA\textit{k}EL with $k=2$, $n=3$ and $\tau=0.5$ is shown in Table~\ref{rakelClassification}. 
\begin{table}[ht!]
\centering
\caption{RA\textit{k}EL example.}
 \centering
  \begin{tabular}{|l||c|c|c|}
      \hline
      \multirow{2}{*}{Classifiers}&\multicolumn{3}{c|}{Prediction} \\\cline{2-4}
      &$l_1$& $l_2$ &$l_3$\\ \hline
      $h_{LP_{1,2}}$&$-1$&$+1$&-\\ \hline
      $h_{LP_{1,3}}$&$-1$&-&$+1$ \\ \hline
      $h_{LP_{2,3}}$&-&$+1$&$-1$ \\ \hline
      \multicolumn{1}{|l||}{Average vote}&$0/2$&$2/2$&$1/2$ \\ \hline
      \multicolumn{1}{|l||}{$y_{new}$}&$-1$&$+1$&$+1$\\ \hline
  \end{tabular}
 \label{rakelClassification}  
\end{table}
\end{subsection}
 
 \begin{subsection}{Algorithm Adaption Approach}
 This category of algorithms modifies the well-known learning algorithms to tackle multi-label classification problem directly. There are several algorithm adaption  techniques proposed in the literature such as AdaBoost.MH and AdaBoost.MR~\cite{schapire2000boostexter}, Multi-Label Decision Tree~\cite{clareknowledge} and Multi-Label \textit{k}-Nearest Neighbor~\cite{zhang2007ml}. Here we review these fundamental methods.
 
 \noindent \textbf{Multi-Label \textit{k}-Nearest Neighbor (ML-\textit{k}NN)}~\cite{zhang2007ml}: 
 It is a multi-label lazy learning approach derived from the traditional $k$-Nearest Neighbor (kNN) algorithm. For each unseen instance $x_{new}$, ML-kNN first identifies its $k$ nearest neighbors $N(x_{new})$ in the training set. For every label $j$, ML-\textit{k}NN then computes the count $ C(j)$ which records the number of $x_{new}$ neighbors with label $j$. The membership counting vector for  $x_{new}$ can be defined as follows.
 \begin{equation}
  C(j) = \sum_{x_a \in N(x_{new})} y_a(j) == 1, j \in \{1, 2, \dots, L\}
 \end{equation}
 where $y_a(j)$ takes the value of 1 if instance $x_a$ is associated with label $j$ and 0 otherwise.  The following maximum a posteriori (MAP) principle is then utilized to determine the label set $y_{new}$ for the unseen instance $x_{new}$. 
 \begin{equation}
  y_{new}(j) = \underset{b\in\{-1,+1\}}{arg~max}~P(H_{b}^{j}|E^{j}_{C(j)}), j \in \{1, 2, \dots, L\}
 \end{equation}
 where $H_{+1}^{j}$  be the event that $x_{new}$ is associated with label $j$ and $H_{-1}^{j}$ be the event that $x_{new}$  is not associated with label $j$. $E^{j}_{C(j)}$ is the event that exactly  $C(j)$ number of $x_{new}$ neighbors has label $j$. Using the  Bayes theorem, the same can be re-written as follows.
 \begin{equation}
  y_{new}(j) = \underset{b\in\{-1,+1\}}{arg~max}~P(H_{b}^{j}|)P(E^{j}_{C(j)}|H_{b}^{j}), j \in \{1, 2, \dots, L\}
 \end{equation}
 
 \noindent \textbf{Multi-Label Decision Tree (ML-DT)}~\cite{clareknowledge}:  ML-DT extend the decision tree algorithm designed for multi-class classification to the multi-label setting. In ML-DT, the decision tree is constructed recursively in the top down manner. At every node of the tree a feature (F) is chosen which best classifies the remaining training examples. The best feature (F) is decided by considering the following information gain (IG) measure. 
 \begin{equation}
  IG(X',Y',F) = entropy(X',Y') - \sum_{v\in F} \frac{|X'_v|}{|X'|}*entropy(X'_v,Y'_v) 
 \end{equation} 
 where F is the feature under consideration, $(X',Y')$ is the set of training examples present at the current node, $(X'_v,Y'_v)$ is the subset of examples with value $v$ for feature $F$ and $entropy(X',Y')$ is defined as follows.
 \begin{equation}
  entropy(X',Y') = -\sum_{j}^{L} P(j)\log P(j) + (1-P(j)) \log (1-P(j))
 \end{equation}
 where $P(j)$ is the proportion of examples $(X',Y')$ associated with label $j$. After the learning process is over, each leaf node will be associated with a set of class labels. For any unseen instance $x_{new}$, the tree is traversed from root to the leaf along the path until
 reaching a leaf node. The label associated with the leaf node is predicted as a label set for $x_{new}$.

 \noindent \textbf{Tree Based Boosting}~\cite{schapire2000boostexter}:  AdaBoost.MH and AdaBoost.MR are the multi-label version of the popular ensemble method AdaBoost which creates a strong classifier by combining many base or weak classifiers. AdaBoost.MH tries to minimize the \textit{hamming loss}  and AdaBoost.MR is designed to find hypotheses based on label ranks. In AdaBoost.MH, each training example $(x_i, y_i)$ is presented as $L$ binary examples of the form $\{([x_i,j],\varphi(y_i,j))|1 \le j \le L\}$. AdaBoost.MH maintains a distribution over examples and labels $(X \times Y)$ and re-weights the examples at each boosting round $t$. The example-label pairs that were misclassified in the previous round have higher weight. Unlike the Adaboost.MH which tries to minimize the \textit{hamming loss}, Adaboost.MR tries to find hypothesis that \textit{ranks} the labels of any instance in such as a way that correct labels place at the top of the ranking.
 
 \end{subsection}

 \section{Embedding based Approach }
 \label{problemFormulation_mlc_hmf}
 Most of the conventional algorithms for multi-label classification perform well on relatively small sized data but have difficulty in dealing with data where feature and label space are sufficiently large. For example, simple methods such as Binary Relevance (BR), that treat each label as a separate binary classification problem fail miserably. In many real-world applications of multi-label classification, often feature and label space are assumed to be extremely large. Under such conditions, the scalability of learning algorithms is of major concern, calling for an effective data management and the use of appropriate data structures for time- and space-efficient implementations. To cope with the challenge of exponential-sized output space,  modelling feature- and label-space correlation has been the major thrust of research in the recent years~\cite{li2016joint,huang2012multi,bi2014multilabel}. Various approaches have been proposed to exploit the intrinsic information in feature- and label-space. The CC~\cite{read2009classifier} discussed in Section~\ref{mlc_hmf_review} captures the label correlation by extending the feature space of each binary classifier with all prior binary relevance predictions in the chain. In~\cite{zhu2005multi}, a maximum entropy method for multi-label classification is proposed in which mutual correlations among data are explicitly considered in the model. In~\cite{ghamrawi2005collective}, the dependencies between an individual feature and labels are modelled using the conditional random field. Zhang et al.~\cite{zhang2010multi} model the joint distribution of the label space conditioned on the feature space using Bayesian Networks. In~\cite{huang2012multi2}, multiple boosted learners are trained simultaneously, one for each label with the assumption that the hypothesis generated for one label can be helpful for the other; each hypothesis not only looks into its own single-label task but also reuses the trained hypotheses from other labels. These methods of exploiting correlation information can be quite effective in multi-label learning, but they are computationally expensive over the exponentially large feature- and label-space.
 
 In recent years much attention on multi-label classification have been devoted to embedding based approach where the prime focus is to reduce the effective number of features and labels~\cite{hsu2009multi, yu2014large,  bhatia2015sparse, jing2015semi}. The embedding based approach assumes that there exists a low-dimensional space onto  which the given set of feature vectors and/ or label vectors can be embedded. The intrinsic relationship among feature- and label-space is invariant when constraining embedded dimension to be significantly lower than original space~\cite{yu2016semisupervised, tai2012multilabel, yu2014large}. The embedding strategies can be grouped into two categories namely; (1) Feature space embedding; and (2) Label space embedding. Feature space embedding aims to design a projection function which can map the instance in original feature space to label space. Feature space embedding can be further divided into two categories -   the first is to learn a projection function which directly maps instances from feature space to label space and simultaneously preserve intrinsic relationship in the original space through regularization~\cite{huang2015learning,huang2016learning}. 
 The second approach models inter-label correlation implicitly using low-rank constraints on embedding ~\cite{yu2014large,cabral2011matrix}. The aim here is to design a projection function which can map the instance in original feature space to a reduced dimensional space and at the same time preserve the intrinsic information in the original feature space. A mapping is then learnt from reduced dimensional space to label space. On the other hand, the label space embedding approach transform the label vectors to an embedded space via linear or local non-linear embeddings, followed by the association between feature vectors and embedded label space for classification purpose. With proper decoding process that maps the projected data back to the original label space, the task of multi-label prediction is achieved~\cite{hsu2009multi,rai2015large,tai2012multilabel}. Some researchers suggest  a simultaneous embedding of feature and label space onto the same space. Embedding of both features and labels to  a single low-rank space is no way obvious and cannot be a routine extension of feature-only embedding or label-only embedding. It is necessary to retain the intrinsic relationship while  mapping feature and label vectors to a single space. In the following subsections, we present the brief literature review of feature- and label-space embedding.

 \begin{subsection}{Feature Space Embedding (FE)}  
 Given $N$ training examples in the form of a pair of feature matrix $X$ and  label matrix $Y$, the goal of FE is to transform each $D$-dimensional feature vector (a row of matrix $X$) from original feature space to a $L$-dimensional label vector (corresponding row in $Y$) by an embedding function $\mathcal{F}:X\rightarrow Y$.  
\begin{figure}[ht!]
	\centering
	\includegraphics[width=3in,height=1in]{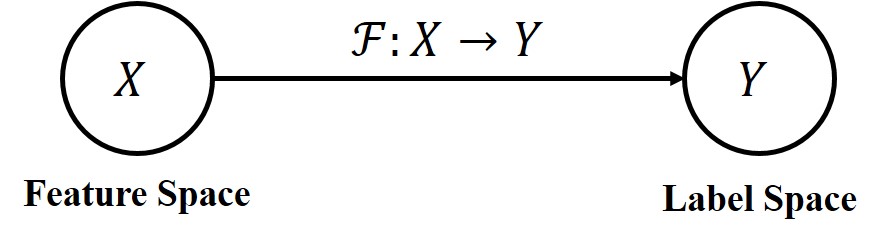}
    \caption{Feature space embedding}
    \label{FEembedding1}
\end{figure}
 There are two strategies of FE. As illustrated in Figure~\ref{FEembedding1}, the first approach is to learn a mapping which directly transforms instance from feature space to label space. Usually, the mapping is achieved by a transformation matrix $W = [w_{.1}, w_{.2}, \dots, w_{.L}] \in \mathbb{R}^{D \times L}$ which directly maps instances from feature space to label space. The vector $w_{.l} \in \mathbb{R}^D$, i.e., the $l$th column of a $W$ can be seen as a transformation vector for the $l$th label. Most of the approaches falling into this category assume that the two strongly correlated labels share more features with each other than two uncorrelated labels and hence their corresponding columns in $W$ will be similar~\cite{huang2015learning,huang2016learning}.
 \begin{figure}[ht!]
	\centering
	\includegraphics[width=5.5in,height=1in]{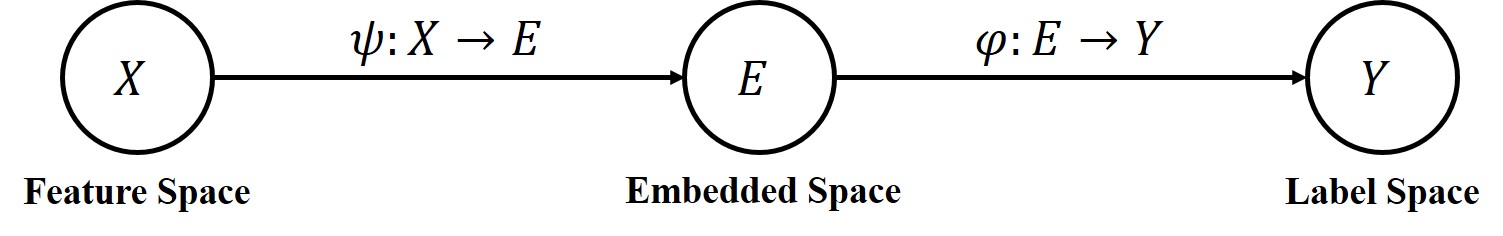}
    \caption{Feature space embedding}
    \label{FEembedding2}
\end{figure} 
 The second approach illustrated in  Figure~\ref{FEembedding2} assume that the label matrix $Y$ is low-rank due to the presence of similar labels and thus model the inter-label correlation implicitly using low-rank constraints on the transformation matrix $W$. The algorithms falling into this category first embed each $D$-dimensional feature vector $x$ to a $d$-dimensional ($d \ll D)$ vector $e \in E \subseteq \mathbb{R}^{d}$ in a latent space by an embedding function $\psi:X \rightarrow E$. Then, the algorithm train a predictive
 model $\varphi:E \rightarrow Y$. For an unseen instance $x_{new}$, a low-dimensional embeding $e_{new} = \psi(x_{new})$ is firstly obtained and then the label vector $y_{new}$ is predicted by the decoding function $\varphi(e_{new})$. We briefly review the major approaches of FE. 
 
 As discussed previously, the goal of FE is to learn a transformation matrix $W$ and a common formulation is the following optimization problem.	
 \begin{equation} \label{basic-FE-formulation}
		\underset{W}{min} \; \ell(Y, XW) + \lambda R(W)\\
 \end{equation}
 where $W\in \mathbb{R}^{D \times L}$, $\ell(\cdot)$ is a loss function that measures how well $XW$ approximates $Y$, $R(\cdot)$ is a regularization function that promotes various desired properties in $W$ (low-rank, sparsity, group-sparsity, etc.)  and the constant $\lambda \ge 0$ is the regularization parameter which controls the extent of regularization. Huang et al. ~\cite{huang2015learning} assume that each label is associated with a subset of features from the original feature set and two strongly correlated labels share more features with each other. The label specific features (a subset of original feature space) learning problem is modelled by linear regression framework with sparsity constraints on the regression parameter $W$. The problem can be formulated as follows.
 \begin{equation} \label{basic-sparse-formulation}
		\underset{W}{min} \; \ell(Y, XW) + \frac{\alpha}{2}\phi(W) + \lambda\|W\|_1
 \end{equation}
 where $\|\cdot\|_{1}$ is $\ell_1$ norm, the second term in Eq. (\ref{basic-sparse-formulation})  models inter-label correlation with the constant $\alpha \ge 0$ to control the extent of correlation.  
 
 The above approaches require $D \times L$ parameters to model the classification problem, which will be expensive when $D$ and $L$ are large~\cite{yu2014large}. To reduce the training cost, a generic empirical risk minimization (ERM) framework with low-rank constraint on linear parametrization $W = UV$, where $U \in \mathbb{R}^{D \times d}$ and $V \in \mathbb{R}^{d \times L}$ are of rank $d \ll D$ is proposed in  Yu et al.~\cite{yu2014large}. The $i$th row of matrix $E=XU$ can be seen as a latent space representation of $i$th training instance whereas the $l$th cloumn of matrix $V$ can be visualized as a latent representation of label $l$. 
 The basic assumption here is that although the original space is sufficiently large, the intrinsic relationship in original space can be captured by representing the instances and labels using a small number of latent factors. The problem can be formulated as
 \begin{equation} \label{basic-low-rank-formulation}
		\underset{U, V}{min} \; \ell(Y, XUV) + \frac{\lambda}{2} (\|U\|_F^2 + \|V\|_F^2)\\
 \end{equation}
 where $\|\cdot\|_{F}$ is Frobenius norm,  $\ell(\cdot)$ is a loss function that measures how well $XUV$ approximates $Y$, and the constant $\lambda \ge 0$ is the regularization parameter which controls the extent of regularization. A semi-supervised joint learning framework in which dimensionality reduction and multi-label classification are performed simultaneously is proposed in~\cite{yu2016semisupervised}. To guide the multi-label learning process, it uses local invariance properties that if two instances are similar in original space then their low-dimensional representation will also be similar. The problem is formulated as  
 \begin{equation} \label{SSJDR-MLL}
		\underset{U, V}{min} \; \sum_{l=1}^{L}{\sum_{i=1}^{N}}h(y_{il}( x_iUV_{.l})) +  \frac{\alpha}{2}\sum_{i,j=1}^{N}e_{i,j}\|x_iU - x_jU\|_F^2 + \frac{\lambda}{2} \|V\|_F^2 \\
 \end{equation}
 where $V_{.l}$ is the $l$th column of $V$, $h(\cdot)$, $\lambda$, $\alpha$ and $\|\cdot\|_F$ are same as defined previously and $e_{i,j}$ is the similarity between instance $x_i$ and $x_j$ in original space. The second term in Eq.~(\ref{SSJDR-MLL}) makes the transformation matrix $U$ neighborhood aware i.e., if two instances $x_i$ and $x_j$ are close in the original space, then their low-dimensional representation are required to be close.
 
 \end{subsection}

\begin{subsection}{Label Space Embedding (LE)}
 Given $N$ training examples in the form of a pair of feature matrix $X$ and  label matrix $Y$, the goal of LE is to transform each $L$-dimensional label vector (a row of matrix $Y$) from original label space to a $d$-dimensional embedded vector $e \in E \subseteq \mathbb{R}^d$ by an embedding function $\Phi:Y\rightarrow E$. Then, a predictive model $\psi: X \rightarrow E$ is trained from original feature space to embedded space. With proper decoding process  $\varphi:E\rightarrow Y$ that maps the projected data back to the original label space, the task of multi-label prediction is achieved~\cite{bi2013efficient,furnkranz2008multilabel,hsu2009multi}. Figure~\ref{LEembedding} illustrate the basic principle of LE. We briefly review the major approaches of LE.
  \begin{figure}[ht!]
	\centering
	\includegraphics[width=5.5in,height=1in]{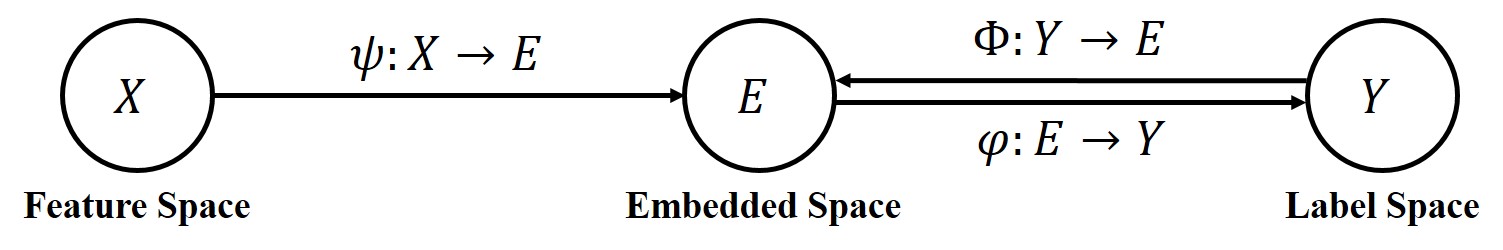}
    \caption{Label space embedding}
    \label{LEembedding}
\end{figure} 

The approach of Hsu et al.~\cite{hsu2009multi} projects the label vectors to a random low-dimensional space, fits a regression model in this space, then projects these predictions back to the original label space. In~\cite{tai2012multilabel}, principal component analysis (PCA) is employed on the label covariance matrix to extract a low-dimensional latent space. In~\cite{balasubramanian2012landmark},  a sparsity-regularized least square reconstruction objective is used to select a small set of  labels that can predict the remaining labels. Recently, Yu et al.~\cite{yu2014large} and Jing et al.~\cite{jing2015semi} proposed to use trace norm regularization to identify a low-dimensional representation of the original large label space. Mineiro et al.~\cite{karampatziakis2015scalable} use randomized dimensionality reduction to learn a low-dimensional embedding that explicitly captures correlations between the instance features and their labels. Some methods work with label or feature similarity matrices, and seek to preserve the local structure of the data in the low-dimensional latent space.  Prabhu et al.~\cite{prabhu2014fastxml} propose a method to train a classification tree by minimizing the Normalized Discounted Cumulative Gain. Rai et al.~\cite{rai2015large} assume that the label vectors are generated by sampling from a weighted combination of label topics, where the mixture coefficients are determined by the instance features. Based on the assumption that all the output labels can be recovered by a small subset, multi-label classification via column subset selection approach (CSSP) is  proposed in~\cite{bi2013efficient}. Given a matrix $Y$, CSSP seeks to find a column index set $C \subset \{1,\dots, L\} $ with cardinality $d~(d  \ll L)$ so that the columns with indices in $C$ can approximately span $Y$. The subset of columns is selected using a randomized sampling procedure. The problem is formulated as follows.    
\begin{equation}
    \label{cssp}
    \underset{C}{min}~\|Y - Y_{C}Y_{C}^{\dag}Y\|_F
\end{equation}
where $\|\cdot\|_{F}$ is Frobenius norm, $Y_C$ denotes the submatrix consisting of columns of $Y$ with indices in $C$ and $Y_{C}Y_{C}^{\dag}$ is the projection matrix onto the $d$-dimensional space spanned by columns of $Y_C$. Alternatively, there have been emerging interests in recent multi-label methods that take the correlation information as prior knowledge while modeling the embedding (encoding). These methods can be efficient when the mapped label space has significantly lower dimensionality than the original label space~\cite{hsu2009multi}.
		
In recent years, matrix factorization (MF) based approach is frequently used to achieve the LE which aims at determining two matrices $U \in \mathbb{R}^{ N \times d}$ and $V \in \mathbb{R}^{ d \times L}$. The matrix $U$ can be viewed as the \emph{basis matrix}, while the matrix $V$ can be treated as the \emph{coefficient matrix} and a common formulation  is the following optimization problem. 	
\begin{equation} \label{basic-LE-formulation}
	\underset{U, V}{min} \; \ell(Y, U, V) + \lambda R(U, V)\\
\end{equation}
where  $\ell(\cdot)$ is a loss function that measures how well $UV$ approximates $Y$, $R(\cdot)$ is a regularization function that promotes various desired properties in $U$ and $V$ (sparsity, group-sparsity, etc.)  and the constant $\lambda \ge 0$ is the regularization parameter which controls the extent of regularization. In~\cite{lin2014multi}, a MF based approach is used to learn the label encoding and decoding matrix simultaneously. The problem is formulated as follows.
\begin{equation}
		\label{faie}
		\underset{U, V}{min}~\|Y - UV\|_{F}^{2} + \alpha\Psi(X, U)
\end{equation}
where $U \in \mathbb{R}^{ N \times d}$ is the code matrix, $V \in \mathbb{R}^{ d \times L}$ is the decoding matrix, $\Psi(X,U)$ is used to make $U$ feature-aware by considering correlations between $X$ and $U$ as side information and the constant $\alpha \ge 0$ is the trade-off parameter. In order to reduce the noisy information in the label space, the method proposed in~\cite{jian2016multi} decompose the original space to a low-dimensional space. Instead of globally projecting onto a linear low-rank subspace, the method proposed in~\cite{bhatia2015sparse} learns embeddings which non-linearly capture label correlations by preserving the pairwise distances between only the closest (rather than all) label vectors. 
\end{subsection}
	
\section{Outline of the Proposed Approach}
\label{outline_mlc_hmf}
In this section, we introduce the underlying principle of our proposed method, termed as MLC-HMF (\textit{Multi-label classification using hierarchical embedding}). As discussed in Section~\ref{Introduction_mlc_hmf}, the proposed method draws motivation from embedding based approach of multi-label classification. The embedding approach aims at modelling the intrinsic information existing in the original space such as label correlation, instance correlation etc., to assist the learning process. The basic assumption underlying the embedding based approach is that the label matrix is low-rank and inter-label correlation can be modelled implicitly by embedding the feature/ label vectors to a reduced low-dimensional latent space.  There are two major approaches for embedding namely \textit{feature space embedding} and \textit{label space embedding}. A detailed discussion of these approaches is given in Section~\ref{problemFormulation_mlc_hmf}. 
A common characteristic of both the approaches is that they try to transform the data from original space to a more manageable (low-dimensional) space such that the learning process can be tackled efficiently without significant loss of prediction performance. One encouraging property of this low-dimensional space is that most of the structures in the original output label space can be explained and recovered.  The latent low-dimensional space can be achieved via a \textit{linear} or \textit{non-linear} transformation and most often it is proposed to use linear transformation due to its attractive computational properties and high interpretability~\cite{li2015multi,bi2013efficient,tai2012multilabel,hsu2009multi}. However, some researchers~\cite{kimura2016simultaneous,li2015multi} suggest a natural extension of their proposal of linear embedding to nonlinear cases but no detailed study is undertaken in this direction.  

In many real-word applications, the low-rank assumption made by embedding methods is violated due to several reasons such as the diversity of the training set, the correlation among labels, the feature-label relationship, the learning algorithm to determine the mapping $h$ and most importantly, the presence of \textit{tail} labels, i.e., the infrequent labels associated with only few training examples~\cite{xu2016robust,bhatia2015sparse}. Hence, embedding with a global projection (a common transformation for all feature/ label vectors) can be complicated and may not well model the optimal mapping. However, on the other hand, we observe through experimental analysis, that it is possible to have efficient embedding for subsets of training samples. Our research is based on the fact that a single linear embedding $h$ may not take us very far in finding accurate multi-label classification. We hypothesize that feature vectors which conform to similar embedding are similar in some sense. Thus, unlike the traditional embedding approach found in the literature, we propose a piecewise-linear embedding of feature-space that generates optimal embeddings for a subset of training examples. Our method is novel in the sense that it judiciously selects a subset of training examples for training and then it assigns a suitable subset of the training set to an embedding. Finally, an unseen instance is classified using the multiple embeddings and their assigned training sets. In this section, we outline the basic principle of the proposed MLC-HMF, an approach based on feature-space embedding.

We start with the formulation given in Eq.~(\ref{basic-low-rank-formulation}). For exploiting correlations in the labels, one way is to factor the matrix $W = UV$ where $U \in \mathbb{R}^{D \times d}$ can be interpreted as an embedding of the features $X$ into a $d$ dimensional latent space and  the $l$th column of $V \in \mathbb{R}^{d \times L}$ is a linear classifier corresponding to label $l$ on this space. Regularization is provided by constraining the dimensionality of the latent space. The minimization in $U$ and $V$ is unfortunately non-convex, and Fazel et al. \cite{fazel2001rank} discovered the nuclear norm (sum of singular values) heuristic for matrix rank minimization, which is the convex relaxation of the rank minimization problem. Since $XUV$ yields continuous values and $Y$ is discrete, a natural choice is to use the well-known principle of maximum margin matrix factorization (MMMF)~\cite{rennie2005fast,kumar2017collaborative}.  For a subset of training examples, the process of determining the embedding using principle of MMMF is described below.

\noindent \textbf{Computing $U, V$:} Let S $\subseteq$ \{1, 2, \dots, n\} be the indices of current set  $X^S$ of training examples and the corresponding label vectors is submatrix $Y^S$ of $Y$.  We use smooth hinge loss function to determine $U^S$ and $V^S$ for a given training set $(X^S, Y^S)$. For sake of simplicity, we drop the suffix $S$. The problem can be formulated as the following minimization problem.
\begin{equation}
\label{MLC-HMF-eq}
\underset{U, V}{min} \;J(U, V)= \sum_{l=1}^{L}\sum_{i \in S}h(y_{il}(x_iUV_{.l}))  + \frac{\lambda}{2} (\|U\|^{2}_{F} + \|V\|^{2}_{F})
\end{equation} 
where $V_{.l}$ is $l$th column of $V$, $\|\cdot\|$, $\lambda$ are same as defined previously and $h(\cdot)$ is smooth hinge loss function defined as
\begin{equation}
 h(z) = 
\begin{cases}
0 & \text{if z $\geq$ 1;} \\
\frac{1}{2}(1 - z)^{2} & \text{if $0 < z < 1$;} \\
\frac{1}{2} - z & \text{otherwise.}
\end{cases}
\end{equation}

\noindent The detailed discussion about smooth hinge loss is given in Section~\ref{bi_level_MMMF}. The gradient of the variables to be optimized is determined as follows.
\begin{align}
\frac{\partial J}{\partial U_{pq}} &= \lambda U_{pq} + \sum_{l=1}^{L}\sum_{i=1}^{n}y_{il}h'(y_{il}(x_iUV_{.l}))x_{iq}V_{lp}\\
\frac{\partial J}{\partial V_{rs}} &= \lambda V_{rs} + \sum_{i=1}^{n}y_{ir}h'(y_{ir}(x_iUV_{.r}))x_{i}U{.s}
\end{align}
where $U_{.s}$ is the $s$th column of $U$. Gradient descent algorithm and its variants such as conjugate gradient  descent and stochastic gradient descent start with random $U$ and $V$ and these are iteratively updated using the equations given in (\ref{one}) and (\ref{two}). Suffixes $t$ and $t+1$ are used to indicate current values and updated values, respectively. 
\begin{align}
U_{pq}^{t+1} &= U_{pq}^{t} - c \frac{\partial J}{\partial U_{pq}^t} \label{one} \\ 
V_{rs}^{t+1} &= V_{rs}^{t} - c \frac{\partial J}{\partial V_{rs}^t} \label{two}
\end{align}
where c is the step length parameter. It is seen that conjugate gradient descent technique exhibits faster rate of	convergence and we follow this technique in the present study.

\noindent \textbf{Geometrical Interpretation:} As discussed in Section~\ref{F-MMMF}, application of maximum margin factorization  has an interesting geometrical interpretation in the present context. The matrix $U$ maps each $D$-dimensional feature vector to  $d$-dimensional space and each row of $V$ defines a decision hyperplane associated with the respective label. An accurate embedding implies that  each of the decision hyperplanes in $d$-dimension classifies the embedded feature points  conforming to the respective rows of the label-matrix $Y$. Prompted by the above observations, we propose a novel algorithm MLC-HMF which is discussed in the following section. 

\section{MLC-HMF: The Proposed Method}
\label{proposedMethod_mlc_hmf}
In this section, a novel method for multi-label classification is proposed which is essentially a hierarchical matrix factorization method. Our algorithm initially separates out training examples into two disjoint components (the process of grouping is discussed later in the section). Then for each component, $U$ and $V$ are learnt. The $U$ and $V$ so computed are used to test the training vectors in the respective components so that instances which are not classified correctly would be further processed for the next round of recursion. The MLC-HMF returns subsets of the training set and the associated $(U, V)$. It also generates residue of training examples which are not used for further classification. In a sense, our algorithm hierarchically selects training set with different degree of suitability for multi-label classification. The depth of the hierarchy determines the finer level of residue. 

\noindent \textbf{Identification of Subsets of Training Instances:} The instance group are not given explicitly and it is necessary to learn from the feature matrix $X$. Thus, before the application of MMMF, we use \textit{k-means} clustering~\cite{jain1999data} with Euclidean metric to cluster the training set into two clusters and each of the clusters are used separately to determine the corresponding embedding. We do not claim that the above process is the best possible practice to embed the subsets of training samples which are similar in some sense. Actually, the grouping can be implemented in different ways based on the clustering algorithm, the number of clusters, or even more sophisticated distance measure other than the Euclidean metric. Nevertheless, our simple construction process of grouping yield competitive performance as shown in Section~\ref{experimentalSection_mlc_hmf}.

\noindent \textbf{Guiding Embedding Process Through Classification Error:}  As discussed previously, $U$ and $V$ are determined through an optimization process and application of any of the gradient-based algorithms may end up in a local minimizing point. As a result, the resultant embedding may not yield accurate classification even for the training instances that are used for optimization. We consider the classification error during the learning process to associated the matrices $U$ and $V$ with a subset of instances. In the proposed method, we assign the training examples that are classified to an acceptable accuracy to the current embedding and separate out wrongly classified example for further processing recursively.

Algorithm~\ref{algotrain:1} outlines the main flow of the proposed method. For each node in the tree, a joint learning framework given in Eq.~(\ref{MLC-HMF-eq}) with low-rank constraint on the parametrization (embedding) and multi-label classification is performed simultaneously. At every node, we maintain the mapping $U$ and the label feature matrix $V$ along with the training examples whose hamming loss is less than the threshold $\mathcal{T}$. The remaining training instances are divided into two parts according to the \textit{k-means} clustering. This process is repeated recursively until either the number of instances in the node is too small or the depth of the tree exceeds a given threshold.  	
\begin{algorithm}[ht!]
	\SetAlgoLined
	\SetKwData{Left}{left}\SetKwData{This}{this}\SetKwData{Up}{up}
	\SetKwFunction{Union}{Union}\SetKwFunction{FindCompress}{FindCompress}
	\SetKwInOut{Input}{input}\SetKwInOut{Output}{output}
	\Input{Data Matrix: $X$, Label Matrix: $Y$, Size of Reduced Dimension Space: $d$, Threshold:  $\mathcal{T}$, Depth of the Hierarchy: $h$}
	\Output{Tree with Mapping $U$ and Label Feature Matrix $V$ at Each Node}
	\BlankLine
	\setcounter{AlgoLine}{0}
	 Divide $X$ into $X^1$ and $X^2$ using $kmeans$ clustering;\\
	\For{i $\in$ \{1,2\}}{
		\If {$|X^i|$ is small  or depth is exceed $h$}{
			Let its corresponding node as leaf node; \\
			\Return
		}			
		Learn the mapping $U$ and label feature matrix $V$ for $X^i$ using Eq.~(\ref{MLC-HMF-eq}); \\
		Let $\bar{X}\subseteq X^i$ is the set of instances whose hamming loss is less than the threshold $\mathcal{T}$ and $\bar{Y}$ is their corresponding label matrix; \\
		Maintain $U$, $V$ and  $\bar{X}$ at the current node; \\
		MLC-HMF ( $X^i \setminus \bar{X}$, $Y^i \setminus \bar{Y}$, $k$, $\mathcal{T}$, $h$)
	}
	\caption{MLC-HMF ( $X$, $Y$, $d$, $\mathcal{T}$, $h$)}
	\label{algotrain:1}
\end{algorithm}

\noindent \textbf{Classification:} The outcome of the learning process described in MLC-HMF has a tree structure. At each non-leaf node of the tree, there is a disjoint subset of training samples along with embedding $U$ and label feature matrix $V$. For any unseen instance $x_{new}\in \mathbb{R}^D$, the label vector is predicted by first finding the $U's$ and $V's$ associated with the \textit{k}-nearest neighbor instances present at non-leaf node of tree. Let $(U^i, V^i), 1\le i \le K$, be the $i$th embedding pair. A label vector is predicted with the help of every ($U^i, V^i$)-pair using the rule $sign(x_{new}U^i{V^i})$. Finally, majority voting rule for the fusion	is applied on the labels obtained in the previous step to obtain the final label vector.

\noindent \textbf{Complexity Analysis:} We analyze the computational and spatial complexity of the proposed method. The time complexity of MLC-HMF mainly comprises of two components: clustering and optimization of the problem as given in Eq.~(4) at every node of the hierarchy. For simplicity of representation, we are ignoring the correctly classified instances at the present node and assume that the set of instances are divided equally between its child. Hence, the average number of instances present in a node at level $l$ is $N/2^{l-1}$. In every iteration of conjugate gradient, the computation cost required for calculation of gradients $\frac{\partial J}{\partial U}$ and $\frac{\partial J}{\partial V}$ is $\frac{2N}{2^{l-1}}D^2d^2L$. Let $t_1$ and $t_2$ be the maximum number of iterations required for gradient update and \textit{k-means} clustering, respectively. Then the overall computation cost required at every node at level $l$ is $(\frac{2N}{2^{l-1}}D^2d^2Lt_1 + \frac{2N}{2^{l-1}}Dt_2 )$. Hence, the overall computation required at level $l$ is $2^l(\frac{2N}{2^{l-1}}D^2d^2Lt_1 + \frac{2N}{2^{l-1}}Dt_2 )$. The overall computation cost required by MLC-HMF is   $2^l(h+1)(\frac{2N}{2^{l-1}}D^2d^2Lt_1 + \frac{2N}{2^{l-1}}Dt_2 )$, that is, $\mathcal{O}(nD^2d^2Lt_1h)$. The space complexity of MLC-HMF mainly comprises of maintaining the mapping $U$, label feature matrix $V$ and the set of instances whose hamming loss is less than the threshold $\mathcal{T}$ at every node. The overall space required by MLC-HMF is $(2^{h+1} - 1)(Dd + Ld) + ND$, that is, $\mathcal{O}(2^{h+1} (Dd + Ld))$.

\section{Experimental Analysis}
\label{experimentalSection_mlc_hmf}			
In this section we analyze the performance of MLC-HMF by taking into account factors such as accuracy and efficiency. This section discusses the experimental setup including the data sets and relevant statistics, the experimental protocols, the competing algorithms, the evaluation metrics as well as the parameter settings.  Following this, we discuss the experimental results.

\begin{subsection}{Data Sets}
\label{mlc_hmfDatasets}

We use twelve multi-label benchmark datasets for experiments, and the detailed characteristics of these datasets are summarized in Table \ref{datasetsCharacteristics}. All of these datasets can be downloaded from \textit{labic}\footnote{ \url{http://computer.njnu.edu.cn/Lab/LABIC/LABIC_Software.html}}, \textit{meka}\footnote{ \url{http://meka.sourceforge.net}} and \textit{mulan}\footnote{ \url{http://mulan.sourceforge.net/datasets-mlc.html}}. 

\begin{table}[h!]
	\renewcommand{\arraystretch}{1.2}
	\centering
	\captionsetup{font=scriptsize,justification=centering}
	\caption{ Description of the experimental datasets}
	\begin{tabular}{llllll}
		\toprule
		Data set&\#instance&\#Feature&\#Label&Domain&LC \\
		\hline
		CAL500&502&68&174&music&26.044\\
		emotions&593&72&6&music&1.869\\
		genbase&662&1185&27&biology&1.252\\
		plant&948&440&12&biology&1.080\\
		medical&978&1449&45&text&1.245\\
		language log&1459&1004&75&text&1.180\\
		human&3108&440&12&biology&1.190\\
		education&5000&550&33&text(web)&1.461\\
		science&5000&743&40&text(web)&1.451\\
		rcv1(subset 2)&6000&944&101&text&2.634\\
		rcv1(subset 5)&6000&944&101&text&2.642\\
		ohsumed&13929&1002&23&text&1.663\\
		\bottomrule
	\end{tabular}
	\label{datasetsCharacteristics}
\end{table}
\end{subsection}

\begin{subsection}{Evaluation Metrics} 
To measure the performance of the different algorithms, we employed six evaluation metrics popularly used in multi-label classification, i.e. \textit{hamming loss, accuracy, exact-match, example based $f_1$ measure, macro $f_1$ and micro $f_1$}~\cite{zhang2014review,sorower2010literature}. Given a test data set $\mathcal{D} = \{x_i, y_i~|~1 \le i\le N\}$, where $y_i \in \{-1, 1\}^L$ is the ground truth labels associated with the $i$th test example, and let $\hat{y_i}$ be its predicted set of labels.	
\newpage
\noindent \textbf{Hamming loss} measures how many times on average, an irrelevant pair (instance, label) is predicted, i.e. a correct label is missed  or an incorrect label is predicted.
\begin{equation*}
    Hamming~loss = \frac{1}{NL}\sum_{i=1}^{N} |y_{i} \ne \hat{y_{i}}|
\end{equation*}

\noindent \textbf{Accuracy} for an instance evaluates the proportion of correctly predicted labels to the total number of active(actual and predicted) labels for that instance. The overall accuracy for a data set is the average across all instances.
\begin{equation*}
	Accuracy = \frac{1}{N}\sum_{i=1}^{N}\frac{|y_{i} \wedge \hat{y_{i}}|}{|y_{i} \vee \hat{y_{i}}|}
\end{equation*}

\noindent \textbf{Subset Accuracy} is an extension of \textit{accuracy} used in single label case to multi-label prediction. For an instance, the prediction is considered to be correct if all the predicted labels are the same as the ground truth labels for that instance. The overall subset-accuracy for a data set is the average across all instances.
\begin{equation*}
	   Subset~Accuracy = \frac{1}{N}\sum_{i=1}^{m}I(y_i =  \hat{y_{i}})
\end{equation*}
where, I is the indicator function.

\noindent \textbf{Example based $F_1$ Measure} is the harmonic mean of precision and recall for each example.
\begin{equation*}
	F_1 = \frac{1}{N}\sum_{i=1}^{N}\frac{2p_ir_i}{p_i + r_i}
\end{equation*}
where $p_i$ and $r_i$ are precision and recall for the $i$th example.
 
\noindent \textbf{Macro $F_1$ }is the harmonic mean of precision and recall for each label.
\begin{equation*}
	Macro F_1 = \frac{1}{L}\sum_{i=1}^{L}\frac{2p_ir_i}{p_i + r_i}
\end{equation*}
where $p_i$ and $r_i$ are precision and recall for the $i$th label.
	
\noindent \textbf{Micro $F_1$} treats every entry of the label vector as an individual instance regardless of label distinction.
\begin{equation*}
	Micro~F_1 = \frac{2\sum_{i=1}^{L}TP_i}{2\sum_{i=1}^{L}TP_i + \sum_{i=1}^{L}FP_i + \sum_{i=1}^{L}FN_i}
\end{equation*}
where $TP_i$, $FP_i$ and $FN_i$ are true positive, false positive and false negative for the $i$th label, respectively.
\end{subsection}

\begin{subsection}{Comparing Algorithms} 
We conducted \textit{ten-fold cross validation} on each data set and the mean value and standard deviation of each evaluation criterion was recorded. The following six well-known state-of-the-art algorithms were
considered for comparison:
\begin{itemize}
	\item \textbf{BSVM}~\cite{boutell2004learning}: This is one of the representative algorithm of problem transformation methods, which treat each label as a separate binary classification problem. For every label, an independent binary classifier is trained by considering the examples with the given class label as positive and others as negative. 
	\item \textbf{BP-MLL}~\cite{zhang2006multilabel}: In this method, a modified loss function and back-propagation are used to account for multi-label data. The number of hidden neurons is set to be $20\%$ of the input dimensionality and the maximum number of training epochs is set to be 100 for all datasets~\cite{zhang2006multilabel,zhang2015lift}.
	\item \textbf{ML-kNN}~\cite{zhang2007ml}: It is a multi-label lazy learning approach derived from the traditional $k$-Nearest Neighbor (kNN) algorithm. For each unseen instance, ML-kNN first identifies its $k$ nearest neighbors in the training set. Based on statistical information gained from the label sets of these neighboring instances, maximum a posteriori (MAP) principle is utilized to determine the label set for the unseen instance. The number of nearest neighbors considered is set to be 10 for all datasets~\cite{zhang2007ml,zhang2015lift}.
	\item \textbf{LIFT}~\cite{zhang2015lift}: LIFT constructs features specific to each label by conducting clustering analysis on its positive and negative instances. After that, for every label, a binary classifier is trained with label specific feature. The ratio parameter $r$ is set to be $0.1$ for all datasets~\cite{zhang2015lift,huang2015learning}.
	\item \textbf{SSJDR-MLL}~\cite{yu2016semisupervised}: This method aims to learn a linear transformation which can reduce the dimension of the original instance and at the same time preserve the inherent property. The reduced dimension is fixed to $100$, and the balanced parameter $\gamma$ is set to be 1.0 for all datasets~\cite{yu2016semisupervised}.
	\item \textbf{LLSF}~\cite{huang2015learning}: This method addresses the inconsistency problem in multi-label classification by learning label specific features for the discrimination of each class label. The hyper-parameters $\alpha$, $\beta$, $\gamma$ and the threshold $\mathcal{T}$ are set to be $0.1$, $0.1$, $0.01$, and $0.5$ respectively for all datasets~\cite{huang2015learning}.	
\end{itemize}
LIBSVM~\cite{CC01a} is employed as the binary learner for classifier induction to instantiate BSVM, LIFT and SSJDR-MLL~\cite{huang2015learning,zhang2015lift}.
\end{subsection}

\begin{subsection}{Parameter Setting} 
 Most of the MLC-HMF hyper-parameters were fixed. The number of neighbors  $K$ considered during the prediction stage were fixed to $5$, the number of reduced latent dimension space $k$ were fixed to $\lceil0.5\mathcal{L}\rceil$. The termination conditions required to repeat the recursive process such as the depth of the hierarchy and the minimum number of instances in the node were set to $5$. The remaining two hyper-parameters are tuned by conducting \textit{ten-fold cross validation}. The regularization parameter $\lambda$ for each data set is tuned from the candidate set $\{10^{\frac{i}{10}}\}, \forall i \in \{ 1, 5, \dots , 20\}$ and the threshold $\mathcal{T}$ is tuned in the range $[0, 0.1]$ with step size 0.02.

\begin{figure}[ht!]
	\centering
	\includegraphics[width=5.5in,height=4.1in]{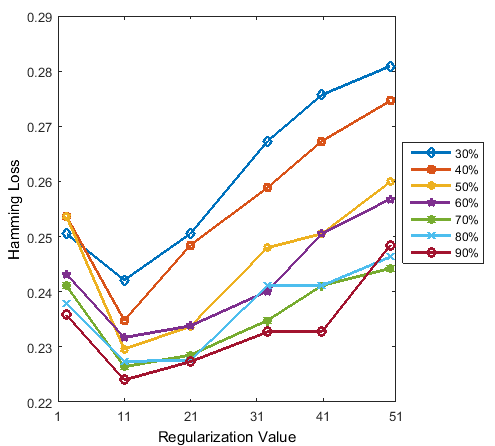}
	\captionof{figure}{Hamming loss for different values of $\lambda$ on emotions data set with varying training size percentage}
	\label{perLambdaComp}
\end{figure}

In  MLC-HMF, there is a need to select the regularization parameter $\lambda$ at every node of the hierarchy as the size of training instances varies at every node of the hierarchy. We conducted  experiments to analyze the effect of $\lambda$ in Eq. (\ref{MLC-HMF-eq}) by varying the size of training data. For this experiment, we first randomly held out $10\%$ of instances from the overall data set which is later used to evaluate the performance. From the remaining instances we selected different percentage of instances to train the model. We repeated the selection process (train/ validation set) $3$ times for every candidate $\lambda$ and calculated the \textit{hamming loss} on the held out set. Figure \ref{perLambdaComp} depicts the variation in \textit{hamming loss} for different values of $\lambda$. It can be seen from Figure~\ref{perLambdaComp} that the data size is not playing any crucial role in the selection of the best $\lambda$. Based on our preliminary experiments, the value of $\lambda$ is fixed for every node in the hierarchy. 
\end{subsection}

\begin{table*}[ht!]
	\renewcommand{\arraystretch}{1.2}
	\centering
	\captionsetup{font=scriptsize,justification=centering}
	\caption{Experimental results of comparison of algorithms (mean$\pm$std rank)  in terms of Hamming Loss, Accuracy, and Subset Accuracy}
\begin{adjustbox}{max width=\textwidth}
		\begin{tabular}{lclclclclclclcl}
			\toprule
			\multirow{2}{*}{Data set}&\multicolumn{14}{c}{Hamming Loss~$\downarrow$}\hspace{20pt}\\ \cline{2-15}
			&BSVM&&BP-MLL&&ML-kNN&&LIFT&&SSJDR-MLL&&LLSF&&MLC-HMF& \\
			\hline
			cal500&\textbf{0.137$\pm$0.006}&\textbf{1.5}&0.140$\pm$0.008&6&0.139$\pm$0.006&5&0.138$\pm$0.005&3.5&0.138$\pm$0.006&3.5&0.146$\pm$0.007&7&\textbf{0.137$\pm$0.005}&\textbf{1.5} \\
			emotions&0.197$\pm$0.029&5&0.209$\pm$0.019&7&0.193$\pm$0.019&3&0.184$\pm$0.015&2&0.195$\pm$0.023&4&0.207$\pm$0.025&6&\textbf{0.182$\pm$0.026}&\textbf{1} \\
			genbase&\textbf{0.002$\pm$0.001}&\textbf{2}&0.008$\pm$0.002&7&0.006$\pm$0.003&6&0.005$\pm$0.002&5&0.003$\pm$0.001&4&\textbf{0.002$\pm$0.001}&\textbf{2}&\textbf{0.002$\pm$0.001}&\textbf{2} \\
			plant&0.090$\pm$0.002&5&0.092$\pm$0.005&6&0.088$\pm$0.003&4&0.085$\pm$0.004&2.5&\textbf{0.084$\pm$0.004}&\textbf{1}&0.123$\pm$0.006&7&0.085$\pm$0.005&2.5 \\
			medical&\textbf{0.011$\pm$0.001}&\textbf{1.5}&0.018$\pm$0.001&6&0.016$\pm$0.002&5&0.012$\pm$0.002&3&0.015$\pm$0.002&4&0.019$\pm$0.002&7&\textbf{0.011$\pm$0.001}&\textbf{1.5} \\
			language log&\textbf{0.016$\pm$0.001}&\textbf{3}&0.020$\pm$0.001&6&\textbf{0.016$\pm$0.001}&\textbf{3}&\textbf{0.016$\pm$0.001}&\textbf{3}&\textbf{0.016$\pm$0.001}&\textbf{3}&0.047$\pm$0.004&7&\textbf{0.016$\pm$0.001}&\textbf{3} \\
			human&0.083$\pm$0.002&4&0.085$\pm$0.002&5.5&0.082$\pm$0.002&3&\textbf{0.078$\pm$0.002}&\textbf{1.5}&\textbf{0.078}$\pm$\textbf{0.002}&\textbf{1.5}&0.086$\pm$0.003&6&0.085$\pm$0.004&5.5 \\
			education&\textbf{0.037$\pm$0.001}&\textbf{1}&0.053$\pm$0.002&7&0.039$\pm$0.002&5&0.038$\pm$0.002&3&0.040$\pm$0.001&6&0.038$\pm$0.001&3&0.038$\pm$0.002&3\\
			science&\textbf{0.030$\pm$0.002}&1&0.043$\pm$0.002&7&0.033$\pm$0.001&5&0.031$\pm$0.001&2&0.034$\pm$0.001&6&0.032$\pm$0.001&3.5&0.032$\pm$0.002&3.5\\
			rcv1(subset2)&\textbf{0.023$\pm$0.001}&\textbf{3}&0.026$\pm$0.001&7&0.024$\pm$0.001&6&\textbf{0.023$\pm$0.001}&\textbf{3}&\textbf{0.023$\pm$0.001}&\textbf{3}&0\textbf{.023$\pm$0.001}&\textbf{3}&\textbf{0.023$\pm$0.001}&\textbf{3} \\
			rcv1(subset5)&\textbf{0.023$\pm$0.001}&\textbf{3}&0.025$\pm$0.001&7&0.024$\pm$0.001&6&\textbf{0.023$\pm$0.001}&\textbf{3}&\textbf{0.023$\pm$0.001}&\textbf{3}&\textbf{0.023$\pm$0.001}&\textbf{3}&\textbf{0.023$\pm$0.001}&\textbf{3} \\
			ohsumed&0.058$\pm$0.002&3.5&0.078$\pm$0.001&7&0.071$\pm$0.001&6&0.056$\pm$0.001&2&0.067$\pm$0.001&5&\textbf{0.055$\pm$0.001}&\textbf{1}&0.058$\pm$0.001&3.5 \\
			\bottomrule
			\multirow{2}{*}{Data set}&\multicolumn{14}{c}{Accuracy~$\uparrow \hspace{25pt}$}\\ \cline{2-15}
			&BSVM&&BP-MLL&&ML-kNN&&LIFT&&SSJDR-MLL&&LLSF&&MLC-HMF& \\
			\hline
			cal500&0.197$\pm$0.010&5&0.216$\pm$0.015&3&0.196$\pm$0.014&6&0.198$\pm$0.008&4&0.192$\pm$0.010&7&\textbf{0.228$\pm$0.013}&\textbf{1}&0.217$\pm$0.009&2 \\
			emotions&0.495$\pm$0.061&7&0.548$\pm$0.032&2.5&0.538$\pm$0.043&4&0.548$\pm$0.043&2.5&0.512$\pm$0.058&5&0.511$\pm$0.049&6&\textbf{0.550$\pm$0.060}&\textbf{1} \\
			genbase&0.973$\pm$0.013&3&0.905$\pm$0.019&7&0.923$\pm$0.029&6&0.929$\pm$0.021&5&0.953$\pm$0.018&4&\textbf{0.981$\pm$0.007}&\textbf{1}&0.978$\pm$0.006&2 \\
			plant&0.037$\pm$0.014&6&0.012$\pm$0.031&7&0.079$\pm$0.021&5&0.150$\pm$0.028&4&0.163$\pm$0.033&3&\textbf{0.206$\pm$0.023}&\textbf{1}&0.173$\pm$0.033&2 \\
			medical&\textbf{0.732$\pm$0.014}&\textbf{1}&0.629$\pm$0.027&5&0.570$\pm$0.051&7&0.659$\pm$0.036&3&0.620$\pm$0.042&6&0.657$\pm$0.043&4&0.702$\pm$0.031&2 \\
			language log&0.062$\pm$0.018&5&\textbf{0.161$\pm$0.024}&\textbf{1}&0.017$\pm$0.009&7&0.109$\pm$0.024&3&0.030$\pm$0.014&6&0.095$\pm$0.024&4&0.121$\pm$0.025&2 \\
			human&0.076$\pm$0.011&6&0.000$\pm$0.001&7&0.094$\pm$0.013&5&0.201$\pm$0.012&3&\textbf{0.210$\pm$0.016}&\textbf{1}&0.193$\pm$0.020&4&0.207$\pm$0.016&2 \\
			education&0.270$\pm$0.019&4&\textbf{0.327$\pm$0.017}&\textbf{1}&0.225$\pm$0.015&6&0.255$\pm$0.016&5&0.158$\pm$0.011&7&0.293$\pm$0.017&3&0.294$\pm$0.021&2\\
			science&0.280$\pm$0.018&3&\textbf{0.311$\pm$0.020}&\textbf{1}&0.198$\pm$0.013&6&0.280$\pm$0.010&3&0.136$\pm$0.013&7&0.264$\pm$0.015&5&0.280$\pm$0.018&3\\
			rcv1(subset2)&0.325$\pm$0.017&2&0.324$\pm$0.011&3&0.188$\pm$0.015&7&0.272$\pm$0.014&5&0.205$\pm$0.016&6&0.281$\pm$0.013&4&\textbf{0.343$\pm$0.014}&\textbf{1}\\
			rcv1(subset5)&0.324$\pm$0.013&2&0.321$\pm$0.014&3&0.192$\pm$0.017&7&0.285$\pm$0.015&4&0.219$\pm$0.013&6&0.275$\pm$0.013&5&\textbf{0.352$\pm$0.012}&\textbf{1}\\
			ohsumed&0.293$\pm$0.014&5&\textbf{0.417$\pm$0.005}&\textbf{1}&0.053$\pm$0.011&7&0.337$\pm$0.013&4&0.131$\pm$0.007&6&0.356$\pm$0.013&2&0.344$\pm$0.011&3 \\
			\bottomrule
			\multirow{2}{*}{Data set}&\multicolumn{14}{c}{Subset Accuracy~$\uparrow\hspace{20pt}$}\\ \cline{2-15}
			&BSVM&&BP-MLL&&ML-kNN&&LIFT&&SSJDR-MLL&&LLSF&&MLC-HMF& \\
			\hline
			cal500&\textbf{0.000$\pm$0.000}&\textbf{4}&\textbf{0.000$\pm$0.000}&\textbf{4}&\textbf{0.000$\pm$0.000}&\textbf{4}&\textbf{0.000$\pm$0.000}&\textbf{4}&\textbf{0.000$\pm$0.000}&\textbf{4}&\textbf{0.000$\pm$0.000}&\textbf{4}&\textbf{0.000$\pm$0.000}&\textbf{4} \\
			emotions&0.271$\pm$0.074&5&0.259$\pm$0.056&7&0.292$\pm$0.054&3&\textbf{0.319$\pm$0.053}&\textbf{1}&0.290$\pm$0.068&4&0.265$\pm$0.050&6&0.305$\pm$0.077&2 \\
			genbase&0.961$\pm$0.015&3&0.837$\pm$0.021&7&0.891$\pm$0.043&6&0.897$\pm$0.031&5&0.936$\pm$0.024&4&\textbf{0.967$\pm$0.019}&\textbf{1}&0.962$\pm$0.012&2 \\
			plant&0.036$\pm$0.014&6&0.012$\pm$0.031&7&0.077$\pm$0.019&5&0.143$\pm$0.025&3&0.150$\pm$0.028&2&0.141$\pm$0.021&4&\textbf{0.162$\pm$0.034}&\textbf{1} \\
			medical&\textbf{0.653$\pm$0.027}&\textbf{1}&0.464$\pm$0.032&7&0.499$\pm$0.055&5&0.586$\pm$0.044&3&0.528$\pm$0.064&4&0.498$\pm$0.052&6&0.630$\pm$0.043&2 \\
			language log&0.193$\pm$0.028&4&0.232$\pm$0.032&2&0.159$\pm$0.026&7&0.230$\pm$0.037&3&0.167$\pm$0.020&6&0.168$\pm$0.026&5&\textbf{0.233$\pm$0.030}&\textbf{1} \\
			human&0.069$\pm$0.012&6&0.000$\pm$0.001&7&0.085$\pm$0.015&5&0.181$\pm$0.011&2&\textbf{0.190$\pm$0.015}&\textbf{1}&0.154$\pm$0.020&4&0.166$\pm$0.018&3 \\
			education&0.226$\pm$0.018&3&0.165$\pm$0.013&6&0.189$\pm$0.015&5&0.212$\pm$0.014&4&0.131$\pm$0.009&7&\textbf{0.240$\pm$0.015}&\textbf{1.5}&\textbf{0.240$\pm$0.020}&\textbf{1.5}\\
			science&\textbf{0.237$\pm$0.018}&\textbf{1}&0.189$\pm$0.020&5&0.170$\pm$0.013&6&0.233$\pm$0.014&2&0.115$\pm$0.015&7&0.219$\pm$0.015&4&0.226$\pm$0.022&3 \\
			rcv1(subset2)&\textbf{0.207$\pm$0.015}&\textbf{1}&0.088$\pm$0.012&7&0.146$\pm$0.015&5&0.166$\pm$0.010&3&0.140$\pm$0.011&6&0.165$\pm$0.014&4&0.200$\pm$0.015&2\\
			rcv1(subset5)&\textbf{0.201$\pm$0.011}&\textbf{1}&0.082$\pm$0.009&7&0.144$\pm$0.013&5&0.174$\pm$0.012&4&0.135$\pm$0.011&6&0.152$\pm$0.012&3&0.194$\pm$0.009&2\\
			ohsumed&0.189$\pm$0.011&4&0.132$\pm$0.008&5&0.033$\pm$0.009&7&0.215$\pm$0.009&2&0.084$\pm$0.009&6&\textbf{0.222$\pm$.011}&\textbf{1}&0.209$\pm$0.011&3 \\
			\bottomrule
	\end{tabular}
\end{adjustbox}
	\label{Err:HlossAccuSubsetAccu}
\end{table*}

\begin{table*}[ht!]
	\renewcommand{\arraystretch}{1.2}
	\centering
	\captionsetup{font=scriptsize,justification=centering}
	\caption{Experimental results of comparison of algorithms (mean$\pm$std rank)  in terms of Example Based $F_1$, Macro $F_1$, and Micro $F_1$}
	\begin{adjustbox}{max width=\textwidth}
		\begin{tabular}{lclclclclclclcl}
			\toprule
			\multirow{2}{*}{Data set}&\multicolumn{14}{c}{Example based $F_1$~$\uparrow\hspace{5pt}$}\\ \cline{2-15}
			&BSVM&&BP-MLL&&ML-kNN&&LIFT&&SSJDR-MLL&&LLSF&&MLC-HMF& \\
			\hline
			cal500&0.325$\pm$0.014&5&0.350$\pm$0.020&2.5&0.323$\pm$0.020&6&0.326$\pm$0.011&4&0.316$\pm$0.013&7&\textbf{0.362$\pm$0.018}&\textbf{1}&0.350$\pm$0.013&2.5 \\
			emotions&0.565$\pm$0.059&7&\textbf{0.647$\pm$0.026}&\textbf{1}&0.620$\pm$0.041&4&0.624$\pm$0.043&3&0.583$\pm$0.058&6&0.591$\pm$0.052&5&0.629$\pm$0.057&2 \\
			genbase&0.976$\pm$0.013&3&0.924$\pm$0.021&7&0.933$\pm$0.025&6&0.938$\pm$0.018&5&0.958$\pm$0.017&4&\textbf{0.985$\pm$0.005}&\textbf{1}&0.983$\pm$0.006&2 \\
			plant&0.038$\pm$0.014&6&0.012$\pm$0.031&7&0.080$\pm$0.022&5&0.153$\pm$0.029&4&0.167$\pm$0.034&3&\textbf{0.230$\pm$0.026}&\textbf{1}&0.180$\pm$0.032&2 \\
			medical&\textbf{0.759$\pm$0.012}&\textbf{1}&0.686$\pm$0.028&4&0.595$\pm$0.051&7&0.684$\pm$0.036&5&0.653$\pm$0.037&6&0.715$\pm$0.041&3&0.728$\pm$0.028&2 \\
			language log&0.067$\pm$0.018&5&\textbf{0.188$\pm$0.025}&\textbf{1}&0.018$\pm$0.009&7&0.118$\pm$0.025&4&0.033$\pm$0.0140&6&0.132$\pm$0.026&3&0.133$\pm$0.026&2 \\
			human&0.079$\pm$0.011&6&0.001$\pm$0.001&7&0.097$\pm$0.013&5&0.214$\pm$0.013&3&0.217$\pm$0.016&2&0.206$\pm$0.021&4&\textbf{0.221$\pm$0.015}&\textbf{1} \\
			education&0.286$\pm$0.019&4&\textbf{0.384$\pm$0.019}&\textbf{1}&0.238$\pm$0.015&6&0.271$\pm$0.018&5&0.167$\pm$0.012&7&0.312$\pm$0.017&3&0.314$\pm$0.022&2\\
			science&0.295$\pm$0.019&4&\textbf{0.357$\pm$0.021}&\textbf{1}&0.207$\pm$0.013&6&0.298$\pm$0.012&3&0.143$\pm$0.013&7&0.279$\pm$0.016&5&0.300$\pm$0.017&2 \\
			rcv1(subset2)&0.370$\pm$0.019&3&\textbf{0.430$\pm$0.014}&\textbf{1}&0.204$\pm$0.015&7&0.313$\pm$0.016&5&0.230$\pm$0.018&6&0.326$\pm$0.012&4&0.400$\pm$0.015&2\\
			rcv1(subset5)&0.374$\pm$0.014&3&\textbf{0.427$\pm$0.016}&\textbf{1}&0.210$\pm$0.019&7&0.331$\pm$0.017&4&0.254$\pm$0.013&6&0.326$\pm$0.014&5&0.415$\pm$0.014&2\\
			ohsumed&0.331$\pm$0.015&5&\textbf{0.521$\pm$0.006}&\textbf{1}&0.060$\pm$0.013&7&0.382$\pm$0.015&4&0.149$\pm$0.007&6&0.404$\pm$.014&2&0.393$\pm$0.013&3 \\
			\bottomrule
			\multirow{2}{*}{Data set}&\multicolumn{14}{c}{Macro $F_1$~$\uparrow\hspace{10pt}$}\\ \cline{2-15}
			\multirow{3}{*}{}&BSVM&&BP-MLL&&ML-kNN&&LIFT&&SSJDR-MLL&&LLSF&&MLC-HMF& \\
			\hline
			cal500&0.046$\pm$0.003&6&0.053$\pm$0.008&3.5&0.053$\pm$0.007&3.5&0.049$\pm$0.003&7&0.041$\pm$0.004&5&\textbf{0.107$\pm$0.010}&\textbf{1}&0.079$\pm$0.006&2 \\
			emotions&0.584$\pm$0.059&7&0.626$\pm$0.029&3&0.622$\pm$0.033&5&\textbf{0.652$\pm$0.035}&\textbf{1}&0.623$\pm$0.046&4&0.617$\pm$0.048&6&0.649$\pm$0.057&2\\
			genbase&0.649$\pm$0.062&2.5&0.534$\pm$0.047&6&0.520$\pm$0.026&7&0.584$\pm$0.052&5&0.631$\pm$0.061&4&\textbf{0.666$\pm$0.076}&\textbf{1}&0.649$\pm$0.053&2.5 \\
			plant&0.042$\pm$0.019&6&0.011$\pm$0.022&7&0.055$\pm$0.016&5&0.086$\pm$0.023&4&0.124$\pm$0.034&3&\textbf{0.164$\pm$0.031}&\textbf{1}&0.126$\pm$0.027&2\\
			medical&0.320$\pm$0.030&2&0.245$\pm$0.020&5&0.206$\pm$0.026&7&0.267$\pm$0.032&4&0.223$\pm$0.022&6&\textbf{0.329$\pm$0.039}&\textbf{1}&0.285$\pm$0.025&3\\
			language log&0.031$\pm$0.007&5&0.051$\pm$0.006&2&0.007$\pm$0.004&7&0.047$\pm$0.015&3.5&0.026$\pm$0.007&6&\textbf{0.074$\pm$0.014}&\textbf{1}&0.047$\pm$0.01&3.5\\
			human&0.041$\pm$0.008&6&0.000$\pm$0.000&7&0.065$\pm$0.016&5&0.119$\pm$0.010&3&\textbf{0.140$\pm$0.010}&\textbf{1}&0.106$\pm$0.008&4&0.122$\pm$0.012&2\\
			education&\textbf{0.140$\pm$0.016}&\textbf{1}&0.094$\pm$0.006&6&0.110$\pm$0.017&5&0.136$\pm$0.021&2&0.084$\pm$0.013&7&0.116$\pm$0.014&3&0.112$\pm$0.009&4\\
			science&0.167$\pm$0.022&2&0.111$\pm$0.012&6&0.12$\pm$0.018&5&\textbf{0.182$\pm$0.019}&\textbf{1}&0.067$\pm$0.013&7&0.132$\pm$0.014&4&0.134$\pm$0.011&3\\
			rcv1(subset2)&\textbf{0.198$\pm$0.014}&\textbf{1}&0.083$\pm$0.003&5&0.082$\pm$0.010&6&0.128$\pm$0.013&3&0.074$\pm$0.006&7&0.117$\pm$0.012&4&0.172$\pm$0.009&2\\
			rcv1(subset5)&0.186$\pm$0.015&2&0.073$\pm$0.003&6.5&0.085$\pm$0.010&5&0.129$\pm$0.020&3&0.073$\pm$0.006&6.5&0.098$\pm$0.009&4&\textbf{0.188$\pm$0.016}&\textbf{1}\\
			ohsumed&0.238$\pm$0.010&5&\textbf{0.436$\pm$0.012}&\textbf{1}&0.041$\pm$0.007&7&0.275$\pm$0.012&4&0.082$\pm$0.005&6&0.332$\pm$0.01&2&0.279$\pm$0.010&3\\
			\bottomrule
			\multirow{2}{*}{Data set}&\multicolumn{14}{c}{Micro $F_1$~$\uparrow\hspace{10pt}$}\\ \cline{2-15}
			&BSVM&&BP-MLL&&ML-kNN&&LIFT&&SSJDR-MLL&&LLSF&&MLC-HMF& \\
			\hline
			cal500&0.321$\pm$0.015&5&0.347$\pm$0.019&3&0.320$\pm$0.020&6&0.323$\pm$0.011&4&0.311$\pm$0.012&7&\textbf{0.367$\pm$0.018}&\textbf{1}&0.348$\pm$0.012&2\\
			emotions&0.636$\pm$0.055&7&0.670$\pm$0.028&3&0.667$\pm$0.043&4&0.679$\pm$0.033&2&0.655$\pm$0.046&5&0.639$\pm$0.049&6&\textbf{0.682$\pm$0.051}&\textbf{1}\\
			genbase&0.981$\pm$0.009&2&0.917$\pm$0.016&7&0.930$\pm$0.031&6&0.947$\pm$0.018&5&0.966$\pm$0.013&4&\textbf{0.984$\pm$0.007}&\textbf{1}&0.978$\pm$0.009&3\\
			plant&0.067$\pm$0.024&6&0.000$\pm$0.000&7&0.133$\pm$0.033&5&0.234$\pm$0.043&4&0.258$\pm$0.051&3&\textbf{0.279$\pm$0.023}&\textbf{}1&0.267$\pm$0.042&2 \\
			medical&\textbf{0.804$\pm$0.020}&\textbf{1}&0.706$\pm$0.022&5&0.676$\pm$0.040&7&0.758$\pm$0.029&3&0.713$\pm$0.029&4&0.701$\pm$0.028&6&0.775$\pm$0.025&2\\
			language log&0.122$\pm$0.025&5&\textbf{0.264$\pm$0.029}&\textbf{1}&0.000$\pm$0.000&7&0.199$\pm$0.041&3&0.063$\pm$0.024&6&0.157$\pm$0.022&4&0.215$\pm$0.031&2\\
			human&0.126$\pm$0.016&6&0.000$\pm$0.001&7&0.152$\pm$0.018&5&0.299$\pm$0.017&2&0.298$\pm$0.018&3&0.283$\pm$0.023&4&\textbf{0.302$\pm$0.024}&\textbf{1}\\
			education&0.377$\pm$0.024&4&\textbf{0.400$\pm$0.019}&\textbf{1.5}&0.330$\pm$0.022&6&0.362$\pm$0.022&5&0.260$\pm$0.018&7&0.397$\pm$0.023&3&\textbf{0.400$\pm$0.026}&\textbf{1.5}\\
			science&0.376$\pm$0.022&2&0.368$\pm$0.021&4&0.286$\pm$0.022&6&\textbf{0.377$\pm$0.016}&\textbf{1}&0.199$\pm$ 0.019&7&0.357$\pm$0.017&5&0.370$\pm$0.020&3\\
			rcv1(subset2)&0.403$\pm$0.021&3&0.419$\pm$0.011&2&0.242$\pm$0.019&7&0.347$\pm$0.016&5&0.268$\pm$0.020&6&0.361$\pm$0.013&4&\textbf{0.421$\pm$0.013}&\textbf{1}\\
			rcv1(subset5)&0.413$\pm$0.012&3&0.430$\pm$0.012&2&0.253$\pm$0.023&7&0.375$\pm$0.016&4&0.298$\pm$0.016&6&0.366$\pm$0.014&5&\textbf{0.436$\pm$0.012}&\textbf{1}\\
			ohsumed&0.407$\pm$0.014&5&\textbf{0.539$\pm$0.007}&\textbf{1}&0.088$\pm$0.017&7&0.456$\pm$0.012&4&0.204$\pm$0.009&6&0.484$\pm$.012&2&0.463$\pm$0.011&3 \\
			\bottomrule
	\end{tabular}
\end{adjustbox}
	\label{Err:exampleBasedF1MacroF1MicroF1}
\end{table*}

\begin{subsection}{Results and Discussion} 
Table~\ref{Err:HlossAccuSubsetAccu} and~\ref{Err:exampleBasedF1MacroF1MicroF1} gives the comparative analysis of the proposed method MLC-HMF against state-of-the-art algorithms on  twelve datasets. Each result is composed of $mean$, $std$ and $rank$. For any data set and given evaluation metric where two or more algorithms obtain the same performance, the rank of these algorithms are assigned with the average result of them. For each evaluation criterion, $\uparrow (\downarrow)$ indicates the larger (smaller) the value, the better the performance. Furthermore, the best results among all comparing algorithms are highlighted in boldface. 
\begin{table}[ht!]
	\renewcommand{\arraystretch}{1}
	\centering
	\caption{Summary of the Friedman statistics $F_F(\mathcal{K}=7,\mathcal{N}=12)$ and the	critical value in terms of each evaluation metric($\mathcal{K}$: \# Comparing Algorithms; $\mathcal{N}$: \# Data Sets).}
	\begin{tabular}{llc}
		\toprule
		Metric &$F_F$&Critical Value ($\alpha = 0.05$)\\
		\toprule
		Hamming Loss&5.730&\multirow{6}{*}{2.239}\\ 
		Accuracy&4.910&\\
		Subset Accuracy&5.584&\\
		Example Base $F_1$&5.338&\\
		Macro $F_1$&2.767&\\
		Micro $F_1$&4.958&\\
		\bottomrule
	\end{tabular}
	\label{ffTest_mlc-hmf}
\end{table}

To conduct statistical performance analysis among the algorithms being compared, we employed Friedman test which is a favorable statistical test for comparisons of more than two algorithms over multiple datasets~\cite{demvsar2006statistical}. Table \ref{ffTest_mlc-hmf} provides the Friedman statistics $F_F$ and the corresponding critical value in terms of each evaluation metric. As shown in Table \ref{ffTest_mlc-hmf} at significance level $\alpha = 0.05$, Friedman test rejects the null hypothesis of equal performance for each evaluation metrics. This leads to the use of post-hoc tests for pairwise comparisons. The Nemenyi test~\cite{demvsar2006statistical} is employed to test whether our proposed method MLC-HMF achieves a competitive performance against the comparing algorithms. The performance of two classifiers is significantly different if the corresponding average ranks differ by at least the critical difference $CD = q_{\alpha}\sqrt{ \frac{\mathcal{K}(\mathcal{K}+1)}{6\mathcal{N}}}$. At significance level $\alpha = 0.05$, the value of $q_{\alpha}= 2.949$, for Nemenyi test with $\mathcal{K}=7$~\cite{demvsar2006statistical}, and thus $CD=2.601$. Figure~\ref{CDDiagram_mlc-hmf} gives the CD diagrams~\cite{demvsar2006statistical}
for each evaluation criterion, where the average rank of each comparing algorithm is marked along the axis (lower ranks to the left). It can be seen from the Figure~\ref{CDDiagram_mlc-hmf} that the proposed method MLC-HMF achieve  better performance than other comparing algorithms in terms of each evaluation metric.
\noindent
\begin{figure}[ht!]
	\begin{tabular}{ll}	
		\begin{minipage}{0.5\textwidth}
			\centering
			\includegraphics[width=2.5in,height=1.5in]{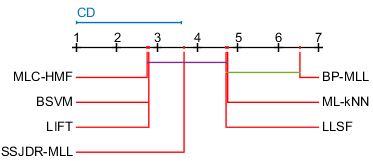}
			\captionof{subfigure}{(a) Hamming loss}
			\label{fig:sub1_MLC_HMF}
		\end{minipage}
		&
		\begin{minipage}{0.5\textwidth}
			\centering
			\includegraphics[width=2.5in,height=1.5in]{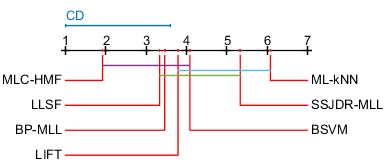}
			\captionof{subfigure}{(b) Accuracy}
			\label{fig:sub2_MLC_HMF}
		\end{minipage}
		\\
		\begin{minipage}{0.5\textwidth}
			\centering
		\includegraphics[width=2.5in,height=1.5in]{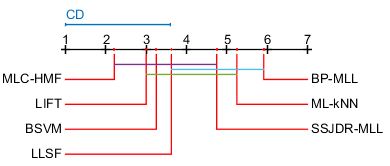}
		\captionof{subfigure}{(c) Subset Accuracy}
		\label{fig:sub3_MLC_HMF}
		\end{minipage}
		& 
		\begin{minipage}{0.5\textwidth}
			\centering
		\includegraphics[width=2.5in,height=1.5in]{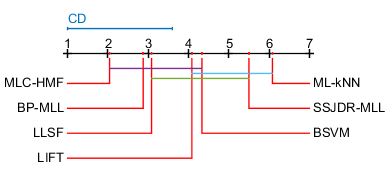}
		\captionof{subfigure}{(d) Example based F$\textsubscript{1}$ }
		\label{fig:sub4_MLC_HMF}
		\end{minipage}
		\\ 
		\begin{minipage}{0.5\textwidth}
			\centering
			\includegraphics[width=2.5in,height=1.5in]{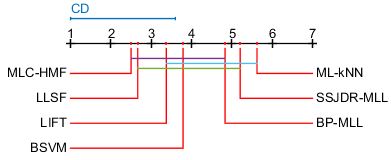}
			\captionof{subfigure}{(e) Macro F$\textsubscript{1}$ }
			\label{fig:sub5_MLC_HMF}%
		\end{minipage}
		&
		\begin{minipage}{0.5\textwidth}
		\centering
		\includegraphics[width=2.5in,height=1.5in]{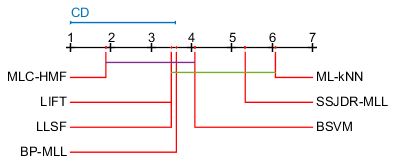}
		\captionof{subfigure}{(f) Micro F$\textsubscript{1}$ }
		\label{fig:sub6_MLC_HMF}
		\end{minipage}
	\end{tabular}
\caption{CD diagrams of the comparing algorithms under each evaluation criterion.}
\label{CDDiagram_mlc-hmf}
\end{figure}
\end{subsection}

\section{Conclusions and Discussion}
	\label{conclusion_mlc_hmf}
	This chapter presented a new multi-label classification method, called MLC-HMF, which learns  piecewise-linear embedding with a low-rank constraint on parametrization to capture nonlinear intrinsic relationships that exist in the original feature and label space. Extensive comparative studies validate the effectiveness of MLC-HMF against the state-of-the-art multi-label learning approaches. 
	
	In multi-label classification problem, infrequently occurring (tail) labels are associated with few training data and are harder to predict than frequently occurring ones but might also be more informative and rewarding. Due to the presence of tail labels the low-rank label matrix assumption fails to hold in embedding methods in real-world applications. We feel that our proposal of hierarchical embedding may pave way to overcome this difficulty and can handle tail labels more efficiently. 
	MLC-HMF is the first ever attempt for piece-wise linear embedding in the context of multi-label learning. It is worthwhile  to carryout an in depth investigation of different ways of embedding such as quasi-linear embeddings and their advantages over linear and non-linear embedding. We plan to carryout this line of investigation in future. Our experimental analysis provides evidence that hierarchical embedding is able to yield more accurate results for multi-label classification. This motivates a new line of future research in which computational complexity of the proposed algorithm can be investigated. The core part of the algorithm is a search based on gradient descent. It is worthwhile to investigate different algorithmic strategies of gradient descent to improve the computational efficiency.

\newpage
\thispagestyle{empty}
\setcounter{chapter}{5}
\chapter{Group Preserving Label Embedding for Multi-Label Classification}
\label{GroPLEChapter}
In the previous chapter, we discussed a joint learning framework called MLC-HMF in which the feature space embedding and multi-label classification are performed simultaneously. The MLC-HMF approach is motivated by the fact that the low-rank assumption made by the embedding methods is violated in most of the real-world applications due to several reasons. Reasons such as the diversity of the training set, the correlation among labels, the feature-label relationship, the mapping to be determined by the learning algorithm and most importantly, the presence of \textit{tail} labels to mention a few. However, on the other hand, it is possible to have a fraction of training instances share the same subset of label correlations. The underlying principle of MLC-HMF is to learn a piecewise-linear embedding of feature-space that generates optimal embeddings for a subset of training instances. As discussed in Chapter~\ref{mlchmfChapter}, there are two different approach of embedding namely, feature space embedding and label space embedding. In this chapter,  we propose a novel label embedding based approach for multi-label classification.

\section{Introduction}      
In this chapter, we discuss a novel label embedding approach for multi-label classification inspired by the extensive applications of data mining algorithms, where features or data items are inherently organized into groups. In the context of  multi-label classification, there are few proposals which model group information but no detailed study in this direction has so far been undertaken. To exploit label-correlations in the data locally, it is assumed in~\cite{huang2012multi} that the training data are  in  groups with instances in the same group sharing the same label correlations. In~\cite{sun2016multi}, highly correlated labels are grouped together using the information from label and instance spaces and for every group, sparse meta-label-specific features are learnt. 

The present research starts with the assumption that there exists a low-rank space onto which the given set of feature vectors and label vectors can be embedded. Feature vectors can be embedded as  points and label vectors correspond to linear predictors, as decision hyperplanes, in this embedded space. There are similarities among labels belonging to the same group such that their low-rank representations share the same sparsity pattern. For example, the set of labels in \textit{corel5k} data set can be grouped into \textit{landscape-nature}, \textit{humans}, \textit{food} etc.~\cite{fakhari2013combination}. The features such as \text{eye} and \textit{leg} are specific to the \textit{humans} whereas the feature like \textit{ridge} is specific to the group \textit{landscape-nature}. Given the label matrix $Y$, it is necessary to find points in a space of reduced dimension and to determine decision hyperplanes such that the classification thereof is compatible with $Y$. While doing so, it is desirable that the process must retain group information of labels.  We achieve this by a matrix factorization based framework in which the label matrix $Y$ is approximated using the product of two matrices $U \in \mathbb{R}^{N \times d}$ and $V \in \mathbb{R}^{d \times L}$.  In a sense the row of matrix $U$ can be viewed as point in a reduced dimension and the column of $V$ defines a set of decision hyperplanes. If there are any dependency properties in labels of $Y$ (column  of $Y$), this is retained in dependencies in decision hyperplanes (column of $V$) and not in embedded points. 
	We use the $\ell_{2,1}$ norm regularization on $V$ to exploit the shared sparsity pattern among the label groups. The second sub-objective is to learn a linear mapping that maps the feature vectors onto the same set of points which are obtained as a result of factorization of label matrix. We achieve this by a separate optimization problem. We make use of correlation coefficients to capture the similarity relation and $\ell_1$ norm for regularization. 
	We use FISTA~\cite{beck2009fast} type of method to learn the label embedding and subsequently mapping from feature space to the embedded label space.  Thus, in this chapter, we develop a novel multi-label classification method. To the best of our knowledge, there has not been any earlier attempt in this direction. We feel that this approach will eventually provide a robust classification technique as demonstrated by our experimental results which looks very promising. 
	
	The rest of the chapter is organized as follows. We introduce our proposed method, termed as GroPLE \textit{(Group Preserving Label Embedding for Multi-Label Classification)} in Section~\ref{proposedMethod}. Experimental analysis of the proposed method is reported in Section~\ref{experimentalSection}. Finally, we conclude with Section~\ref{conclusion} and indicates several issues for future work.
	
	\section{GroPLE: The Proposed Method}   
	\label{proposedMethod}
	In this section, a novel method of multi-label classification is proposed. The proposed method GroPLE has three major stages namely (1) Identification of groups of labels; (2) Embedding of label vectors to a low rank-space so that the sparsity characteristic of individual groups remains invariant; and (3) Determining a linear mapping that embeds the feature vectors onto the same set of points, as in stage 2, in the low-rank space.
	
	\noindent \textbf{Identification of Groups of Labels:} The label groups are not given explicitly and it is necessary to learn from the label matrix $Y$. One approach is to cluster the columns of $Y$. There are several clustering algorithms proposed in the literature such as \textit{k-means}~\cite{jain1999data,jain2010data}, \textit{hierarchical clustering}~\cite{johnson1967hierarchical} and \textit{spectral clustering}~\cite{zelnik2004self,ng2002spectral,von2007tutorial}. We adopt \textit{spectral clustering}. We do not claim that \textit{spectral clustering} is the best option. We first construct a graph $G = <\mathcal{V}, E>$ in the label space, where $\mathcal{V}$ denotes the vertex/ label set, and $E$ is the edge set containing edges between each label pair. We adopt heat kernel weight with self-tuning technique (for parameter $\sigma$) as edge weight if two labels are connected $A_{i,j} = \exp(\frac{(-\|Y_{.i} - Y_{.j}\|^2)}{\sigma})$ where $Y_{.i}$ and $Y_{.j}$ are the $i$th and $j$th column of matrix $Y$~\cite{zelnik2004self}. Labels can be grouped into $K$ clusters by performing	\textit{k-means} with K largest eigenvectors as seeds  of the normalized affinity matrix $L=D^{-\frac{1}{2}}AD^{-\frac{1}{2}}$, where $D$ is a diagonal matrix with $D_{i,i} = \sum_{j}{}A_{i,j}$.		

	\noindent \textbf{Label Space Embedding:} Given a label matrix $Y$, each column corresponds to a label and our assumption is that related labels form groups.	
	Let the columns of $Y \in \{-1,1\}^{N \times L}$ be divided into K groups as $Y = (Y^1, \dots, Y^K)$, where $Y^k \in  \{-1, 1\}^{N \times L_k}$ and $\sum_{k}^{K} L_k = L$. Matrix factorization based approach of label embedding as discussed in Section~\ref{problemFormulation_mlc_hmf} aims to find latent factor matrices $U \in \mathbb{R}^{ N \times d}$ and $V \in \mathbb{R}^{ d \times L}$ to approximate $Y$. In the present case, where labels are divided into groups, we approximate $Y^k$ using $U$ and $V^k \in \mathbb{R}^{ d \times L_k}$. Ideally, there should be a subset of columns of $U$ associated with any group and hence, the corresponding  vector in $V^k$ of a label should have nonzero values only for the entries which correspond to the subset of columns of $U$ associated with the group. More concretely, we expect that for deciding any label groupss all the features are not important and each label in that group can be decided by linear combination of fewer group features. To achieve this,  we add a $\ell_{2, 1}$-norm regularization on $V^k$ that encourages row sparsity of $V^k$. Then the sub-objective to learn the embedding from original label space is given by	
	\begin{equation}
	\label{GroPLE}
	\underset{U, V^1, \dots, V^K}{min}~f(U, V^1, \dots, V^K) = \sum_{k=1}^{K}\|Y^{k}-UV^{k}\|^2_F + \lambda_1\|U\|_F^2 + \lambda_2\sum_{k=1}^{K}\|V^k\|_{2,1}
	\end{equation}
	where for a given matrix $A \in \mathbb{R}^{n \times m}$, $\|A\|_F^2 = \sum_{i=1}^{n}{\sum_{j=1}^{m}{A_{ij}^2}}$ and  $\|A\|_{2,1} = \linebreak \sum_{i=1}^{n}{\sqrt{\sum_{j=1}^{m}A_{ij}^2}}$. 	
	
	We can solve Eq. (\ref{GroPLE}) by alternating minimization scheme that iteratively optimizes each of the factor matrices keeping the other fixed. For simplicity of notation, the matrix formed by arranging the columns of $V^k$, $1\le k \le K$, according to indices of columns in $Y$ will be referred to as $V$ in subsequent discussions. $f(U, V)$ is written as $f_V(U)$ when $V$ is held constant and  $f_U(V)$ when $U$ is held constant. For given $V$, the factor matrix $U$ can be obtained by solving the following subproblem.	
	\begin{equation}
	\label{eqUpdateU}
	\underset{U}{min}~f_V(U)=~\|Y-UV\|^2_F + \lambda_1\|U\|_F^2 + c
	\end{equation}
	where $c \ge 0$ is a constant. The subproblem given in Eq.~(\ref{eqUpdateU}) has a closed form solution. Taking the derivative of $f_V(U)$ w.r.t $U$, and setting the derivative (in matrix notation) to zero, we have
	\begin{align}
		\nabla f_V(U) &= 2((Y-UV)(-V^T) + \lambda_1 U)= 0 \nonumber \\
		&\Rightarrow U = YV^T \inv{(VV^T+\lambda_1 I)}
	\end{align}
	For fixed $U$, the matrix $V^k$, $k \in \{1, \dots, K\}$, can be obtained by solving the following subproblem
	\begin{equation}
	\label{subproblemVg}
	\underset{V^k}{min}~f_{U}(V^k) = \|Y^{k}-UV^{k}\|^2_F  + \lambda_2\|V^k\|_{2,1} + c
	\end{equation}
	The above objective function is a composite convex function involving the sum of a smooth and a non-smooth function  of the form
	\begin{equation}
	\label{compositeFun}
	\underset{V^k}{min}~f_{U}(V^k) = g(V^k) + h(V^k)
	\end{equation}
	where $g(V^k) = \|Y^{k}-UV^{k}\|^2_F $ is convex and differentiable and $h(V^k) = \lambda_2\|V^k\|_{2,1}$ is closed, convex but non-differentiable. 
	
	We further show that for any two matrices ${V^{'}}^{k}$, ${V^{''}}^{k} \in \mathbb{R}^{d \times L_k}$, the function $g(V^k) $ is Lipschitz continuous. 
	The gradient of $g(V^k)$ (in matrix notation) is given by
	\begin{align*}
		\nabla g(V^k) = 2(U^{T}UV^k - U^{T}Y)
	\end{align*}
	For any two matrices ${V^{'}}^{k}$ and ${V^{''}}^{k}$, we have
	\begin{align*}
	\|\nabla g({V^{'}}^{k}) - \nabla g({V^{''}}^{k}) \|_F^2 &= \|2(U^{T}U{V^{'}}^{k} - U^{T}Y) - 2(U^{T}U{V^{''}}^{k} - U^{T}Y)\|^2_F \\
	&=\|2U^{T}U({V^{'}}^{k}-{V^{''}}^{k})\|_F^2\\
	&\le\|2U^{T}U\|^2_F~\|{V^{'}}^{k} - {V^{''}}^{k}\|_F^2	
	\end{align*}	
	Therefore, the Lipschitz constant is
	\begin{equation}
		L_g = \sqrt{\|2U^{T}U\|^2_F} 
	\end{equation}	
	
	We employ Accelerated Proximal Gradient search~\cite{toh2010accelerated} which is specifically tailored to minimize the optimization problem given in Eq.~(\ref{compositeFun}). Such an optimization strategy is suitable in the present situation as the computation of the proximal operation is inexpensive. The optimization step of Accelerated Proximal Gradient iterates as follows.
	\begin{align}
	G_t &= {V^k}^{(t)} + \frac{b_{t-1}-1}{b_t}({V^k}^{(t)} - {V^k}^{(t-1)})\\
	{V^k}^{(t)} &= prox_{h}(G_t - \frac{1}{L_g}\nabla g(G_t)) 
	\end{align}

	It is shown in~\cite{toh2010accelerated} that setting $b_t$ satisfying $b_t^2 - b_t \le b_{t-1}^2$ can improve convergence rate to $O(\frac{1}{t^2})$.       ${V^k}^{(t)}$ is the result of $t$th iteration. Proximal mapping of a convex function $h$ is given by
	\begin{equation}
	prox_h(V^k) = \underset{W}{argmin}~\left(h(W) + \frac{1}{2}\|W-V^k\|^2_2~\right)
	\end{equation} 	
	
	In the present situation, where $h(V^k) = \lambda_2\|V^k\|_{2,1}$, $prox_h(V^k)$ is the shrinkage function $S[\cdot]$ and is given by
	\begin{equation}
	S_{\frac{\lambda_2}{L_g}}[V^k] = \bigg [\frac{v^{k}_{i}}{\|v^{k}_{i}\|_2} (\|v^{k}_{i}\|_2- \lambda_2/L_g)_{+}\bigg ]_{i=1}^{i=d}
	\end{equation}
	where $(z)_{+} = max(z,0)$ and $v^k_i$ is the $i$th row of $V^k$. Algorithm~\ref{algo:LE} outlines the main flow of the optimization steps to solve Eq.~(\ref{GroPLE}).   
	\begin{algorithm}
		\SetAlgoLined
		\SetKwData{Left}{left}\SetKwData{This}{this}\SetKwData{Up}{up}
		\SetKwFunction{Union}{Union}\SetKwFunction{FindCompress}{FindCompress}
		\SetKwInOut{Input}{input}\SetKwInOut{Output}{output}
		\Input{Label Matrix: $Y$, Size of Latent Dimension Space: $d$, Number of Groups: $K$, Regularization Parameters: $\lambda_1$ and $\lambda_2$}
		\Output{Basis Matrix: $U$, Coefficient Matrix: $V$}
		\BlankLine
		\textbf{initialize :} $U$\\
		Form label groups $Y^1$, $Y^2$, \dots, $Y^K$ \\
		\Repeat{stop criterion reached}{
			\For{k $\in$ \{1,\dots, K\}}{
				$V^k \leftarrow$ APG($U$, $Y^k$, $\lambda_2$)
			}
		    $V \leftarrow Combine(V^1, V^2, \dots, V^k)$ \\
		    $U \leftarrow YV^T \inv{(VV^T+\lambda_1 I)}$ 
		}
		\caption{Label-Embedding ( $Y$, $d$, $K$, $\lambda_1$, $\lambda_2$)}
		\label{algo:LE}
	\end{algorithm}    

    \begin{algorithm}
    	\SetAlgoLined
    	\SetKwData{Left}{left}\SetKwData{This}{this}\SetKwData{Up}{up}
    	\SetKwFunction{Union}{Union}\SetKwFunction{FindCompress}{FindCompress}
    	\SetKwInOut{Input}{input}\SetKwInOut{Output}{output}
    	\Input{Basic Matrix: $U$, Label Matrix: $Y^k$ and Regularization Parameters: $\lambda_2$}
    	\Output{Coefficient Matrix: $V^k$}
    	\BlankLine
    	\textbf{initialize :} \\
    	$b_0$, $b_1 \leftarrow 1$, $V^k_0$, $V^k_1 \leftarrow$ \inv{($U^TU + \gamma I $)}$U^TY^k$\\
    	\Repeat{stop criterion reached}{
    		$G_t \leftarrow {V^k}^{(t)} + \frac{b_{t-1}-1}{b_t}({V^k}^{(t)} - {V^k}^{(t-1)})$ \\
    		${V^k}^{(t)} = S_{\frac{\lambda_2}{L_g}}[G_t - \frac{1}{L_g}\nabla g(G_t)]$\\
    		$b_t \leftarrow \frac{1+ \sqrt{1 + 4b_t^2}}{2}$ \\
    		$t \leftarrow t+1$
    	}
        $V^k \leftarrow V^k_t$
    	\caption{APG ($U$, $Y^k$, $\lambda_2$)}
    	\label{Algo:APG}
    \end{algorithm} 

	\noindent \textbf{Feature Space Embedding:}
	The $U$ matrix computed above as a result of the learning process described
	in Algorithm~\ref{algo:LE} represents a set of points and it is desired that these points, in some sense, represent the training objects. We assume that there exists a linear embedding $Z \in \mathbb{R}^{D \times d}$ that maps the feature matrix $X$ to $U$. Thus, we justify our hypothesis that there exists a low-rank space where both $X$ and $Y$ are embedded and this embedding retains the intrinsic feature-label relation as well as the group information. In order to achieve the embedding of feature vectors, we try to capture the correlation among the embedded representation of $Y$ and formulate the objective function as follows. 
	\begin{equation}
	\label{featureEmbedding}
	\underset{Z}{min}~ \|XZ - U\|_{F}^{2} + \alpha \sum_{j=1}^{d}R_{ij}Z_{i}^TZ_{j}  + \beta\|Z\|_{1}
	\end{equation}
	where $Z_i$ is the $i$th column of matrix $Z$ and $R_{ij} = 1 - C_{ij}$, where $C_{ij}$ represent the correlation coefficient between $i$th and $j$th column of matrix $U$. The above objective function is of the form defined in Eq.~(\ref{compositeFun}) where $g(Z) = \|XZ - U\|_{F}^{2} + \alpha \Tr(RZ^TZ)$ is convex and Lipschitz continuous and $h(Z) = \beta\|Z\|_{1}$ is closed, convex but non-differentiable. The Accelerated Proximal Gradient technique described previously is used to solve Eq.~(\ref{featureEmbedding}). The optimization step of Accelerated Proximal Gradient initializes $Z_0 = Z_1 = \inv{(X^TX + \gamma I)} $ and iterates as follows.	
	\begin{align}
	G_t &= {Z}^{(t)} + \frac{b_{t-1}-1}{b_t}(Z^{(t)} - Z^{(t-1)})\\
	Z^{(t)} &= prox_{h}(G_t - \frac{1}{L_g}\nabla g(G_t)) \label{proxz}
	\end{align}
	where $Z^{(t)}$ is the result of $t$th iteration, the Lipschitz constant $L_g$ and $prox_h(Z)$ is given below. The gradient of $g(Z)$ (in matrix notation) is given by
	\begin{equation*}
		\nabla g(Z) = 2X^T(XZ-U) + \alpha ZR
	\end{equation*}
	For any two matrices $Z^{'}$, $Z^{''} \in \mathbb{R}^{D \times d}$, we have    
	\begin{align*}
	\|\nabla g(Z^{'}) - \nabla g(Z^{''}) \|_F^2 &= 
	\|2(X^{T}XZ^{'} - X^{T}U) +  \alpha Z^{'}R - 2(X^{T}XZ^{''} - X^{T}U) -  \alpha Z^{''}R\|_F^2 \\
	&=\|2X^TX(Z^{'}-Z^{''}) + \alpha(Z^{'}-Z^{''})R\|^2_F   \\
	&\le \|2X^TX\|^2_F~\|Z^{'}-Z^{''}\|^2_F + \|\alpha R\|^2_F~\|Z^{'}-Z^{''}\|^2_F  \\
	&= \|2X^TX + \alpha R\|^2_F~\|Z^{'}-Z^{''}\|^2_F 
	\end{align*}
	Therefore, the Lipschitz constant is
	\begin{equation}
	L_g = \sqrt{\|2X^TX + \alpha R\|^2_F} 
	\end{equation}		
	The proximal mapping of $h(Z) = \beta\|Z\|_1$ is a shrinkage function $S[\cdot]$ and is given by	
	\begin{equation}
	prox_h(z) = S_\frac{\beta}{L_g}[Z] =
	\begin{cases}
	Z_{ij} - \beta / L_g, & ~~\text{$Z_{ij} > \beta / L_g$} \\ 
	0, & \text{$-\beta / L_g\le Z_{ij} \le \beta / L_g$}\\
	Z_{ij} +\beta / L_g, & ~~\text{$Z_{ij} < \beta / L_g$}
	\end{cases}
	\end{equation}
	where, $1\le i \le d$ and $1\le j \le k$.\\
  
   \noindent \textbf{Complexity Analysis:} We analyze the computational complexity of the proposed method. The time complexity of GroPLE mainly comprises of three components: formation of label groups and the optimization of the problem given in Eq.~(\ref{GroPLE}) and~(\ref{featureEmbedding}). The formation of label groups has two parts: construction of neighbourhood graph and spectral decomposition of a graph Laplacian. This part takes $O(NL^2 + L^3)$. For each iteration in Algorithm~\ref{algo:LE}, updating $U$ requires $O(NLd + d^3 + Nd^2)$. For simplicity of representation, we are ignoring the number of groups $K$ and using the total number of labels $L$. Hence, the updation of $V$ takes  $O(NLd + d^3 + Nd^2 + Ld^2)$. Let $t_1$ be the maximum number of iterations required for gradient update, then overall computation required in LE process is $O(NL^2 + L^3) + O(t_1(NLd + d^3 + Nd^2)) + O(t_1(NLd + d^3 + Nd^2 + Ld^2))$, that is, $O(t_1(NLd + d^3 + Nd^2 + Ld^2))$. Similarly, the complexity of feature space embedding is $O(ND^2 + D^3) + O(2NDd + Dd^2)$, that is, $O(t_2(2NDd + Dd^2))$, where $t_2$ is the number of iteration. Hence the overall computation required by GroPLE is $O(t_1(NLd + d^3 + Nd^2 + Ld^2) + t_2(2NDd + Dd^2))$.
	
	\section{Experimental Analysis}
	\label{experimentalSection}		
	To validate the proposed GroPLE, we perform experiments on twelve commonly used multi-label benchmark data sets. The detailed characteristics of these data sets are summarized in Table \ref{datasetsCharacteristics_GroPLE}. All the data sets are publicly available and can be downloaded from \textit{meka}\footnote{ \url{http://meka.sourceforge.net}} and \textit{mulan}\footnote{ \url{http://mulan.sourceforge.net/datasets-mlc.html}}.
	
	\begin{table}[h!]
		\renewcommand{\arraystretch}{1.2}
		\centering
		\captionsetup{font=scriptsize,justification=centering}
		\caption{Description of the experimental datasets.}
		\begin{tabular}{llllll}
			\toprule
			Data set&\#instance&\#Feature&\#Label&Domain&LC \\
			\hline
	    	genbase&662&1185&27&biology&1.252\\
			medical&978&1449&45&text&1.245\\
			CAL500&502&68&174&music&26.044\\
			corel5k&5000&499&374&image&3.522\\
			rcv1 (subset 1)&6000&944&101&text&2.880\\
			rcv1 (subset 2)&6000&944&101&text&2.634\\
			rcv1 (subset 3)&6000&944&101&text&2.614\\
			bibtex&7395&1836&159&text&2.402\\
			corle16k001&13766&500&153&image&2.859\\
			delicious&16105&500&983&text(web)&19.020\\
			mediamill&43907&120&101&video&4.376\\
			bookmarks&87856&2150&208&text&2.028\\
			\bottomrule
		\end{tabular}
		\label{datasetsCharacteristics_GroPLE}
	\end{table}    
	\subsection{Evaluation Metrics}
	To measure the performance of different algorithms, we have employed four evaluation metrics popularly used in multi-label classification, i.e. \textit{accuracy, example based $f_1$ measure, macro $f_1$} and \textit{micro $f_1$}~\cite{zhang2014review,sorower2010literature}. Given a test data set $\mathcal{D} = \{x_i, y_i~|~1 \le i\le N\}$, where $y_i \in \{-1, 1\}^L$ is the ground truth labels associated with the $i$th test example, and $\hat{y_i}$ be its predicted set of labels. The detailed discussion of these evaluation metrics is given in Section~\ref{experimentalSection_mlc_hmf}.
	
%
%
	
	\subsection{Baseline Methods}
	
	For performance comparison, we consider seven well-known state-of-the-art algorithms and these are the following.
	
	\begin{itemize}
		\item \textbf{BSVM}~\cite{boutell2004learning}: This is one of the representative algorithms of problem transformation methods, which treat each label as a separate binary classification problem. For every label, an independent binary classifier is trained by considering the examples with the given class label as positive and others as negative. LIBSVM~\cite{CC01a} is employed as the binary learner for classifier induction to instantiate BSVM.
		\item \textbf{PLST}~\cite{tai2012multilabel}: Principal label space transformation (PLST) uses singular value decomposition (SVD) to project the original label space into a low dimensional label space. 
		\item \textbf{CPLST}~\cite{chen2012feature}: CPLST is a feature-aware conditional principal label space transformation which utilizes the feature information during label embedding. 
		\item \textbf{CSSP}~\cite{bi2013efficient}: Using randomized sampling procedure, CSSP sample a small subset of class labels that can approximately span the original label space. Once this subset of labels are selected, a binary classifier is trained for each of them. 
		\item \textbf{FAiE}~\cite{lin2014multi}: FAiE encodes the original label space to a low-dimensional latent space via feature-aware implicit label space encoding. It directly learns a feature-aware code matrix and a linear decoding matrix via jointly maximizing recoverability of the original label space. 
		\item \textbf{LEML}~\cite{yu2014large}:  In this method a framework is developed to model  multi-label classification as  generic empirical risk minimization (ERM) problem with low-rank constraint on linear transformation. It can also be seen as a joint learning framework in which dimensionality reduction and multi-label classification are performed simultaneously. 
		\item \textbf{MLSF}~\cite{sun2016multi}: Based on the assumption that meta-labels with specific features exist in the scenario of multi-label classification, MLSF embed label correlations into meta-labels in such a way that the member labels in a meta-label share strong dependency with each other but have weak dependency with the other non-member labels. 
	\end{itemize}
	 A linear ridge regression model is used in PLST, CPLST, CSSP and FAiE to learn the association between feature space and reduced label space. For PLST, CPLST, CSSP, FAiE  and LEML the number of reduced dimensions $d$ is searched in  $\{\lceil0.1L\rceil, \linebreak \lceil0.2L\rceil,\dots, \lceil0.8L\rceil\}$.   The regularization parameter in ridge regression, the parameter $\alpha$ in FAiE and the parameter $\lambda$ in LEML are searched in the range $\{ 10^{-4}, 10^{-3}, \dots, 10^{4}\}$. For MLSF, the number of meta-labels $K$ is searched in  $\{\lceil L/5 \rceil, \lceil L/10 \rceil, \lceil L/15 \rceil, \lceil L/20 \rceil \}$ and the parameters $\gamma$ and $\rho$ are tuned from the candidate set $\{ 10^{-4}, 10^{-3}, \dots, 10^{4}\}$. The remaining hyper-parameters  were kept fixed across all datasets as was done in~\cite{sun2016multi}.	Implementations of PLST, CPLST, CSSP, FAiE, LEML, MLSF were provided by the authors.
	
	\subsection{Results and Discussion}

    We first demonstrate the effect of group sparsity regularization on the recovered matrix $V$ on \emph{Medical} data set. The regularization parameter $\lambda_1$ and $\lambda_2$ in Eq.~(\ref{GroPLE}) are selected using Cross-validation. The label matrix $Y$ is divided into five groups using the method described in Section~\ref{proposedMethod}. 
     \begin{figure}[ht!]
    	\centering
    	\includegraphics[width=3.3in,height=2.8in]{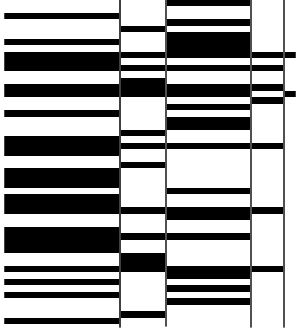}
    	\caption{Latent factor matrix $V^k$ recovered with five label groups.} 
    	\label{grayScaleImgofLatentFactor}
    \end{figure}
    The recovered feature matrix $V^k$ for each group is depicted in Figure~\ref{grayScaleImgofLatentFactor}. The shaded color represents the non-zero rows in the recovered matrix $V^k$ and a separation line is artificially drawn to distinguish between the latent factor matrix of the different groups. It is evident from Figure~\ref{grayScaleImgofLatentFactor} that the feature matrix recovered for different groups exhibits a sparsity pattern and the labels corresponding to objects in the same group have similar sparsity.     
        
    \begin{figure}[ht!]
		\begin{minipage}{1\linewidth}
			\centering
			\includegraphics[width=3.8in,height=3.5in]{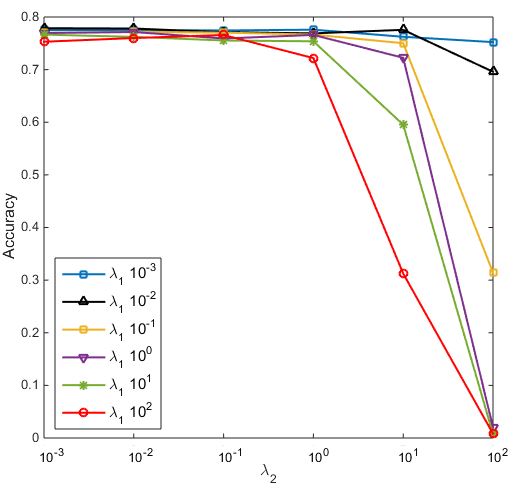}
			\captionof{subfigure}{(a) Medical dataset}
            \label{fig:lambda1lambda2Accuracy_Medical}
        \end{minipage}%
        \vspace{0.3cm}
		\begin{minipage}{1\linewidth}
			\centering
            \includegraphics[width=3.8in,height=3.5in]{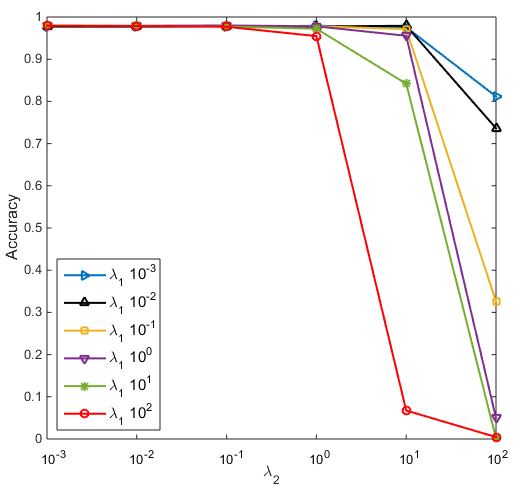}
            \captionof{subfigure}{(b) Genbase dataset}
            \label{fig:lambda1lambda2Accuracy_Genbase}
		\end{minipage}
\caption{Influence of regularization parameters $\lambda_1$ and $\lambda_2$.}		
\label{lambda1Lambda2Analysis}
\end{figure}         
To study the sensitivity of GroPLE with respect to the regularization parameters $\lambda_1$ and $\lambda_2$, we conducted experiments  on \textit{Medical} and \textit{Genbase} data sets. We perform \textit{five-fold cross validation} on each data set and the mean value of accuracy is recorded. In this experiment, the latent dimension space $d$ is fixed to $100$, the number of groups $K$ is fixed to $5$ and the regularization parameters $\lambda_1$ and $\lambda_2$ are searched in  $\{10^{-3}, 10^{-2}, \dots, 10^{2}\}$. For each $(\lambda_1, \lambda_2)$ -pair, the regularization parameters $\alpha$, $\beta$ are searched in $\{ 10^{-4}, 10^{-3}, \dots, 10^{4}\}$. Figure~\ref{lambda1Lambda2Analysis} report the influence of parameters $\lambda_1$ and $\lambda_2$ on  \textit{Medical} and \textit{Genbase} data set.   It can be seen that in most cases: (a) GroPLE perform worse when the value of $\lambda_1$ is large; (b) The performance of GroPLE is stable with the different value of group sparsity regularization $\lambda_2$, but the larger value such as $\lambda_2 \ge 1$ is often harmful. Therefore, we fixed the regularization parameter $\lambda_1$ and $\lambda_2$ to $0.001$ and $1$, respectively, for the subsequent experiments.
   
\begin{figure}[ht!]
		\begin{minipage}{1\linewidth}
			\centering
			\includegraphics[width=4in,height=3.4in]{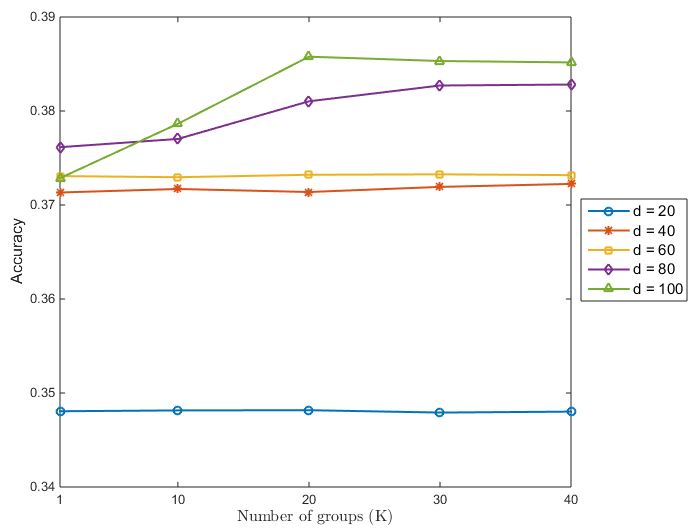}
            \captionof{subfigure}{ (a) Accuracy}
            \label{fig:groupVsNoGroupAccuracy}
        \end{minipage}%
        \vspace{0.2cm}
		\begin{minipage}{1\linewidth}
			\centering
            \includegraphics[width=4in,height=3.4in]{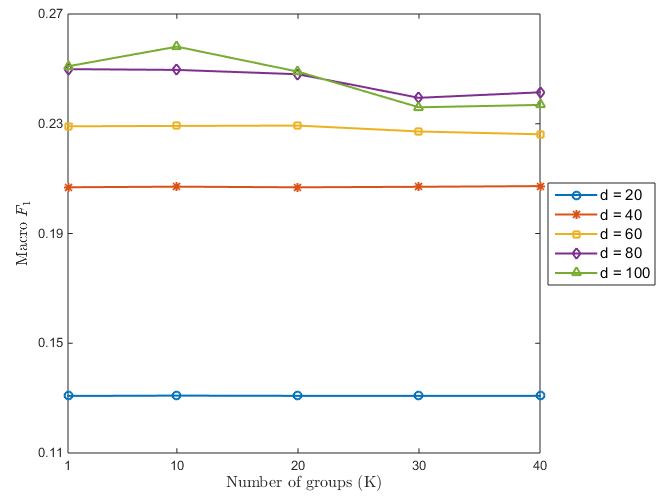}
           \captionof{subfigure}{(b) Macro $F_1$}
            \label{fig:groupVsNoGroupMacroF1}
		\end{minipage}
\caption{Performance of GroPLE on \textit{rcv1 (subset 1)} data set with different group size.}
\label{groupVsNoGroup}
\end{figure}
We have also analyzed the performance of GroPLE with respect to the latent dimension $d$ and the number of groups $K$ on \textit{rcv1 (subset 1)} data set.  We have conducted \textit{five-fold cross validation} and the mean value of accuracy is recorded. The latent dimension $d$ is varied from $[20,100]$ with step size $20$ and the number of groups $K$ is selected from $\{1, 10, 20, 30, 40\}$. The regularization parameters $\alpha$ and $\beta$ are tuned in the range given previously. The plots of Figure~\ref{groupVsNoGroup} shows the classification performance of GroPLE in terms of \textit{accuracy} and \textit{Macro $F_1$}. It can be seen from  Figure~\ref{groupVsNoGroup} that the classification performance of GroPLE is nearly constant for different group size when the latent dimension $d$ is small which is also obvious as there are less number of features to differentiate between one group from others. The classification performance improved as we increased the number of groups $K$ for sufficiently large $d$. It can also be seen that the performance in terms of \textit{Macro $F_1$} degrade for sufficiently large group size $K$. Hence, by considering the balance between latent dimension $d$ and number of groups $K$, we fixed the value of $K$ and $d$ to $10$ and $100$, respectively, for subsequent experiments.
   
\begin{table*}[ht!]
 	\renewcommand{\arraystretch}{1.2}
	\centering	
	\captionsetup{font=scriptsize,justification=centering}
	\caption{Experimental results of each comparing algorithm (mean$\pm$std rank)  in terms of Accuracy, Example Based $F_1$, Macro $F_1$, and Micro $F_1$. Method that cannot be run  with available resources are denoted as ``-".}
	\adjustbox{max width=\linewidth}{
		\begin{tabular}{lcccccccc}
			\toprule
			\multirow{2}{*}{Data set}&\multicolumn{8}{c}{Accuracy}\\ \cline{2-9}
			&GroPLE&PLST&CPLST&CSSP&FAiE&LEML&MLSF&BSVM \\
			\hline
			genbase&0.972 $\pm$ 0.014&0.968 $\pm$ 0.013&0.971 $\pm$  0.013&0.931 $\pm$  0.042&0.965 $\pm$  0.014&0.971 $\pm$  0.014&\textbf{0.974 $\pm$ 0.011}&0.973 $\pm$ 0.010\\
			medical&0.769 $\pm$ 0.031&0.762 $\pm$ 0.034&0.763 $\pm$ 0.033&0.735 $\pm$ 0.041&0.765 $\pm$ 0.035&0.690 $\pm$ 0.025&\textbf{0.779 $\pm$ 0.029}&0.753 $\pm$ 0.034\\
			CAL500&0.234 $\pm$ 0.006&0.224 $\pm$ 0.007&0.224 $\pm$ 0.007&0.224 $\pm$ 0.006&\textbf{0.235 $\pm$ 0.007}&0.222 $\pm$ 0.006&0.177 $\pm$ 0.010&0.202 $\pm$ 0.007\\
			corel5k&\textbf{0.151 $\pm$ 0.002}&0.054 $\pm$ 0.002&0.054 $\pm$ 0.002&0.052 $\pm$ 0.004&0.089 $\pm$ 0.003&0.053 $\pm$ 0.002&0.088 $\pm$ 0.008&0.081 $\pm$ 0.002\\
			rcv1 (subset 1)&\textbf{0.335 $\pm$ 0.003}&0.231 $\pm$ 0.009&0.232 $\pm$ 0.008&0.218 $\pm$ 0.004&0.265 $\pm$ 0.008&0.223 $\pm$ 0.006&0.322 $\pm$ 0.071&0.288 $\pm$ 0.010\\
			rcv1 (subset 2)&\textbf{0.371 $\pm$ 0.011}&0.299 $\pm$ 0.014&0.299 $\pm$ 0.013&0.297 $\pm$ 0.013&0.328 $\pm$ 0.011&0.286 $\pm$ 0.013&0.365 $\pm$ 0.010&0.348 $\pm$ 0.012\\
			rcv1 (subset 3)&\textbf{0.375 $\pm$ 0.016}&0.289 $\pm$ 0.014&0.289 $\pm$ 0.014&0.276 $\pm$ 0.017&0.319 $\pm$ 0.018&0.280 $\pm$ 0.014&0.373 $\pm$ 0.009&0.347 $\pm$ 0.009\\
			bibtex&\textbf{0.329 $\pm$ 0.004}&0.284 $\pm$ 0.009&0.288 $\pm$ 0.007&0.266 $\pm$ 0.010&0.310 $\pm$ 0.004&0.288 $\pm$ 0.007&0.328 $\pm$ 0.005&0.326 $\pm$ 0.008\\
			corel16k001&\textbf{0.158 $\pm$ 0.006}&0.039 $\pm$ 0.001&0.038 $\pm$ 0.001&0.035 $\pm$ 0.003&0.080 $\pm$ 0.002&0.039 $\pm$ 0.002&0.034 $\pm$ 0.006&0.025 $\pm$ 0.002\\
			delicious&\textbf{0.185 $\pm$ 0.001}&0.109 $\pm$ 0.002&0.109 $\pm$ 0.002&0.107 $\pm$ 0.002&0.134 $\pm$ 0.001&0.093 $\pm$ 0.001&0.118 $\pm$ 0.006&0.130 $\pm$ 0.001\\
			mediamill&\textbf{0.434 $\pm$ 0.004}&0.414 $\pm$ 0.003&0.414 $\pm$ 0.003&0.406 $\pm$ 0.016&0.425 $\pm$ 0.004&0.411 $\pm$ 0.003&0.395 $\pm$ 0.012&0.393 $\pm$ 0.003\\
			bookmarks&\textbf{0.283 $\pm$ 0.002}&0.174 $\pm$ 0.002&&0.169 $\pm$ 0.007&-&0.175 $\pm$ 0.002&0.163 $\pm$ 0.004&-\\
			\hline
			\multirow{2}{*}{Data set}&\multicolumn{8}{c}{Example based $F_1$}\\ \cline{2-9}
			\multirow{3}{*}{}&GroPLE&PLST&CPLST&CSSP&FAiE&LEML&MLSF&BSVM \\
			\hline
			genbase&0.978 $\pm$ 0.013&0.975 $\pm$ 0.012&0.977 $\pm$ 0.013&0.939 $\pm$ 0.043&0.973 $\pm$ 0.012&0.977 $\pm$ 0.013&0.978 $\pm$ 0.015&\textbf{0.980 $\pm$ 0.010}\\
			medical&0.800 $\pm$ 0.031&0.793 $\pm$ 0.030&0.794 $\pm$ 0.033&0.766 $\pm$ 0.039&0.796 $\pm$ 0.034&0.738 $\pm$ 0.027&\textbf{0.807 $\pm$ 0.026}&0.783 $\pm$ 0.032\\
			CAL500&\textbf{0.370 $\pm$ 0.008}&0.357 $\pm$ 0.009&0.358 $\pm$ 0.009&0.358 $\pm$ 0.007&0.369 $\pm$ 0.009&0.355 $\pm$ 0.008&0.296 $\pm$ 0.014&0.330 $\pm$ 0.010\\
			corel5k&\textbf{0.230 $\pm$ 0.003}&0.078 $\pm$ 0.002&0.078 $\pm$ 0.002&0.075 $\pm$ 0.005&0.127 $\pm$ 0.004&0.076 $\pm$ 0.003&0.126 $\pm$ 0.005&0.114 $\pm$ 0.004\\
			rcv1 (subset 1)&\textbf{0.447 $\pm$ 0.002}&0.298 $\pm$ 0.009&0.299 $\pm$ 0.009&0.283 $\pm$ 0.006&0.339 $\pm$ 0.009&0.288 $\pm$ 0.006&0.399 $\pm$ 0.006&0.379 $\pm$ 0.012\\
			rcv1 (subset 2)&\textbf{0.466 $\pm$ 0.011}&0.346 $\pm$ 0.015&0.346 $\pm$ 0.014&0.343 $\pm$ 0.014&0.380 $\pm$ 0.011&0.331 $\pm$ 0.014&0.429 $\pm$ 0.011&0.423 $\pm$ 0.013\\
			rcv1 (subset 3)&\textbf{0.471 $\pm$ 0.016}&0.335 $\pm$ 0.015&0.336 $\pm$ 0.015&0.321 $\pm$ 0.019&0.372 $\pm$ 0.019&0.325 $\pm$ 0.014&0.434 $\pm$ 0.015&0.422 $\pm$ 0.008\\
			bibtex&\textbf{0.415 $\pm$ 0.004}&0.335 $\pm$ 0.009&0.339 $\pm$ 0.008&0.314 $\pm$ 0.010&0.367 $\pm$ 0.004&0.338 $\pm$ 0.008&0.399 $\pm$ 0.009&0.383 $\pm$ 0.008\\
			corel16k001&\textbf{0.227 $\pm$ 0.006}&0.053 $\pm$ 0.002&0.053 $\pm$ 0.002&0.047 $\pm$ 0.005&0.111 $\pm$ 0.002&0.053 $\pm$ 0.002&0.035 $\pm$ 0.011&0.034 $\pm$ 0.002\\
			delicious&\textbf{0.293 $\pm$ 0.001}&0.169 $\pm$ 0.002&0.169 $\pm$ 0.002&0.167 $\pm$ 0.003&0.209 $\pm$ 0.001&0.142 $\pm$ 0.002&0.181 $\pm$ 0.010&0.201 $\pm$ 0.002\\
			mediamill&\textbf{0.552 $\pm$ 0.004}&0.533 $\pm$ 0.004&0.533 $\pm$ 0.004&0.524 $\pm$ 0.018&0.544 $\pm$ 0.004&0.530 $\pm$ 0.003&0.507 $\pm$ 0.019&0.515 $\pm$ 0.003\\
			bookmarks&\textbf{0.324 $\pm$ 0.006}&0.179 $\pm$ 0.002&-&0.173 $\pm$ 0.007&-&0.180 $\pm$ 0.002&0.166 $\pm$ 0.005&-\\
			\hline
			\multirow{2}{*}{Data set}&\multicolumn{8}{c}{Macro $F_1$}\\ \cline{2-9}
			\multirow{3}{*}{}&GroPLE&PLST&CPLST&CSSP&FAiE&LEML&MLSF&BSVM \\
			\hline
			genbase&0.711 $\pm$ 0.086&0.709 $\pm$ 0.068&0.711 $\pm$ 0.075&0.673 $\pm$ 0.069&0.682 $\pm$ 0.066&0.705 $\pm$ 0.078	&0.736 $\pm$ 0.084&\textbf{0.737 $\pm$ 0.082}\\
			medical&0.374 $\pm$ 0.034&0.373 $\pm$ 0.025&0.378 $\pm$ 0.030&0.360 $\pm$ 0.026&\textbf{0.383 $\pm$ 0.025}&0.342 $\pm$ 0.025&0.385$\pm$ 0.036&0.383 $\pm$ 0.037\\
			CAL500&\textbf{0.129 $\pm$ 0.007}&0.110 $\pm$ 0.006&0.104 $\pm$ 0.005&0.107 $\pm$ 0.005&0.118 $\pm$ 0.004&0.110 $\pm$ 0.005&0.034 $\pm$ 0.011&0.057 $\pm$ 0.001\\
			corel5k&\textbf{0.049 $\pm$ 0.002}&0.016 $\pm$ 0.002&0.016 $\pm$ 0.001&0.015 $\pm$ 0.002&0.026 $\pm$ 0.001&0.016 $\pm$ 0.002&0.046 $\pm$ 0.002& 0.044 $\pm$ 0.002\\
			rcv1 (subset 1)&\textbf{0.263 $\pm$ 0.006}&0.126 $\pm$ 0.006&0.125 $\pm$ 0.005&0.119 $\pm$ 0.006&0.163 $\pm$ 0.009&0.125 $\pm$ 0.005&0.257 $\pm$ 0.014&0.257 $\pm$ 0.005\\
			rcv1 (subset 2)&\textbf{0.247 $\pm$ 0.003}&0.111 $\pm$ 0.006&0.112 $\pm$ 0.006&0.109 $\pm$ 0.006&0.148 $\pm$ 0.004&0.111 $\pm$ 0.006&0.241 $\pm$ 0.006&0.240 $\pm$ 0.007\\
			rcv1 (subset 3)&\textbf{0.244 $\pm$ 0.004}&0.109 $\pm$ 0.003&0.109 $\pm$ 0.003&0.101 $\pm$ 0.005&0.144 $\pm$ 0.007&0.108 $\pm$ 0.004&0.235 $\pm$ 0.013&0.231 $\pm$ 0.012\\
			bibtex&0.304 $\pm$ 0.014&0.197 $\pm$ 0.005&0.208 $\pm$ 0.008&0.178 $\pm$ 0.007&0.236 $\pm$ 0.007&0.208 $\pm$ 0.008&0.326 $\pm$ 0.006&\textbf{0.327 $\pm$ 0.004}\\
			corel16k001&\textbf{0.088 $\pm$ 0.004}&0.015 $\pm$ 0.001&0.015 $\pm$ 0.002&0.014 $\pm$ 0.002&0.023 $\pm$ 0.002&0.015 $\pm$ 0.002&0.040 $\pm$ 0.002&0.036 $\pm$ 0.006\\
			delicious&0.089 $\pm$ 0.000&0.048 $\pm$ 0.001&0.048 $\pm$ 0.002&0.046 $\pm$ 0.002&0.059 $\pm$ 0.002&0.048 $\pm$ 0.002&\textbf{0.102 $\pm$ 0.001}&0.100 $\pm$ 0.003\\
			mediamill&\textbf{0.087 $\pm$ 0.001}&0.045 $\pm$ 0.001&0.045 $\pm$ 0.001&0.043 $\pm$ 0.002&0.055 $\pm$ 0.001&0.043 $\pm$ 0.001&0.041 $\pm$ 0.001&0.032 $\pm$ 0.001\\
			bookmarks&\textbf{0.140 $\pm$ 0.003}&0.057 $\pm$ 0.001&-&0.053 $\pm$ 0.004&-&0.059 $\pm$ 0.001&0.040 $\pm$ 0.001&-\\
			\hline
			\multirow{2}{*}{Data set}&\multicolumn{8}{c}{Micro $F_1$}\\ \cline{2-9}
			\multirow{3}{*}{}&GroPLE&PLST&CPLST&CSSP&FAiE&LEML&MLSF&BSVM \\
			\hline
			genbase&0.967 $\pm$ 0.031&0.969 $\pm$ 0.012&0.972 $\pm$ 0.013&0.951 $\pm$ 0.022&0.967 $\pm$ 0.013&0.973 $\pm$ 0.013&0.978 $\pm$ 0.012&\textbf{0.979 $\pm$ 0.008}\\
			medical&0.821 $\pm$ 0.028&0.821 $\pm$ 0.030&0.822 $\pm$ 0.029&0.806 $\pm$ 0.029&\textbf{0.823 $\pm$ 0.030}&0.739 $\pm$ 0.019&0.822 $\pm$ 0.033&0.812 $\pm$ 0.027\\
			CAL500&\textbf{0.375 $\pm$ 0.007}&0.360 $\pm$ 0.008&0.361 $\pm$ 0.009&0.361 $\pm$ 0.006&0.374 $\pm$ 0.008&0.359 $\pm$ 0.007&0.290 $\pm$ 0.038&0.327 $\pm$ 0.009\\
			corel5k&\textbf{0.242 $\pm$ 0.003}&0.105 $\pm$ 0.003&0.105 $\pm$ 0.002&0.101 $\pm$ 0.006&0.162 $\pm$ 0.004&0.103 $\pm$ 0.003&0.154 $\pm$ 0.007&0.143 $\pm$ 0.005\\
			rcv1 (subset 1)&\textbf{0.463 $\pm$ 0.005}&0.350 $\pm$ 0.010&0.351 $\pm$ 0.010&0.336 $\pm$ 0.005&0.384 $\pm$ 0.009&0.344 $\pm$ 0.008&0.405 $\pm$ 0.006&0.394 $\pm$ 0.009\\
			rcv1 (subset 2)&\textbf{0.455 $\pm$ 0.009}&0.373 $\pm$ 0.016&0.373 $\pm$ 0.015&0.372 $\pm$ 0.015&0.403 $\pm$ 0.012&0.362 $\pm$ 0.015&0.413 $\pm$ 0.012&0.411 $\pm$ 0.009\\
			rcv1 (subset 3)&\textbf{0.460 $\pm$ 0.010}&0.366 $\pm$ 0.009&0.366 $\pm$ 0.009&0.353 $\pm$ 0.012&0.396 $\pm$ 0.013&0.360 $\pm$ 0.010&0.415 $\pm$ 0.011&0.409 $\pm$ 0.008\\
			bibtex&0.417 $\pm$ 0.015&0.390 $\pm$ 0.006&0.396 $\pm$ 0.007&0.371 $\pm$ 0.007&0.420 $\pm$ 0.004&0.396 $\pm$ 0.007&\textbf{0.424 $\pm$ 0.007}&\textbf{0.424 $\pm$ 0.002}\\
			corel16k001&\textbf{0.256 $\pm$ 0.007}&0.071 $\pm$ 0.003&0.070 $\pm$ 0.003&0.063 $\pm$ 0.008&0.137 $\pm$ 0.004&0.071 $\pm$ 0.004&0.050 $\pm$ 0.013&0.049 $\pm$ 0.003\\
			delicious&\textbf{0.231 $\pm$ 0.004}&0.194 $\pm$ 0.003&0.194 $\pm$ 0.003&0.191 $\pm$ 0.003&0.241 $\pm$ 0.002&0.172 $\pm$ 0.002&0.208 $\pm$ 0.011&0.226 $\pm$ 0.003\\
			mediamill&\textbf{0.581 $\pm$ 0.003}&0.547 $\pm$ 0.003&0.547 $\pm$ 0.003&0.539 $\pm$ 0.012&0.562 $\pm$ 0.003&0.543 $\pm$ 0.002&0.524 $\pm$ 0.012&0.520 $\pm$ 0.002\\
			bookmarks&\textbf{0.278 $\pm$ 0.008}&0.201 $\pm$ 0.003&-&0.192 $\pm$ 0.012&-&0.202 $\pm$ 0.002&0.180 $\pm$ 0.004&-\\
			\hline
		\end{tabular}
	}
	\label{Err:AccuExamFMacroFMicroF}
\end{table*}

Table~\ref{Err:AccuExamFMacroFMicroF} gives the comparative analysis of the proposed method GroPLE against state-of-the-art algorithms on  eleven data sets. We have conducted \textit{five-fold cross validation} and the $mean$ and $std$ is recorded. The best results among all the algorithms being compared are highlighted in boldface. For each data set, the number of latent dimension space $d$ is fixed to $100$ and the number of groups $K$ is set to $10$.  The regularization parameter $\lambda_1$ and $\lambda_2$ are fixed to $0.001$ and $1$, respectively, and the parameters $\alpha$ and $\beta$ are searched in the range given previously.

\begin{figure}[ht!]
	\begin{tabular}{ll}	
		\begin{minipage}{0.5\textwidth}
			\includegraphics[width=2.5in,height=1.8in]{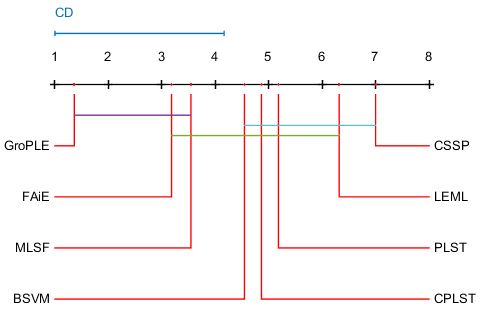}
			\captionof{subfigure}{(a) Accuracy}
			\label{fig:sub2_GroPLE}
		\end{minipage}
		& 
		\begin{minipage}{0.5\textwidth}
		\includegraphics[width=2.5in,height=1.8in]{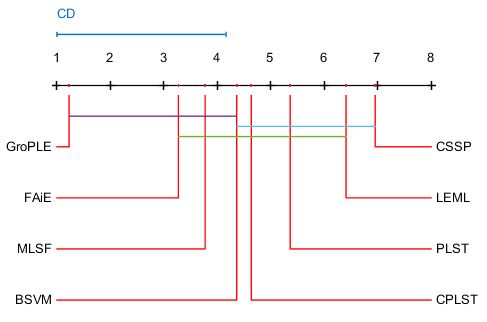}
		\captionof{subfigure}{(b) Example based F$\textsubscript{1}$ }
		\label{fig:sub4_GroPLE}
		\end{minipage}
		\\ 
		\begin{minipage}{0.5\textwidth}
			\includegraphics[width=2.5in,height=1.8in]{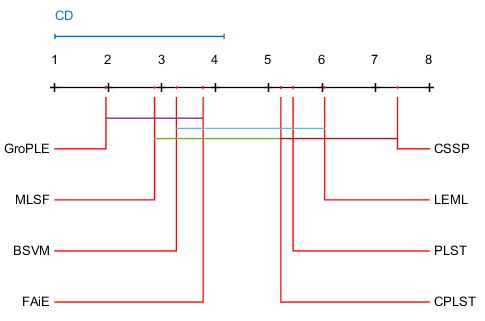}
			\captionof{subfigure}{(c) Macro F$\textsubscript{1}$ }
			\label{fig:sub5_GroPLE}%
		\end{minipage}
		&
		\begin{minipage}{0.5\textwidth}
		\includegraphics[width=2.5in,height=1.8in]{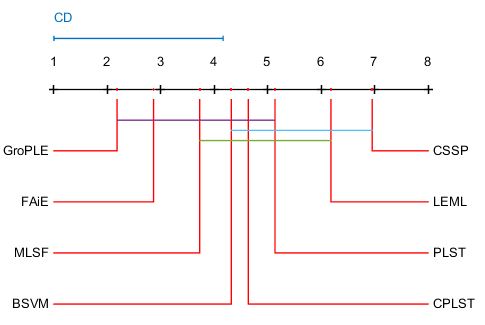}
		\captionof{subfigure}{(d) Micro F$\textsubscript{1}$ }
		\label{fig:sub6_GroPLE}
		\end{minipage}
	\end{tabular}
\caption{CD diagrams of the comparing algorithms under each evaluation criterion.}
\label{CDDiagramGroPLE}
\end{figure}

\begin{table}[ht]
	\centering
	\caption{Summary of the Friedman statistics $F_F(\mathcal{K}=8,\mathcal{N}=11)$ and the critical value in terms of each evaluation metric($\mathcal{K}$: \# Comparing Algorithms; $\mathcal{N}$: \# Data Sets).}
	\begin{tabular}{llc}
		\toprule
		Metric &$F_F$&Critical Value ($\alpha = 0.05$)\\
		\toprule
		Accuracy&11.692
		&\multirow{4}{*}{2.143}\\
		Example Base $F_1$&12.353&\\
		Macro $F_1$&12.686&\\
		Micro $F_1$&7.470&\\
		\bottomrule
	\end{tabular}
	\label{ffTest}
\end{table}
 To conduct statistical performance analysis among the algorithms being compared, we employed \textit{Friedman test}\footnote{Results of bookmarks data set is not included for this test.} which is a favorable statistical test for comparing more than two algorithms over multiple data sets~\cite{demvsar2006statistical}. Table \ref{ffTest} provides the \textit{Friedman statistics} $F_F$ and the corresponding critical value in terms of each evaluation metric. As shown in Table \ref{ffTest} at significance level $\alpha = 0.05$, Friedman test rejects the null hypothesis of equal performance for each evaluation metric. This leads to the use of post-hoc tests for pairwise comparisons. The Nemenyi test~\cite{demvsar2006statistical} is employed to test whether our proposed method GroPLE achieves a competitive performance against the algorithms  being compared. The performance of two classifiers is significantly different if the corresponding average ranks differ by at least the critical difference $CD = q_{\alpha}\sqrt{ \frac{\mathcal{K}(\mathcal{K}+1)}{6\mathcal{N}}}$. At significance level $\alpha = 0.05$, the value of $q_{\alpha}= 3.031$, for Nemenyi test with $\mathcal{K}=8$~\cite{demvsar2006statistical}, and thus $CD=3.166$. Figure~\ref{CDDiagramGroPLE} gives the CD diagrams~\cite{demvsar2006statistical} for each evaluation criterion, where the average rank of each comparing algorithm is marked along the axis (lower ranks to the left). It can be seen from the Figure~\ref{CDDiagramGroPLE} that the proposed method GroPLE achieve  better performance as compared to the other algorithms in terms of each evaluation metric.  
 \noindent

\section{Conclusions}
	\label{conclusion}
	This chapter presented a new multi-label classification method, called GroPLE, which embeds the original label vectors to  a low-rank space by retaining the group dependencies. We ensure that the labels belonging to the same group share the same sparsity pattern in their low-rank representations. In order to achieve the embedding of feature vectors, a linear mapping is then learnt that maps the feature vectors onto the same set of points which are obtained as a result of label embedding phase. We achieve this by a separate optimization problem.  Extensive comparative studies validate the effectiveness of GroPLE against the state-of-the-art multi-label learning approaches.
	
	In the future, it is interesting to see whether GroPLE can be further improved by considering side information from feature space while label embedding. Furthermore, designing other ways to fulfill the strategy of group formation and modeling group-specific label embedding is a direction worth studying.

 \newpage
\thispagestyle{empty}

\chapter{Conclusions and Future Work}

In this thesis, we focused on the development of novel techniques for collaborative filtering and multi-label classification, which are two important research areas in the domain of recommender systems and machine learning.  We show that the problem of collaborative filtering (CF) and multi-label classification (MLC) can be visualized as a matrix factorization (MF) problem. In Chapter~\ref{hmf_chapter} and Chapter~\ref{pmmmfChapter}, we discussed two novel approaches for CF namely, HMF (\textit{Hierarchical Matrix Factorization}) and PMMMF (\textit{Proximal Maximum Margin Matrix Factorization}), respectively. The basic factorization technique used in Chapters ~\ref{hmf_chapter} and~\ref{pmmmfChapter} is that of MMMF (\textit{Maximum Margin Matrix Factorization}) wherein a multi-class extension of the hinge loss function is proposed to handle the factorization of discrete ordinal rating matrix. Assuming MMMF to be matrix analog of ordinal regression under hinge loss, in Chapter~\ref{hmf_chapter}, we examined whether a set of bi-level maximum margin matrix factorizations can be used to solve matrix completion problem for multi-level ordinal rating matrix in more efficient and effective manner. We proposed a novel method termed as HMF for constructing a hierarchical two-class structure of binary matrix factorization to handle matrix completion of ordinal rating matrix. Our experimental results demonstrate the advantages of the proposed method over other matrix factorization based collaborative filtering techniques. 

It is more than a decade since MMMF has been proposed and till date, to the best of our knowledge, no other alternative criterion has been investigated for matrix factorization other than the maximum margin criterion. In Chapter~\ref{pmmmfChapter}, we proposed a different criterion, namely \textit{proximity}, motivated by the advent of Proximal SVMs (PSVMs) for binary classification, where two \textit{parallel} planes are generated  one for each class, unlike the standard SVMs. We conducted experiments on real and synthetic datasets to validate the effectiveness of PMMMF.

In the later part of the thesis we extended the concept of matrix factorization for multi-label classification and proposed two novel techniques namely, MLC-HMF (\textit{Multi-label Classification Using Hierarchical Embedding}) and  GroPLE (\textit{Group Preserving Label Embedding for Multi-Label Classification}) in Chapters~\ref{mlchmfChapter} and~\ref{GroPLEChapter}, respectively.
In Chapter~\ref{mlchmfChapter}, we developed a matrix factorization based feature space embedding method termed as MLC-HMF for multi-label classification.  MLC-HMF is an embedding based approach which learns  piecewise-linear embedding with a low-rank constraint on \linebreak parametrization to capture nonlinear intrinsic relationships that exists in the original feature and label space. We conducted extensive comparative studies which validate the effectiveness of the proposed method MLC-HMF. In Chapter~\ref{GroPLEChapter}, we assume that the input data form groups and as a result, the label matrix exhibits a sparsity pattern and hence the labels corresponding to objects in the same group have similar sparsity. We study the embedding of labels together with the group information with an objective to build an efficient multi-label classification. Extensive comparative studies validate the effectiveness of GroPLE against the state-of-the-art multi-label learning approaches.

There are many interesting directions in which the research work depicted in this thesis can be carried out in the future. As discussed earlier, HMF and PMMMF are novel matrix factorization techniques developed for collaborative filtering with ordinal preferences. HMF is a stage-wise matrix completion technique that makes use of several bi-level MMMFs in a hierarchical fashion. Investigating different ways in which a hierarchy can be built is one important direction that can be pursued. The core part of HMF is a binary loss function and studying the impact of different binary loss functions can be another direction to pursue future research. As discussed in Chapter~\ref{pmmmfChapter}, in every iteration of the gradient descent method PMMMF requires less number of variables to be updated as compared to that of MMMF. But on the other hand, the updation step in PMMMF takes more time because of the repeated computation of the \emph{mean} for obtaining the \emph{thresholds} for every user. To reduce the computation cost, one can think of a better approximation of the mean and the thresholds or even think of designing novel optimization techniques that best suit the optimization problem of PMMMF. We realize that there are more general problems of matrix factorization for matrices having discrete entries which are not ordered. Extending the concept of HMF and PMMMF for such matrices is another line of investigation that can be carried out in the future.

In Chapter~\ref{mlchmfChapter} and Chapter~\ref{GroPLEChapter}, we proposed two different embedding approach for multi-label classification namely, MLC-HMF and GroPLE, respectively. MLC-HMF is based on the embedding of feature space whereas GroPLE is a label space embedding approach. We developed novel matrix factorization techniques to achieve the embedding of feature and label space. It is worthwhile to carryout an in depth investigation of different ways of embedding such as quasilinear embeddings and their advantages over linear and non-linear embedding. One can plan to carryout this line of investigation in the future. Reducing the computational complexity of MLC-HMF by investigating the different ways of building an hierarchy or formation of group instances is another direction to pursue. In the future, it is interesting to see whether GroPLE can be further improved by considering the side information from the feature space while label embedding. Furthermore, designing other ways to fulfill the strategy of group formation and modeling group-specific label embedding is a direction worth studying.




%
\begin{singlespace}
  \addcontentsline{toc}{chapter}{REFERENCES}
 \bibliographystyle{plain}
   \bibliography{vikas_thesis}
\end{singlespace}






\end{document}